\documentclass[final]{elsarticle}
\usepackage[margin=2.5cm]{geometry}
\usepackage{rotating}
\usepackage{amsthm}
\usepackage{showkeys}
\usepackage{tabularx}
\usepackage{bbm}
\usepackage{siunitx}
\usepackage{algorithm}
\usepackage{algpseudocode}
\usepackage{subfig}
\usepackage{lineno}
\usepackage{upgreek}
\usepackage{bm}
\usepackage{times}
\usepackage{amsmath}
\usepackage{amsopn}
\usepackage{amssymb}
\usepackage{mathrsfs}
\usepackage{cancel}
\usepackage{mathtools}
\usepackage{xcolor}
\usepackage{natbib}
\usepackage[colorlinks]{hyperref} 
\hypersetup{ 
    colorlinks=true,       
    linkcolor=red,          
    citecolor=blue,        
    filecolor=magenta,      
    urlcolor=cyan           
}
\DeclareMathAlphabet{\mathpzc}{OT1}{pzc}{m}{it}
\newcommand{\Omegagb}{\Omega_0^{\rm{GB}}}
\newcommand{\Omegagbc}{\Omega_0 \backslash \Omegagb}
\newcommand{\wh}{\widehat}
\newcommand{\mcal}{\mathcal}
\newcommand{\BO}{\mathcal B^0}
\newcommand{\PO}{\mathcal P^0}

\newcommand{\F}{\bm F}
\newcommand{\ol}{\overline}
\newcommand{\Fe}{\bm F^{\rm L}}

\newcommand{\Jp}{J^{\rm P}}

\newcommand{\Ce}{\bm C^{\rm L}}

\newcommand{\FeT}{(\bm F^{\rm L})^{\rm T}}
\newcommand{\ReT}{\bm R^{\rm L T}}
\newcommand{\fp}{F^{\rm P}}
\newcommand{\Fp}{\bm F^{\rm P}}
\newcommand{\Rp}{\bm R^{\rm P}}
\newcommand{\Up}{\bm U^{\rm P}}
\newcommand{\up}{U^{\rm P}}
\newcommand{\Lp}{\bm L^{\rm P}}

\newcommand{\dphi}{\nabla \phi}

\newcommand{\invFp}{\bm F^{\rm{p-1}}}
\newcommand{\invfe}{F^{\rm{L-1}}}
\newcommand{\ue}{U^{\rm{L}}}
\newcommand{\invue}{U^{\rm{L-1}}}
\newcommand{\Ue}{\bm U^{\rm{L}}}
\newcommand{\invUe}{\bm U^{\rm{L-1}}}
\newcommand{\invUeT}{\bm U^{\rm{L-T}}}
\newcommand{\UeT}{\bm U^{\rm{LT}}}
\newcommand{\re}{R^{\rm{L}}}
\newcommand{\invre}{R^{\rm{L-1}}}
\newcommand{\invFe}{\bm F^{\rm{L-1}}}

\newcommand{\FpT}{\bm F^{\rm{pT}}}

\newcommand{\Ee}{\bm{\mathbb E}^{\rm L}}

\newcommand{\eref}[1]{(\ref{#1})}
\newcommand{\erefs}[2]{(\ref{#1})--(\ref{#2})}
\newcommand{\sref}[1]{Section~\ref{#1}}
\newcommand{\fref}[1]{Fig.~\ref{#1}}
\newcommand{\frefs}[2]{Figs.~\ref{#1}--\ref{#2}}

\newcommand{\Ph}{\bm{\mathfrak p}}

\newcommand{\psikwc}{\psi^{\rm{KWC}}}

\newtheorem{definition}{Definition}
 
\DeclareMathOperator{\Curl}{Curl}

\DeclareMathOperator{\Div}{Div}
\DeclareMathOperator{\Grad}{\nabla}
\DeclareMathOperator{\tr}{Tr}

\DeclareMathOperator{\curl}{curl}
\DeclareMathOperator{\ocurl}{\ol{curl}}
\modulolinenumbers[5]

\journal{Journal of Mechanics and Physics of Solids}

\begin{document}

\begin{frontmatter}

\title{A unified framework for polycrystal plasticity with grain boundary evolution}


\author[mseaddress]{Nikhil Chandra Admal\corref{mycorrespondingauthor}}
\cortext[mycorrespondingauthor]{Corresponding author}
\ead{admal002@g.ucla.com}

\author[meaddress]{Giacomo Po}

\author[mseaddress,meaddress]{Jaime Marian}

\address[mseaddress]{Materials Science and Engineering, University of California Los Angeles}
\address[meaddress]{Mechanical Engineering, University of California Los Angeles}

\begin{abstract}
    Plastic deformation in polycrystals is governed by the interplay between
    intra-granular slip and grain boundary-mediated plasticity. However, while
    the role played by bulk dislocations is relatively well-understood, the
    contribution of grain boundaries (GBs) has only recently begun to be studied.
    GB plasticity is known to play a key role along with bulk plasticity under a wide range of conditions, such as dynamic recovery,
    superplasticity, severe plastic deformation, etc., and developing models capable of
    simultaneously capturing GB and bulk plasticity has become a topic of high
    relevance.
    In this paper we develop a thermodynamically-consistent polycrystal plasticity
    model capable of simulating a variety of grain boundary-mediated plastic
    processes in conjunction with bulk dislocation slip. The model starts from the
    description of a single crystal and creates lattice strain-free polycrystalline
    configurations by using a specially-designed multiplicative decomposition developed by
    the authors. This leads to the introduction of a particular class of geometrically
    necessary dislocations (GND) that define fundamental GB features
    such as misorientation and inclination. The evolution of the system is based on
    an energy functional that uses a non-standard function of the GND tensor to
    account for the grain boundary energy, as well as for the standard elastic
    energy. Our implementation builds on smooth descriptions of GBs  inspired on
    diffuse-interface models of grain evolution for numerical convenience. We demonstrate
    the generality and potential of the methodology by simulating a wide variety of
    phenomena such as shear-induced GB sliding, coupled GB motion, curvature-induced
    grain rotation and shrinkage, and polygonization via dislocation sub-grain
    formation.
\end{abstract}


\end{frontmatter}


\section{Introduction}
\label{sec:intro}
Engineering materials, particularly metallic alloys, almost always involve
polycrystals, characterized by a conglomerate of grains with different crystal
orientations\footnote{With respect to a global \emph{laboratory frame of
reference}.} separated by grain boundaries (GBs). Deformation
models of polycrystalline materials must then necessarily account for grain boundary
plasticity, such as that produced by shear-induced grain boundary motion, grain rotation and
shrinkage, in addition to bulk or intra-grain plasticity. In fact, both classes of
mechanisms often co-occur, such that it is typically quite difficult to
distinguish between these in the general context of deformation. Never is this
more true than in the well-known cases of recovery, grain growth, and
recrystallization \citep{cotterill1976recrystallization,humphreys2012recrystallization}, which are commonplace processes taking place during
high-stress deformation of metallic alloys --particularly at elevated
temperature--, and where grain boundaries undergo microstructural
transformations contemporaneously with bulk dislocations \citep{van2002grain,evers2004scale}. 
Traditionally, however, both types of transformations have been modeled
separately, as independent processes that are then linked via some
phenomenological coupling (some of these will be discussed below). This has
proven unsatisfactory to capture the full complexity observed during
microstructural evolution at high stress and/or temperature. It thus urges to
take a fresh look at the current theories to explore new avenues to model
polycrystal plasticity with co-occurring grain boundary evolution. 

The advent of highly-accurate and efficient atomistic methods has enabled
the direct simulation of the mechanisms behind grain boundary evolution. Recent applications of
molecular dynamics (MD) simulations has opened a new window into the rich physics 
and complexities associated with grain boundary phenomena 
\citep{cahn2006,upmanyu2006simultaneous,mishin2010,trautt2012,molodov2015}. 
In addition, other methods such as the phase field crystal (PFC) model
\citep{elder2002modeling,elder2004modeling}, a continuum model
which operates at atomic length and diffusive time scales, has
been successfully used to simulate various mechanisms observed in grain boundary
plasticity \citep{yamanaka2017phase}. 
As well, discrete disclinations, a method pioneered by \citet{taupin2015}, has been used 
successfully to obtain the energetics of grain boundaries, although the kinetics
is still an open problem.

Evidently, the spatio-temporal limitations of atomistic, PFC,
and disclination-based models preclude us from studying recovery and
recrystallization on relevant timescales. Yet, new
models can and should benefit from the understanding gained over the last few
years from atomic-level simulations of grain boundaries structure and
properties. Another outstanding limitation of most mesoscale continuum models is that they either cover
the evolution of microstructure or deformation, but not both, lacking the
generality to model annealing and recovery phenomena.
For example, phase field models such as the
Kobayashi--Warren--Carter (KWC) model \citep{kobayashi2000,kobayashi1998},
or the \emph{multiphase} field model \citep{steinbach1999,krill2002}, and
cellular automata \citep{raabe2002} have been specifically
devised to study the kinetics of grain growth, but do not capture
deformation. On the other end of the spectrum are various crystal plasticity
(CP) models for polycrystals which have been successfully applied to
fundamental problems in materials deformation with a
high rate of accuracy \citep{gurtin1,gurtin2,ma2006consideration}. These models are geared towards modeling bulk plasticity, with
fixed grain boundaries playing a surrogate role of describing the variations in
slip planes and elastic moduli, and at times providing the necessary back stress
to resist the build up of bulk dislocations \citep{van2013,WULFINGHOFF201333}, or to incorporate
grain boundary sliding \citep{wei2004grain,gurtin2008nanocrystalline}. 
A notable exception to the above observation is a class of sharp-interface
models developed by
\citet{cahn2006,basak2015,frolov2012}, wherein grain boundaries
move and result in macroscopic deformation. Nevertheless, since these models are not
rooted in crystal plasticity they are less equipped to deal with the interaction
of bulk dislocations and grain boundaries, which plays a key role during the stages of recovery
and recrystallization.

Very interesting advances have been proposed recently by
coupling phase field/level set/cellular automata models to standard crystal plasticity
\citep{raabe2000coupling,abrivard2012a,abrivard2012b,bernacki2011,takaki2008,POPOVA201585} to include
microstructure evolution along with deformation. 
In these coupled models, the plastic strain or the stress are updated using CP and the
resulting dislocation density is passed to the grain boundary evolution model,
which uses it in a penalty function to steer grain boundary evolution to areas
of high accumulated plastic strain. Therefore, within this framework,
deformation is a result of only bulk crystal plasticity, and grain boundary
motion does not contribute to macroscopic deformation. This
implies phenomena such as shear-induced grain boundary
motion, grain sliding and subgrain nucleation are beyond the reach of such 
models.

While the power of these coupled 
formulations to simulate grain boundary kinetics along with deformation at
different levels of accuracy must be recognized, in this paper we propose a model that can
simulate bulk and grain boundary plasticity in unison. The model stems from a unifying framework developed recently by the authors \citep{admal:po:marian:2017}. Unlike in bulk
polycrystal plasticity, where the reference
configuration (represented by the vector $\bm X$) is a strain-free polycrystal (resulting in piecewise-constant slip
systems and elastic moduli) with initial state described as $\Fe=\Fp=\bm I$ (where $\Fe$ and $\Fp$ are the elastic and plastic part of the deformation gradient $\F$, and $\bm I$ is the identity tensor), in this
framework we begin with a strain-free single crystal as the reference
configuration, and grain boundaries emerge naturally as arrangements of geometrically necessary
dislocations (GNDs) that preserve the compatibility of the deformation. This is done using a special decomposition of $\bm
F(\bm X,0)\equiv \bm I$ by having $\Fp(\bm X,0)$
as a smoothened piecewise-constant rotation field and $\Fe=\FpT$.
The above construction results in a lattice strain-free diffuse-interface polycrystal as the initial
state, with grain misorientations arising due to the presence of GNDs. The
unique feature of this framework is that grain boundaries are \emph{not} viewed as a
new class of defects, but an integral part of the microstructure, thus making the interaction of bulk dislocation with grain
boundaries a tractable problem.

This notion of grain boundary GNDs is subsequently used to construct a system
free energy expressed as a non-standard function of the GND density along with the classical
elastic energy. This free energy is then used in a dissipative thermodynamic
framework that leads to evolution equations that can be integrated under a
variety of conditions representative of elementary grain boundary processes.  We
demonstrate the potential of this unified framework and the constitutive law by
simulating coupled grain boundary motion, grain sliding, rotation and
subgrain nucleation during dynamic recovery.

The paper is organized as follows: after this Introduction section we discuss
the phase field approach by \citet{kobayashi1998,kobayashi2000} in
\sref{sec:kwc}, which we use a
starting point to elaborate our theory. We then introduce the kinematic
framework to include grain boundary kinetics in \sref{sec:pc}, followed by the
derivation of balance laws in Sections \ref{sec:virtual} and \ref{sec:energy}. In
\sref{sec:cn}, we develop a thermodynamically-consistent constitutive law,
followed by results in \sref{sec:numerics}, where we discuss the numerical
aspects of the model and show how the model handles the various grain boundary
processes mentioned above, including polygonization. We finalize with a
discussion and the conclusions. We use
standard notation throughout the paper unless noted otherwise. \footnote{All
    scalar, tensor and vector fields in this paper are assumed
    to be infinitely smooth unless otherwise stated. 
    Bold letters are used to
    represent vectors and tensors. The derivative of a field $f(\bullet)$ with
    respect to its argument $\bullet$ is denoted by $f,_\bullet$. 
    $\nabla$ denotes the gradient operator, and its action on vector and scalar
    fields in indicial notation is defined as $[\nabla \bm v]_{ij}:= v_{i,j}$ and
    $[\nabla a]_i:=a,_i$ respectively.
    The divergence operator is denoted by $\Div$, and its action on vector and tensor
    fields in indicial notation (with Einstein summation convention) is defined as
$\Div \bm v := v_{i,i}$ and $[\Div \bm T]_i:= T_{ij,j}$ respectively. The volume
and surface elements in the reference configuration are denoted by $dV$ and $dA$
respectively.}

\section{The Kobayashi-Warren-Carter model for crystal grain evolution}
\label{sec:kwc}
The Kobayashi-Warren-Carter (KWC) model, proposed by \citet{kobayashi1998,kobayashi2000} is a phase-field
model to study grain evolution in
polycrystalline materials. The model for a two-dimensional polycrystal
$\Omega_0 \in \mathbb R^2$, consists of two scalar order parameters $\phi$ and
$\theta$, representing, respectively, phase state and crystal orientation.
$\phi$ ranges between 0 (disordered phase) and 1 (crystalline state), while
$\theta$ represents the orientation\footnote{Since the system is
    two-dimensional, a scalar is sufficient to represent the
orientation of a crystal.} of a crystal. The KWC free energy functional $\mathcal W^{\rm{KWC}}$ is given by
\begin{align}
    \mathcal W^{\rm{KWC}}[\phi,\theta] = \int_{\Omega_0} \psikwc \, dV.
\end{align}
with
\begin{align}
    \psikwc(\phi,\nabla \phi,\nabla \theta) = \frac{\alpha^2}{2}|\nabla \phi|^2 +
    f(\phi) + g(\phi) s |\nabla \theta| +
    \frac{\epsilon^2}{2}|\nabla \theta|^2,
    \label{eqn:response}
\end{align}
and 
\begin{align*}
f(\phi)&=e(\phi-1)^2, \\
g(\phi)&=\phi^2.
\end{align*}
The constants $e$, $\alpha$, $\epsilon$ and $s$ are material
constants. The Euler--Lagrange equations resulting from taking independent
variations of $\psikwc$ with respect to $\phi$ and $\theta$ are given by
\begin{subequations}
    \label{eqn:y_dot}
    \begin{align}
        b^\phi \dot \phi &= \alpha^2 \triangle \phi - f_{\phi} - g_{\phi} s
        |\nabla \theta|,
        \label{eqn:phi_dot}\\
        b^\theta \dot \theta &= \Div \left[ \epsilon^2 \nabla \theta + g s
        \frac{\nabla \theta}{|\nabla \theta|} \right], \label{eqn:theta_dot}
    \end{align}
\end{subequations}
where $b^\phi$ and $b^\theta$ are the inverse mobilities of the phase fields $\phi(\bm X)$
and $\theta(\bm X)$, respectively.
\citet{lobkovsky2001} have shown that the sharp-interface limit of equations
\eref{eqn:phi_dot} and \eref{eqn:theta_dot} gives rise to grain rotation and grain
boundary motion by curvature. 

The term $\nabla \theta/|\nabla\theta|$ makes \eref{eqn:theta_dot} a \emph{singular diffusive equation}\footnote{See \cite{kobayashi1999} for an excellent
introduction to singular diffusive equations}. This is a consequence of having the non-standard weighted total variation term $g(\phi) s|\nabla \theta|$ in \eref{eqn:response}. This term tends to localize the
grain boundary, while $\epsilon^2 |\nabla \theta|^2/2$ tends to diffuse it. 
The opposing nature of the two terms can be easily examined by studying the
steady state solution of a bicrystal (see \citet{kobayashi1999}). If $\epsilon=0$ and $s\ne0$, then the steady-state solution for $\theta(\bm X)$ is a step function, resulting in a
sharp-interface bicrystal. Conversely, for $s=0$
and $\epsilon \ne 0$, $\theta$ is a linear function in steady state. 
The two terms in $|\nabla \theta|$ therefore act together giving rise to grain boundaries with
finite width. In addition to grain boundary regularization, $\epsilon$ plays an
important role in the mobility of the grain boundary, as will be shown below.
\begin{figure}[t]
    \centering
    \subfloat{
        \includegraphics[scale=0.6]{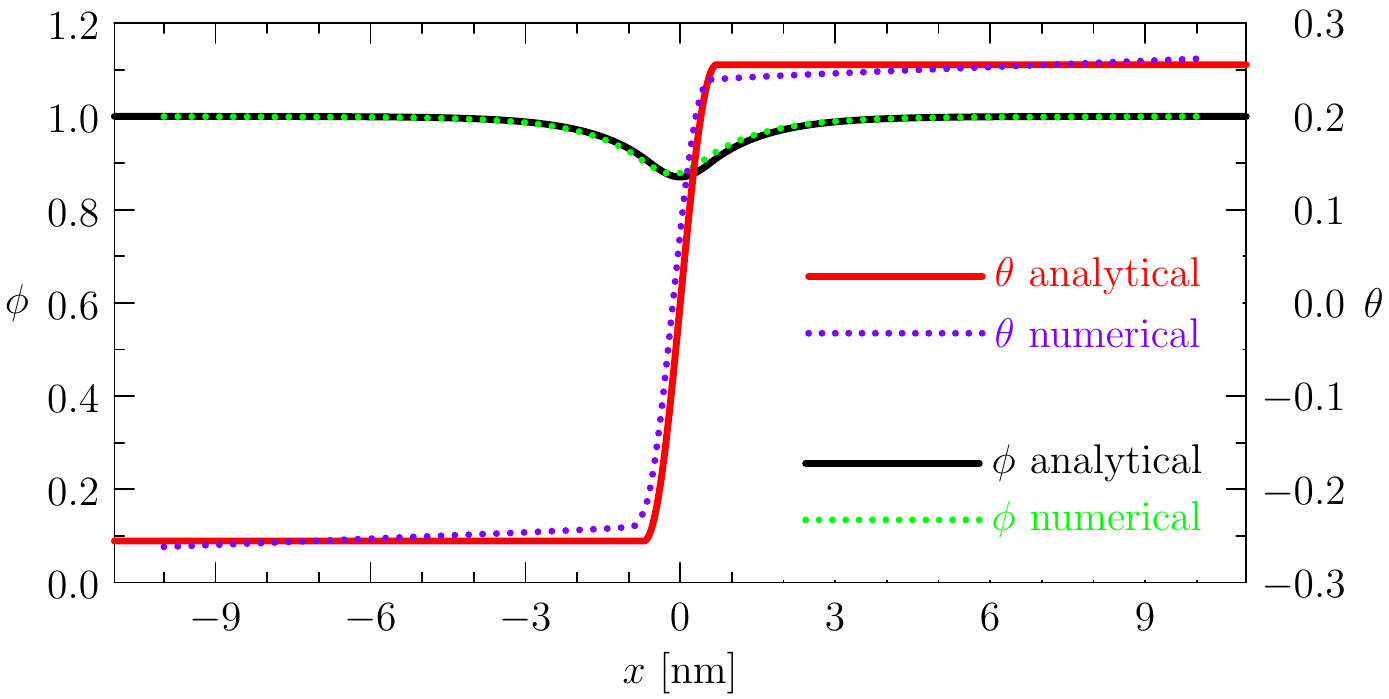}
        \label{fig:kwc_phi_theta}
    }
    \caption{Comparison of the steady state numerical solution to the steady
        state analytical solution given in \ref{sec:kwc_1d}.}
    \label{fig:kwc_phi}
\end{figure}
\begin{figure}
    \centering
    \subfloat{\includegraphics[scale=0.6]{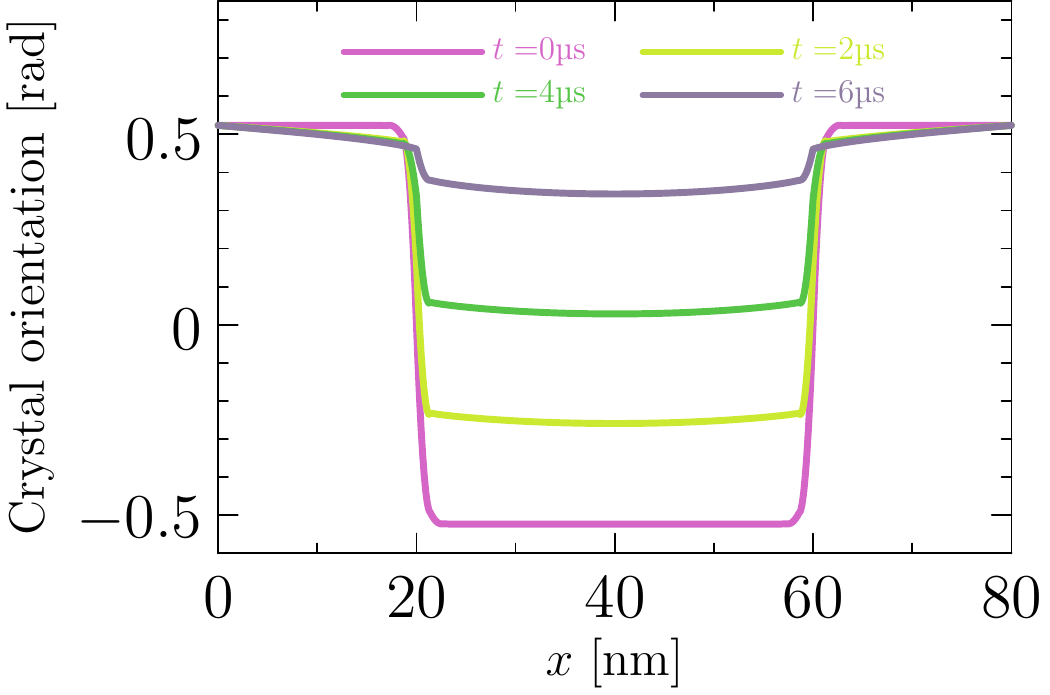}}
    \subfloat{\includegraphics[scale=0.6]{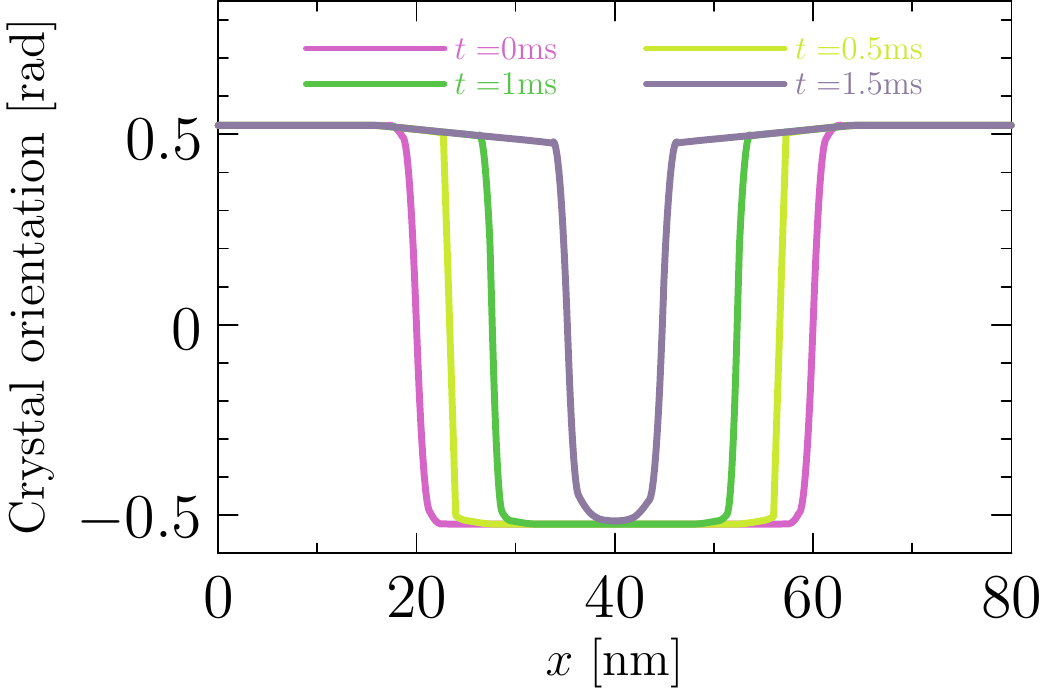}}
    \caption{Evolution of the $\theta$ field during rotation and shrinkage of a circular grain according to the KWC model.}
    \label{fig:kwc_rotshr}
\end{figure}

For the sake of numerical convenience, \citet{kobayashi1999} proposed to replace the singular term in \eref{eqn:theta_dot}, $|\nabla \theta|$, with the non-singular approximation:
\begin{align}
    p(|\nabla \theta|) := \frac{\ln(\cosh(\gamma |\nabla \theta|))}{\gamma}
    \label{eqn:approx}
\end{align}
where $\gamma$ is an adjustable parameter. This term converges to $|\nabla \theta|$ in the limit $\gamma \to \infty$. The resulting Euler-Lagrange equations are now given by
\begin{subequations}
    \label{eqn:y_dot_approx}
    \begin{align}
        b^\phi \dot \phi &= \alpha^2 \triangle \phi - f,_{\phi} - g,_{\phi} s p,
        \label{eqn:phi_dot_approx}\\
        b^\theta \dot \theta &= \Div \left[ \epsilon^2 \nabla \theta + g s
        p,_{|\nabla \theta|} \right].
        \label{eqn:theta_dot_approx}
    \end{align}
\end{subequations}
For a nominal parameter set, the above model results in the steady state
profiles given in \fref{fig:kwc_phi_theta} for $\phi$ and $\theta$. As the
figures show, within each grain $\phi=1$, which symbolizes perfect crystalline
order. $\phi$ drops in value at the grain boundary. For its part, $\theta$
behaves as a regularized step function, with the step height representing the
misorientation. An important aspect of the KWC model is that the evolution of
both phase field variables is linked due to the cross-term in
\eref{eqn:phi_dot}, which attaches the grain boundary to the evolution of both
variables. The numerical details behind these profiles as well as details about
the solution procedure can be found in \ref{sec:kwc_1d}.

One of the most attractive features of the KWC model is that its free energy functional allows for processes such as grain rotation and shrinkage. The two modes can be explored independently by appropriately
choosing the mobilities in eqs.~\ref{eqn:phi_dot} and \ref{eqn:theta_dot}. For example, 
for a circular grain of radius $\SI{20}{\nm}$ with a misorientation of $\ang{60}$ embedded inside a square
domain, the two processes are simulated in \fref{fig:kwc_rotshr}, with rotation manifesting itself as changes in misorientation --without grain boundary displacement--, and shrinkage as a constant misorientation with inward motion of the grain boundaries. 

Despite this great flexibility, the KWC free energy
functional is subjected to two fundamental limitations. 
The first is that it is completely phenomenological, without any connection to the underlying plastic mechanisms.
In second place, because the driving force for grain boundary motion modeled using the KWC
functional arises only due to its curvature, it is insensitive to external
stress. Therefore, the KWC functional cannot model phenomena such as shear-induced
grain boundary motion. The main aim of this paper is to devise a polycrystal
plasticity model that can address these shortcomings to study the general
response of polycrystalline systems to deformation.

\section{Kinematics of polycrystal plasticity}
\label{sec:pc}
The aim of this section is to present the kinematics behind a continuum polycrystal plasticity model
capable of simultaneously modeling bulk deformation and grain boundary evolution. 
First, in \sref{sec:kinematics}, we define the kinematic variables of bulk polycrystal plasticity with fixed grain boundaries and identify the challenges involved in their generalization to include grain boundary evolution. In \sref{sec:framework}, we discuss the
central idea of the paper in order to address the challenges of bulk polycrystal
plasticity by introducing an abstract kinematic framework of constructing grain
boundaries using geometrically necessary dislocations. Using the kinematic
framework introduced in \sref{sec:framework}, a thermodynamically-consistent
polycrystal plasticity model with grain boundary energy is developed in Sections
\ref{sec:virtual}, \ref{sec:energy} and \ref{sec:cn}.

\subsection{Kinematics of bulk polycrystal plasticity}
\label{sec:kinematics}
As is customary in continuum mechanics, a body is represented as an open subset
$\mathcal B$ of the three-dimensional Euclidean space $\mathbb R^3$. The
deformation of the body is described relative to a reference configuration $\BO
\in \mathbb R^3$. A point in $\BO$ is referred to as a \emph{material point},
and it is denoted by $\bm X$. A time-dependent deformation of the body is given
using a one-to-one deformation map $\bm y(\bm X,t)$ such that $\det \bm F \ne
0$, where
\begin{align}
    \bm F(\bm X,t):= \nabla \bm y,
\end{align}
where $\nabla$ is the gradient with respect to the material coordinate. A
central idea in the theory of crystal plasticity is the multiplicative
decomposition of the material deformation gradient into lattice and plastic
components \citep{lee1969elastic,kroner1959allgemeine,reina2014kinematic}, i.e.
\begin{align}
    \bm F = \Fe \Fp,
    \label{eqn:fefp}
\end{align}
where $\Fe$ denotes the lattice distortion, and $\Fp$ denotes the
lattice-invariant plastic distortion. In this paper, $\Fp$ represents
a lattice-invariant plastic shear associated to the mechanism of dislocation
slip. Note that unlike $\bm F$,  $\Fe$ and $\Fp$ are \emph{not} necessarily
gradients of vector fields. $\Fp$ maps the infinitesimally small material
element $d\bm X$ to $\Fp d\bm X$. The collection of distorted material elements
$\Fp d\bm X$ is denoted by lattice configuration. Therefore, $\Fp$ maps the
reference configuration to lattice configuration, and $\Fe$ maps the lattice
configuration to the deformed configuration.

Since $\Fp$ represents plastic distortion due to dislocation slip, its evolution
is given using the available families of dislocation slip planes $\bm s^\alpha$
and their normal $\bm m^\alpha$,\footnote{The vectors $\bm s^\alpha$ and $\bm
    m^\alpha$ exist in the lattice configuration.} along with the slip rates
$v^\alpha$. $\Fp$ is evolved using the flow rule and the initial condition
\begin{subequations}
\begin{align}
    \dot{\bm F}^{\rm P} &= \Lp \Fp, \label{eqn:flow} \\
    \Fp(\bm X,0) &= \bm I, \quad \bm X \in \BO
    \label{eqn:flow_initial}
\end{align}
\end{subequations}
respectively, where $\Lp$ is commonly referred to as the \emph{plastic velocity gradient}.
In single crystal plasticity, $\Lp$ is defined as the linear combination of slip contributions on all available slip systems, denoted by superindex $\alpha$, where each slip system is defined by a unique set of $\bm s^\alpha$ and $\bm m^\alpha$:
\begin{align}
    \Lp:=\sum_{\alpha=1}^A v^\alpha \bm s^\alpha \otimes \bm m^\alpha.
    \label{eqn:Lp}
\end{align}
The \emph{Schmid} tensor $\bm s^\alpha \otimes \bm m^\alpha$ projects the amount of slip on each system w.r.t.~the laboratory frame of reference.
For bulk polycrystal plasticity with fixed grain boundaries,
$\bm s^\alpha$ and $\bm m^\alpha$ in \eref{eqn:Lp} are replaced by $\bm
R^0 \bm s^\alpha$ and $\bm R^0 \bm m^\alpha$ respectively, where $\bm R^0$ is a
fixed piecewise constant rotation tensor describing the relative orientation of
each grain.\footnote{In other words, the slip direction and the slip plane normals
    are piecewise constant in bulk polycrystal plasticity.}

In single crystal plasticity, the elastic free energy density, denoted by $\psi$, is assumed to
be a function of the lattice Lagrangian strain
\begin{align}
    \Ee := (\FeT \Fe-\bm I)/2,
    \label{eqn:Ee}
\end{align}
while in polycrystal plasticity, $\psi(\Ee)$ is replaced by $\psi(\bm R^0 \Ee
\bm R^{0\rm T})$, reflecting the piecewise constant elastic response of each
grain. Therefore, bulk polycrystal plasticity is equivalent to single crystal
plasticity with save for the difference that the slip directions, slip plane normals, and the
elastic moduli are piecewise constant in the former. 

The primary aim of this paper is to generalize the above described bulk
polycrystal plasticity model to
include grain boundary evolution. We first note that an evolving grain
boundary may not only affect bulk deformation but it also
results in plastic distortion. For example, consider a circular grain 
embedded in a larger ambient grain, with a misorientation of $\Delta \theta$
which results in a nonzero grain boundary energy. One possible mechanism to
decrease the internal energy is for the circular grain to shrink thus decreasing
the grain boundary surface area. As the grain boundary sweeps through the
material, the lattice in the swept region rotates by an angle of $\Delta
\theta$, while the rest of the lattice remains unchanged. If $\Fp$
is equal to identity during this process, then this results in an incompatible
$\bm F$. This is a conclusive evidence that $\Fp \not \equiv \bm I$ in the swept
area. In other words, grain boundary motion always results in plastic distortion.
Next, we present our approach for modeling grain boundary evolution within the framework of polycrystal plasticity.

\subsection{An abstract kinematic framework to include grain boundary plasticity}
\label{sec:framework}
The central idea behind our approach is to define grain boundaries as dislocation arrangements with their own unique properties and densities that suffice to specify the key properties of each boundary.
One of the most important advantages of this idea is that plastic distortion due to grain boundary motion
emerges naturally from the original flow rule given in \eref{eqn:flow}, without the need to specify 
extra mechanisms to account for their contributions to plastic deformation. In \sref{sec:numerics} we show that this approach can indeed model phenomena such as shear-induced grain boundary motion, grain boundary
sliding and grain rotation. 

\begin{figure}[t]
    \centering
    \includegraphics[scale=0.3]{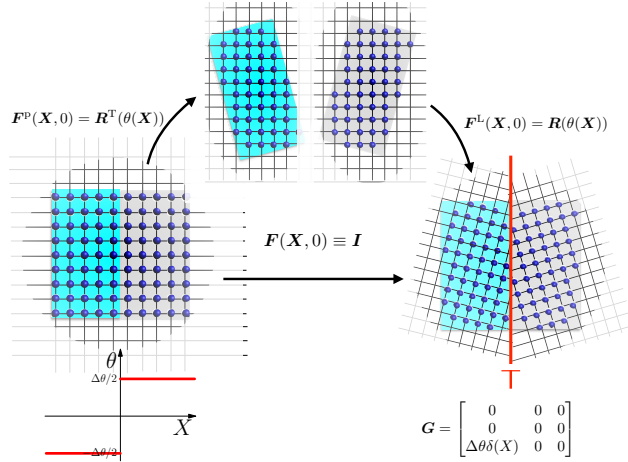}
       \caption{In two-dimensions, the above construction results in exactly two
           non-zero components ($G_{31}$ and $G_{32}$) of $\bm G$. In
           particular, for a symmetric tilt boundary oriented as
           shown above, $G_{32}\equiv0$ when $\theta$ is a step function. 
           }
    \label{fig:decomposition}
\end{figure}
We begin by noting that a polycrystalline material such as that discussed in
\sref{sec:kinematics} is modeled as a sharp-interface system with $\bm R^0(\bm X) \in SO(3)$,
a step function in the space of special orthogonal tensor fields, representing the
lattice rotation field in the polycrystal, with piecewise-constant values in
each grain. In addition, recall that in \sref{sec:kinematics} the reference
configuration is a lattice strain-free polycrystal with the initial state:
\begin{align}
    \bm F(\bm X,0) = \Fe(\bm X,0) = \Fp(\bm X,0) \equiv \bm I.
    \label{eqn:initial}
\end{align}
In the current framework, we start instead with a single crystal as
the reference configuration, and an initial state given by:
\begin{align}
\Fe(\bm X,0) = \bm R^0(\bm
X), && \Fp(\bm X,0) = \bm R^0(\bm X)^{\rm T},
\label{initFeFp}
\end{align}
resulting in 
\begin{align}
    \bm F(\bm X,0) \equiv \bm I.
    \label{initF}
\end{align}
\fref{fig:decomposition} shows the decomposition given in
eqs.\ \eqref{initFeFp} and \eqref{initF} for a single grain boundary in a bicrystal. Recall
that $\Fp$ deforms the material leaving the lattice fixed, while $\Fe$ deforms the lattice resulting in a total
deformation gradient $\bm F$ that is compatible. Comparing the reference and the
final configurations in \fref{fig:decomposition}, this assertion seems
contradictory, since the material is shown to be deformed even though $\bm F \equiv \bm I$.
We have resolved this apparent contradiction resorting to the notion of \emph{weak-convergence} discussed in our recent work
\citep{admal:po:marian:2017}, by which interpreting $\Fp(\bm X,0)=\bm R^{0\rm T}(\bm X)$ and $\bm F \equiv
\bm I$ for a discrete lattice in an average sense, as a sequence of lattice-invariant deformations, proves the correctness of eqs.\ \eqref{initFeFp} and \eqref{initF}.

The piecewise-constant field $\bm R^0(\bm X)$ results in a sharp-interface
polycrystal with surfaces at which $\bm R^0(\bm X)$ jumps signaling the existence of a grain
boundaries. Alternatively, these jumps can be replaced with a smoothened version $\widetilde{\bm R}^0(\bm X) \in
SO(3)$ of $\bm R^0$, resulting in a
\emph{diffuse-interface} grain boundary.\footnote{Note that the components of a
    $\widetilde{\bm R}^0 \in SO(3)$ are not obtained by regularizing the
    components of its piecewise-constant counterpart $\bm R^0(\bm X)$. Instead, it
    has to be constructed by appropriately regularizing the axis and angle
descriptors of $\bm R^0(\bm X)$.} The smoothening results in a more convenient
numerical implementation (see \citet{admal:po:marian:2017}) of the kinematic variables, as they can be discretized 
using continuous finite elements. In addition to numerical simplicity, we will
show in \sref{sec:cn} that a diffuse-grain boundary description enables us to
construct a grain boundary energy inspired from the KWC energy functional.

An important consequence of the decomposition given in eq.\ \eref{initFeFp} is that
the resulting sharp/diffuse-interface polycrystal is lattice strain-free since the
Lagrangian strain, defined in \eref{eqn:Ee}, is equal to zero.
Therefore, eq.~\eqref{initFeFp} describes a polycrystalline state which is
obtained from a reference single crystal by the right amount of slip in each
grain such that grains undergo relative rotation but the underlying lattice in
the polycrystal remains strain-free.

The main advantage of the initial kinematic state constructed using
\eref{initFeFp} is that we have immediate access
to the grain boundary dislocation density content in the form of the
\emph{geometrically necessary dislocation density} $\bm G$ tensor defined as
\begin{align}
    \bm G = \Fp \Curl \Fp,
    \label{eqn:G}
\end{align}
where $\Curl$ denotes the curl\footnote{
    The curl of a tensor field $\bm T$ is defined as
    \begin{align}
        (\Curl T)\bm n := \Curl (\bm T^{\rm T} \bm n),
        \notag
    \end{align}
    where $\bm n$ is an arbitrary constant vector, and the curl on the
    right-hand-side of the above equation is the
    curl of a vector field defined as $(\Curl v)_i = \epsilon_{ijk} v_{j,k}$,
    for any vector field $\bm v$. In indicial notation, it is given by $(\Curl \bm
    T)_{ij} = \epsilon_{ipq} T_{jq,p}$.
} 
of a tensor field with respect to the material/reference coordinate.
See \citet{acharya2008counterpoint} for a discussion on the physical
significance of the above definition, and other alternatives.
For a given normal $\bm n$ in the lattice configuration, the vector $\bm G^{\rm
T}\bm n$ measures the net Burgers vector of dislocation lines per unit area
passing through a plane of normal $\bm n$, in the lattice configuration. 
From the definition of $\bm G$ in \eref{eqn:G}, it is clear that for a
sharp-interface polycrystal, $\bm G$ is not a function but a distribution whose
support is the collection of grain boundary surfaces. For instance, the
bicrystal constructed in \fref{fig:decomposition} results in $G_{31}=\theta_0
\delta(\bm X)$, while all other components are identically equal to zero. On the
other hand, $G_{32}$ is not equal to zero for a flat diffuse grain boundary.
This can be seen using a smoothened step function $\widetilde \theta(X_1)$
describing the orientation of the two-dimensional bicrystal, and $\Fp = \bm R^{0\rm
T}(\widetilde \theta(X_1))$. The GND tensor evaluated using \eref{eqn:G}
results in 
\begin{align}
    G_{31} = -\cos(\widetilde\theta(X_1)) \widetilde \theta'(X_1), \quad 
    G_{32} = -\sin(\widetilde\theta(X_1)) \widetilde \theta'(X_1),
    \label{eqn:nonzeroG}
\end{align}
and rest of the components of $\bm G$ identically equal to zero.
Constructing grain boundaries using GNDs has the advantage that the evolution of
$\bm G$ not only tracks bulk dislocations but also sharp/diffuse grain boundaries.

In \sref{sec:numerics_shear}, we show that constructing grain boundaries using GNDs enables
us to simulate shear-induced grain boundary motion since
dislocations move under applied shear stress. In addition, we show in
\sref{sec:curvature} that grain boundary motion by curvature is also possible
within this framework by including grain boundary energy in the free energy
through a non-standard dependence on $\bm G$ and a phase field variable $\phi$
which has the same significance in this model as in the KWC model. The exact
functional dependence on $\bm G$ and $\phi$ will be made clear in \sref{sec:cn}
when we make the connection with the KWC energy functional. 

Summarising the kinematics of the polycrystal plasticity model, we have the
displacement $\bm u :=\bm y-\bm X$, slip rates $v^\alpha$ ($\alpha=1,\dots,A$), and $\phi$ as
the independent kinematic variables. Corresponding to these kinematic variables,
we introduce conjugate forces, power and a virtual power formulation in the
next section, and derive the necessary momentum balance laws.
\section{Virtual power formulation of the standard and microscopic force
balance}
\label{sec:virtual}
In this section, we develop a virtual power formulation for the kinematic
variables introduced in \sref{sec:pc} based on the framework developed by
\citet{gurtin2}. Let $\PO \subset \BO$ denote an arbitrary part of the body. The
formulation of the principle of virtual work is based on the balance between
external power
$\mathcal W(\PO)$ expended on $\PO$, and the internal power $\mathcal I(\PO)$ expended
within $\PO$. We assume that the internal power is expended by a stress $\bm P$
conjugate of $\F$, a stress vector $\Ph$ conjugate to $\dphi$, a scalar
internal microscopic force $\pi$ power-conjugate to $\dot \phi$, and for
each slip system $\alpha$, a scalar internal microscopic force $\Pi^\alpha$
power-conjugate to the slip $v^\alpha$, and a vector microscopic stress
$\bm \xi^\alpha$ power conjugate to $\nabla v^\alpha$. In other words, the
internal power expended within $\PO$ is given by
\begin{align}
    \mathcal I(\PO) = \int_{\PO} (\bm P \cdot \dot \F + \Ph \cdot
    \nabla \dot\phi + \pi \dot \phi)
    \, dV + \sum_\alpha \int_{\PO} (\Pi^\alpha v^\alpha + \bm \xi^\alpha \cdot \nabla
    v^\alpha) \, dV.
    \label{eqn:ip}
\end{align}
The external power on $\PO$ is assumed to be a result of body forces, and
various traction forces acting on $\partial \PO$. The structure of the work done
by external traction forces is obtained by examining the surface integrals
resulting from the expression for internal power given in \eref{eqn:ip}, i.e.
\begin{align}
    \mathcal I(\PO) &= \int_{\partial \PO} \left [\bm P \bm N \cdot \dot{ \bm y} + 
    \Ph \cdot \bm N \dot\phi + \sum_\alpha (\bm \xi^\alpha \cdot \bm N) v^\alpha \right
    ]\, dA + \notag \\
    &\quad\int_{\PO} \left [
        \dot \phi (\pi-\Div \Ph) - \dot{\bm y} \cdot \Div \bm P 
    \right ]
    \, dV + \sum_\alpha \int_{\PO} (\Pi^\alpha - \Div \bm \xi^\alpha) v^\alpha
    \, dV. \label{eqn:ip_split}
\end{align}
Note that the surface integrals in \eref{eqn:ip_split} consist of a macroscopic
surface traction that is conjugate to $\dot{\bm y}$, and two classes of microscopic
tractions that are conjugate to $\dot\phi$ and slip rates respectively.
This suggests the following form for the external power:
\begin{align}
   \mathcal W(\PO) &= \int_{\partial \PO} \left [
       \bm t(\bm N) \cdot \dot{\bm y} +
       s(\bm N) \dot\phi  +
       \Xi^\alpha(\bm N) v^\alpha
   \right ]\, dA 
\end{align}
where $\bm t$ is the macroscopic traction conjugate to $\dot{\bm
y}$, and $s$ and $\Xi^\alpha$ are microscopic tractions conjugate to
$\dot\phi$ and $v^\alpha$ respectively.

Taking independent variations in $\dot{\bm y}$, $\dot \phi$ and $v^\alpha$
$(\alpha=1,\dots,A)$, we arrive at the necessary macroscopic and microscopic
force balance equations.\\
\textit{\underline{Macroscopic force balance}}
\begin{subequations}
    \label{eqn:macroscopic}
    \begin{align}
        \Div \bm P &= \bm 0 \text{ in $\BO$}, \\
        \bm t &= \bm P \bm N \text{ on }\partial \BO.
    \end{align}
\end{subequations}
\textit{\underline{Microscopic force balance for each slip system
$\alpha=1,\dots,A$}}
\begin{subequations}
    \label{eqn:microscopic1}
    \begin{align}
        \Div \bm \xi^\alpha -\Pi^\alpha &= \bm 0 \text{ in $\BO$}, \\
        \Xi^\alpha &= \bm \xi^\alpha \cdot \bm N \text{ on $\partial \BO$}.
    \end{align}
\end{subequations}
\textit{\underline{Microscopic force balance for $\phi$}}
\begin{subequations}
    \label{eqn:microscopic2}
    \begin{align}
        \Div \Ph-\pi &= \bm 0 \text{ in $\BO$}, \\
        s &= \Ph \cdot \bm N \text{ on $\partial \BO$}.
    \end{align}
\end{subequations}

\section{Energy Balance}
\label{sec:energy}
In this section, we introduce the law of the balance of energy, and the second
law of thermodynamics expressed in the form of the Clausius--Duhem inequality.
Table~\ref{table:notation} list the notation for the various physical quantities
introduced in this section.
\begin{table}
    \centering
    \begin{tabular}[t]{r l}
        $\epsilon$    & energy density,\\
        $\bm q$       & heat flux vector,\\
        $r$           & external heat source,\\
        $\eta$        & entropy density,\\
        $T$           & absolute temperature,\\
        $\psi$        & $=\epsilon-T\eta$ free energy density
    \end{tabular}
    \caption{Notation introduced in \sref{sec:energy}}
    \label{table:notation}
\end{table}
All densities defined in Table~\ref{table:notation} are with respect to the reference volume.

Energy balance for an arbitrary subpart $\PO$ of the body is given by
\begin{align}
    \dot{\overline{\int_{\PO} \epsilon \, dV}} = 
    -\int_{\partial \PO} \bm q \cdot \bm n \, dA + 
    \int_{\PO} r \, dV + \mcal W(\PO),
\end{align}
where $\bm q$ denotes the heat flux vector, and $r$ denotes the external heat
source. By the principle of virtual work, we have $\mcal W(\PO)=\mcal I(\PO)$.
Therefore, substituting \eref{eqn:ip} into the above equation, we obtain the
following in a differential form:
\begin{align}
    \label{eqn:energy}
    \dot \epsilon = -\Div \bm q + r + 
    (\bm P \cdot \dot \F + \Ph \cdot \nabla \dot\phi) + \sum_\alpha (\Pi^\alpha
    v^\alpha + \bm \xi^\alpha \cdot \nabla
    v^\alpha)  + \pi \dot\phi.
\end{align}
It is convenient to express the term $\bm P \cdot \dot{\bm F}$, which is the power
expended due to deformation, in terms of the lattice Lagrangian strain $\Ee$ as
\begin{align}
    \bm P \cdot \dot{\bm F}  &=
    \bm S \cdot \dot{\mathbb E}^{\rm L} +
    \sum_{\alpha=1}^A (\psi,_{\Ee} \bm m^\alpha \cdot \Ce \bm s^\alpha)
    v^\alpha, \notag \\
    &=
    \bm S \cdot \dot{\mathbb E}^{\rm L} +
    \sum_{\alpha=1}^A \tau^\alpha v^\alpha,
\end{align}
where 
\begin{align}
    \bm S:= \invFe \bm P \FpT, \quad 
    \tau^\alpha:= \psi,_{\Ee} \bm m^\alpha \cdot \Ce \bm s^\alpha
\end{align}
are the lattice stress tensor and the resolved shear stress on the $\alpha$ slip
plane respectively.
The second law is written in the form of the Clausius--Duhem inequality
for an arbitrary subpart $\PO$ as
\begin{align}
    \dot{\overline{\int_{\PO} \eta \, dV}} \ge 
    -\int_{\partial \PO} \frac{\bm q \cdot \bm n}{T} \, dA + 
    \int_{\PO} \frac{r}{T}\, dV,
\end{align}
where $\eta(\bm X,t)$ and $T(\bm X,t)$ denote the entropy density and temperature fields
respectively. This implies
\begin{align}
    \dot \eta \ge -\Div \left (\frac{\bm q}{T} \right ) + \frac{r}{T}.
\end{align}
Multiplying the above equation by $T$ results in 
\begin{align}
    T \dot \eta &\ge -\Div \bm q + \frac{\bm q \cdot \Grad T}{T} + r \notag \\
    &= \dot \epsilon -(\bm S \cdot \dot \Ee + \Ph \cdot
    \nabla \dot \phi) - \sum_\alpha (\Pi^\alpha v^\alpha + \tau^\alpha v^\alpha+\bm \xi^\alpha \cdot \nabla
    v^\alpha) - \pi \dot \phi+ \frac{\bm q \cdot \Grad T}{T},
\end{align}
where to arrive at the last equality, we replaced $\Div \bm q$ using the
relation given in \eref{eqn:energy}.
Introducing the free energy density $\psi = \epsilon - T\eta$ into the above
equation, we obtain the following dissipation inequality:
\begin{align}
    \dot \psi + \dot T \eta -(\bm S \cdot \dot \Ee + \Ph \cdot
    \nabla \dot \phi) - \sum_\alpha (\Pi^\alpha v^\alpha + \tau^\alpha v^\alpha+\bm \xi^\alpha \cdot \nabla
    v^\alpha) - \pi \dot\phi + \frac{\bm q \cdot \Grad T}{T} \le 0.
    \label{eqn:dissipation}
\end{align}

\section{Constitutive equations and the Coleman--Noll procedure}
\label{sec:cn}
In this section, we arrive at
thermodynamically-consistent constitutive laws that connect the forces
introduced in \sref{sec:virtual} to the kinematic variables introduced in
\sref{sec:kinematics} using the Coleman--Noll procedure. The guiding principle
here is to include grain boundary
energy in addition to the bulk elastic energy into the free energy density
$\psi$ through its dependence on $s=(T,\nabla T,\Ee,\bm G,\phi,\dot \phi,\nabla
\phi)$, $\bm v = (v^1,\dots,v^\alpha)$ and $\nabla \bm v = (\nabla
v^1,\dots,\nabla v^\alpha)$. We begin with the following constitutive
assumptions:
\begin{align}       
    \psi &= \wh \psi(s,\bm v,\nabla \bm v),
\end{align}
and the fields $\eta$, $\bm q$, $\bm \xi^\alpha$, $\pi$, $\Pi^\alpha$ and $\bm S$
are assumed to be functions of $s$, $\bm v$, $\nabla \bm v$ and $\Fp$.
The dependence of the free energy on $\Ee$ and $\bm G$ instead of $\Fe$, $\Fp$
or $\nabla \Fp$ is a result of the frame-invariance of $\psi$
\citep{berdichevsky2006continuum}.

Using the Coleman--Noll procedure, demonstrated in \ref{sec:cn_app}, we arrive
at the following restrictions on the above functional forms that make them
thermodynamically-consistent. First, $\psi$ does not depend
on $\nabla T$, $\dot \phi$, $\bm v$ or $\nabla \bm v$, and 
\begin{subequations}
    \label{eqn:relations}
    \begin{align}
        \eta &= -\psi,_T,\label{eqn:relations1} \\
        \bm S&= \psi,_{\Ee}, \label{eqn:relations2}\\
        \Ph&= \psi,_{\nabla \phi}. \label{eqn:relations3}
    \end{align}
\end{subequations}
Second, the microscopic stress $\bm \xi^\alpha$ consists of an energy part $\bm
\xi^\alpha_{\rm{en}}$, and a dissipative part $\bm \xi^\alpha_{\rm d}$, i.e.
\begin{align}
    \bm \xi^\alpha = \bm \xi^\alpha_{\rm{en}} + \bm \xi^\alpha_{\rm{d}}.
    \label{eqn:xi_split}
\end{align}
$\bm \xi^\alpha_{\rm{en}}$ can be interpreted as the distributed Peach--Koehler force
due to the pile up of dislocations, and it is given by
\begin{align}
    \bm \xi ^\alpha_{\rm{en}} = \Jp (\Fp)^{-1} (\bm m^\alpha \wedge
    \psi,_{\bm G} \bm s^\alpha),
    \label{eqn:xi_relation}
\end{align}
where $\Jp := \det \Fp$. The term $\bm \xi^\alpha_{\rm d}$ is the
dissipative microstress conjugate to the gradient in slip rate, and given by
\begin{align}
    \bm \xi^\alpha_{\rm d} &= B^\alpha \nabla v^\alpha, \label{eqn:xi_d}
\end{align}
where $B^\alpha$ is a positive-valued inverse mobility associated to $\nabla \bm v$.
Finally, the scalar internal microforces $\Pi^\alpha$ and $\pi^\alpha$, and the
heat flux vector $\bm q$ are given by
\begin{subequations}
    \begin{align}
        \Pi^\alpha &= \psi,_{\bm G} : (\mathbb S^\alpha \bm G + \bm G
        \mathbb S^{\alpha\rm T})-\tau^\alpha+b^\alpha(s,\bm v,\nabla \bm v)
        v^\alpha,\label{eqn:dissipative_relations1}\\
        \bm q &= - \bm K(s,\bm v,\nabla \bm v) \nabla T,
        \label{eqn:dissipative_relations2}\\
        \pi &= \psi,_\phi + b^\phi(s,\bm v,\nabla \bm v) \dot
        \phi,\label{eqn:dissipative_relations3}
    \end{align}
    \label{eqn:dissipative_relations}
\end{subequations}
where the functions $b^\alpha$ and $b^\phi$ are positive-valued inverse
mobilities associated to the slip rate $v^\alpha$ and $\dot \phi$ respectively,
and $\bm K$ is the thermal conductivity tensor. Summarising, we have now
expressed all forces appearing in the governing equations
\ref{eqn:macroscopic}, \ref{eqn:microscopic1}, \ref{eqn:microscopic2} and
\ref{eqn:energy} in terms of the kinematic variables, temperature and its
gradient.
               
\subsection{Bulk and grain boundary energies}
In this section, we construct an explicit free energy density 
$\psi(T,\Ee,\bm G,\phi,\nabla \phi)$ such that it includes temperature-dependent
bulk elastic and grain boundary energies. We assume that $\psi$ is additively
decomposed into bulk elastic and grain boundary energy densities, given by 
\begin{align}
    \psi=\psi^{\rm b}(T,\Ee) + \psi^{\rm{gb}}(T,\phi,\nabla \phi,\bm G),
\end{align}
where we assume that $\psi^{\rm b}$ is the classical
elastic energy that depends only on $T$ and $\Ee$, while $\psi^{\rm{gb}}$ is
independent of $\Ee$. For $\psi^{\rm{gb}}$, we use the following polyconvex
energy density for isotropic materials proposed by \citet{ciarlet1982},
\begin{align}
    \psi^{\rm{b}} = &a(3+2\tr(\Ee)) + b(3+4\tr(\Ee)+2\tr^2(\Ee)-2\tr((\Ee)^2))
    +  \notag \\
    &c \det(2\Ee+\bm I) - \frac{1}{2} d \log \det (2\Ee+\bm I) - (3a+3b+c),
    \label{eqn:psi_el}
\end{align}
where the constants $a$, $b$, $c$ and $d$ are expressed in terms of the
isotropic materials' Lam\'{e} constants $\lambda$ and $\mu$ as.
\begin{subequations}
    \label{eqn:psi_el_constants}
    \begin{align}
        a &= \mu + \frac{1}{2}\left (-\mu-\frac{\lambda}{4}\right), \quad
        b = -\frac{\mu}{2}-\frac{1}{2}\left (-\mu-\frac{\lambda}{4}\right ),\\
        c &= \frac{\lambda}{8}, \quad 
        d = \mu + \frac{\lambda}{2}.
    \end{align}
\end{subequations}
The construction of
$\psi^{\rm{gb}}$ is the most non-trivial part of the constitutive law. Under the
kinematic framework developed in \sref{sec:kinematics}, since the initial
distribution of GNDs describe grain boundaries, it is natural to develop a grain
boundary energy density that is a function of $\bm G$. Moreover, we expect that
a steady state solution yields a grain boundary of finite width, a feature
central to the KWC model. Therefore, the
construction of $\psi^{\rm{gb}}$ is 
inspired by the KWC energy functional given in
\eref{eqn:response}. Recall that the order parameter $\theta$ in the KWC energy
functional describes the orientation of the lattice. Therefore, we intend to
construct $\psi^{\rm{gb}}$ by replacing $\nabla \theta$ in \eref{eqn:response}
with the gradient of the lattice orientation. It is well-known that the GND
tensor $\bm G$ describes the gradient of lattice rotations under
the assumption of small strain gradients \citep{nye1953some}. 
To the best of our knowledge, the only relation that connects the gradient of
lattice rotation to the GND tensor and the lattice strain exists under geometric
linearity, attributed to \citet{kroner1981continuum}. 
In \ref{sec:theorem}, we show under a geometrically nonlinear setting, that the
gradient of lattice rotation can be expressed in terms of the GND tensor, the
lattice stretch tensor, and its gradient as
\begin{align}
    \ReT (\curl \ReT) = (J^{\rm L})^{-1} \left (\Ue \bm G \UeT + \Ue
    \ocurl \Ue \right ), 
    \label{eqn:decompose}
\end{align}
where $\ocurl$ and $\curl$ denote the curl operators with respect to the lattice
and deformed configurations respectively. 

The term $\ReT (\curl \ReT)$ in \eref{eqn:decompose}, which describes
the gradient of lattice rotation, qualifies to replace $\nabla \theta$ in the KWC
energy density due to its frame-indifference.\footnote{Note that the term $\ReT(\curl \ReT)$ maps the
lattice configuration to itself, and since the vectors and tensors
    defined in the lattice configuration are frame-indifferent, the resulting
    $\psi^{\rm{gb}}$ is frame-invariant.}
The resulting $\psi^{\rm{gb}}$ is a function of $\bm G$, $\Ue$ and its gradient,
yielding a lattice strain
gradient model.\footnote{Note the distinction between a lattice strain gradient
    theory and the more commonly used theory of ``strain gradient plasticity'', where
gradient in the latter refers to the gradient in $\Fp$.}
In this paper, we do not pursue such a model in the interest
of computational simplicity.\footnote{A lattice strain gradient theory of grain
    boundaries would involve modifying the principle of virtual work, stated in
    \sref{sec:virtual}, to include power expended due to the kinematic variable
$\nabla \bm F$ and its corresponding hyperstress.}  
Instead, we construct $\psi^{\rm {gb}}$ by replacing $\nabla \theta$
in \eref{eqn:response} with $\bm G$, i.e. 
\begin{align}
    \psi^{\rm{gb}}(T,\phi,\nabla \phi,\bm G) = \frac{\alpha^2}{2}|\nabla \phi|^2 +
    f(\phi) + g(\phi) s |\bm G| + \frac{\epsilon^2}{2}|\bm G|^2,
    \label{eqn:psi_gb}
\end{align}
where $\alpha$ and $\epsilon$ are functions of $T$. In \sref{sec:curvature}, we
show that although \eref{eqn:psi_gb} is not an exact analog
of the KWC energy density, it results in the intended grain boundary motion by
curvature. Moreover, the dependence of $\psi^{\rm{gb}}$ on the norm of $\bm G$
results in a free energy density that depends only on the misorientation, and not on
the inclination of a grain boundary. We defer to future work any generalization to include
the dependence of free energy on inclination.

Recall that the kinematics described in \sref{sec:kinematics} enables us to
construct lattice strain-free sharp- or diffuse-interface grain boundaries. It is
clear from \eref{eqn:psi_gb} that $\psi^{\rm{gb}}$ for a sharp-interface grain
boundary is infinite. Therefore, similar to the KWC model, we expect that the
steady state solution to the governing equations result in a finite grain
boundary thickness which depends on the parameters $\epsilon$ and $\alpha$.

\section{Results}
\label{sec:numerics}
\begin{table}[t]
\begin{tabularx}{\linewidth}{|X|X|}
   \hline
   {\begin{align*}
       &\Div\bm P = \bm 0 \text{ in $\BO$} \\
       &\text{with Dirichlet boundary condition on $\bm u$}
   \end{align*}}
   &
   {\begin{align*}
       \left.
       \begin{array}{rl}
           \Div \bm \xi^\alpha -\Pi^\alpha &= 0 \text{ in $\BO$} \\
           v^\alpha(\partial \BO,t) &\equiv 0
       \end{array}
       \right \}
       \alpha = 1,\dots,A
   \end{align*}}
   \\
   \hline
   {\begin{align*}
       \Div \Ph-\pi &= 0 \text{ in $\BO$} \\
       \phi(\BO,0) & \equiv 1 \\
       \phi(\partial \BO,t) & \equiv 1
   \end{align*}}
   &
   {\begin{align*}
       \dot{\bm F}^{\rm P} &= \Lp \Fp \text{ in $\BO$}\\
       \Fp(\cdot,0) &= \bm R^{0\rm T} \text{ in $\BO$}
   \end{align*}}\\
   \hline
\end{tabularx}
\caption{The governing equations for the Dirichlet BVP for the unknown kinematic
   variables $\bm u$, $v^\alpha$ ($\alpha=1,\dots,A$), $\phi$ and $\Fp$.
   The forces $\bm P$, $\Ph$, $\bm \xi^\alpha$, $\Pi$ and $\pi$ are
   expressed in terms of the kinematic variables through the
   relations given in \erefs{eqn:relations}{eqn:dissipative_relations}, with
   $\Fe=\bm F \invFp$. The short-hand argument $(\partial
   \BO,t)$ is used to described a function on $\partial \BO$ at time $t$.}
\label{table:ge}
\end{table}
The goal
of this section is to demonstrate that the polycrystal plasticity model presented in this paper can
simulate the four elementary grain boundary processes: (i) grain boundary sliding and coupled
motions, (ii) grain rotation, (iii) grain shrinkage, and (iv) interactions between bulk
dislocations and grain boundaries. Although the framework developed in this paper is
applicable to an arbitrary polycrystal in any dimension, we limit our numerical
study to bicrystals in one and two dimensions since our primary focus at this
stage is to demonstrate the validity of the model rather than apply to study
problems in crystal plasticity. In addition, we do not solve the energy balance
equation in this numerical study as all simulations are performed at a constant
temperature.

The Lam\'{e} constants entering the bulk elastic energy density (see
 \eref{eqn:psi_el} and \eref{eqn:psi_el_constants}) are
taken as $\lambda=\SI{9.515e-2}{\femto\joule \per \nm \cubed}$ and
$\mu=\SI{4.477e-2}{\femto\joule \per \nm \cubed}$. The presence of the
linear term $g(\phi) s|\bm G|$ in \eref{eqn:psi_gb}, results in a singular
diffusive term in the microscopic balance equation \eref{eqn:microscopic1}.
Therefore, similar to the KWC model, a numerical implementation of the model
warrants approximating $|\bm G|$ in $g(\phi) s |\bm G|$ with $p(\bm G)$, where
the function $p$ is defined in \eref{eqn:approx}. The material parameters
corresponding to the grain boundary energy density are identical to those used to
simulate the KWC model (see Table \ref{table:kwc}), unless otherwise stated. The
parameter $B^\alpha$ associated with the dissipative microstress $\bm
\xi^\alpha_{\rm d}$ (see \eref{eqn:xi_d}) is equal to
$\SI{1}{\femto\joule\second \per \meter}$ for all the simulations.
We assume that plastic distortion evolves due to the presence of four slip
systems, i.e. $A=4$ in all the simulations. The slip directions  are taken as
\begin{subequations}
    \label{eqn:s}
    \begin{align}
        \bm s^1 &= (1,0),\label{eqn:s1} \\
        \bm s^2 &= (0,1),\label{eqn:s2} \\
        \bm s^3 &= \frac{1}{\sqrt 2}(1,1),\label{eqn:s3} \\
        \bm s^4 &=\frac{1}{\sqrt 2}(-1,1),\label{eqn:s4}
    \end{align}
\end{subequations}
with the corresponding normals perpendicular to the slip direction and the
out-of-plane dislocation line direction. In simulations where we want only a
subset of the above-mentioned slip systems, we deactivate the remaining slip
systems by decreasing the corresponding mobilities.

The initial bicrystal is constructed using a step function $\theta(\bm X)$
representing the orientation of the crystal lattice. As described in \sref{sec:kinematics}, a
diffuse-interface polycrystal is generated by regularizing $\theta(\bm X)$ into
a smooth function
$\widetilde\theta(\bm X)$, and starting with the initial condition $\Fp(\bm X,0)
= \widetilde{\bm R}^{0\rm T}(\bm X)$, where $\widetilde{\bm R}^0$ is the smooth
field in $SO(3)$ corresponding to $\widetilde \theta(\bm X)$.
The governing equations with
Dirichlet boundary conditions, listed in Table \ref{table:ge}, are
numerically solved for the unknowns $\bm u$, $v^\alpha$
($\alpha=1,\dots,A$), $\phi$ and $\Fp$ using the finite element method. 

The three displacement variables $u_1$, $u_2$ and $u_3$, the four slips $v_i$
($i=1,\dots,4$), and the order parameter
$\phi$ are interpolated using the Lagrange quadratic finite elements.
Since $\Fp$ is a smooth rotation field at $t=0$, it satisfies the orthogonality
condition $\FpT \Fp \equiv \bm I$. However, because a Lagrange finite element
interpolation of $\Fp$ does not satisfy such condition, we
express $\Fp$ using its polar decomposition $\Fp=\Rp\Up$, where $\Rp \in SO(3)$,
and $\Up$ is the positive-definite symmetric stretch tensor. Moreover,
since we are limiting ourselves to at most two dimensions, $\Rp$ is a function
of a single variable $\theta^{\rm P}$. Using the above representation, $\Fp$ is
interpolated using the Lagrange quadratic finite element interpolation of
$\theta^{\rm P}$, $\up_{11}$, $\up_{12}$ and $\up_{22}$. This guarantees the
interpolant of $\Fp(\bm X,0)$ to be in $SO(3)$.

The system of equations listed
in Table \ref{table:ge} is evolved in a segregated manner using the MUMPS
direct solver, and BDF (Backward Differential Formula) time stepping algorithm
implemented in \texttt{COMSOL 5.2}.

\subsection{Steady state solution of a flat grain boundary}
\label{sec:numerics_ss}
\begin{figure}[t]
    \centering
    \subfloat[Plastic distortion] {
        \includegraphics[scale=0.55]{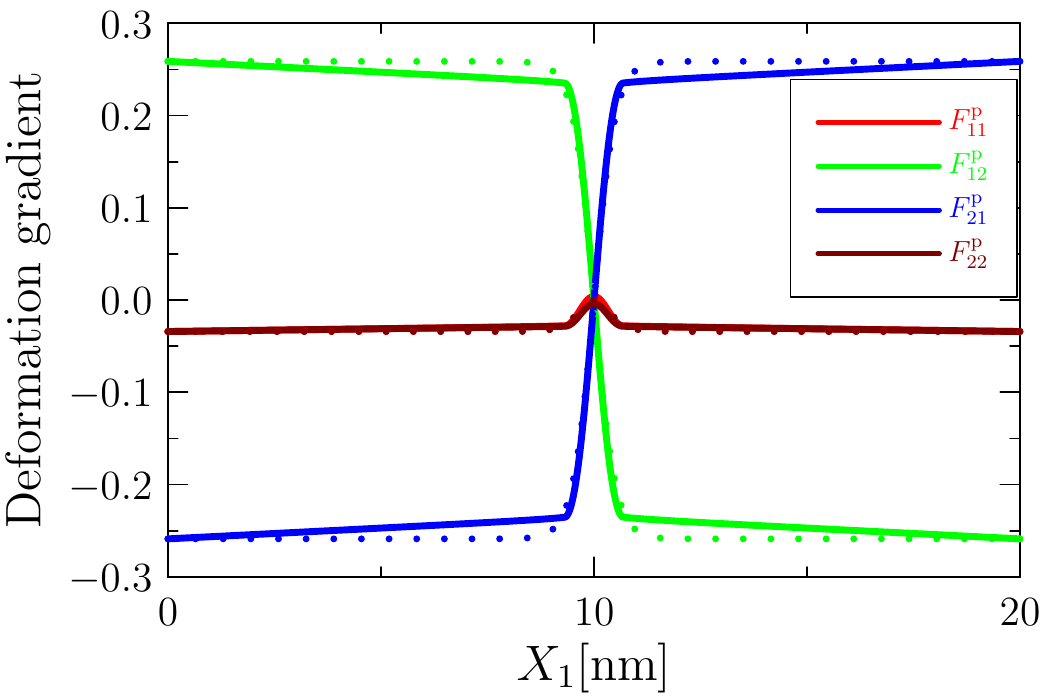}
        \label{fig:Fp}
    }
    \subfloat[Dislocation density] {
        \includegraphics[scale=0.55]{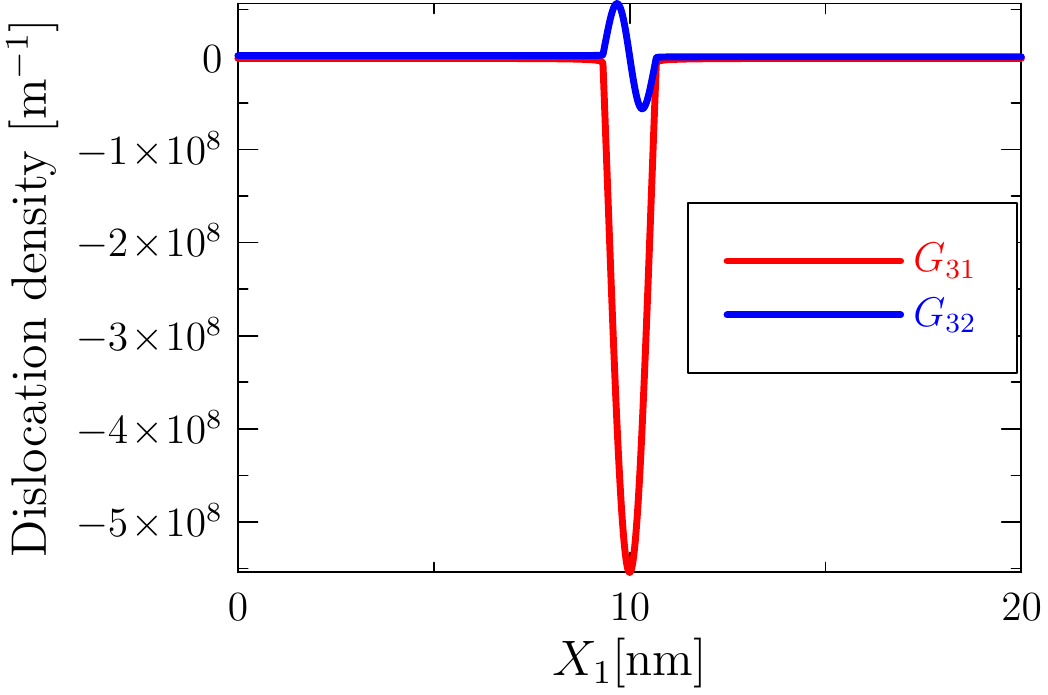}
        \label{fig:G}
    }\\
    \subfloat[Lattice Lagrangian strain] {
        \includegraphics[scale=0.55]{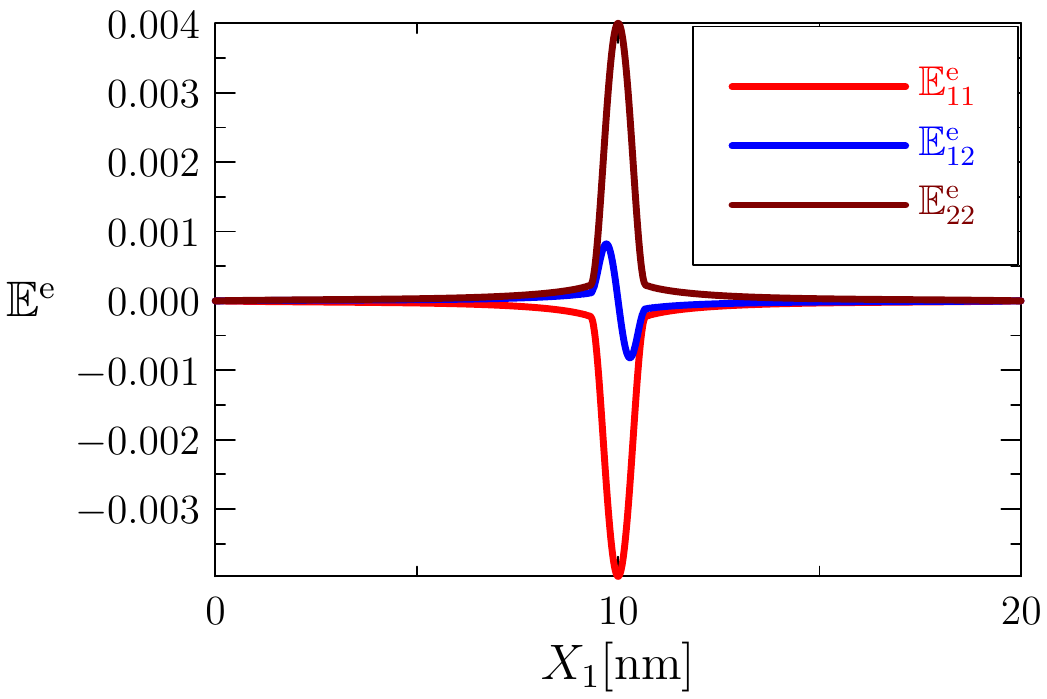}
        \label{fig:Ee}
    }
    \subfloat[Constributions to the total energy density] {
        \includegraphics[scale=0.55]{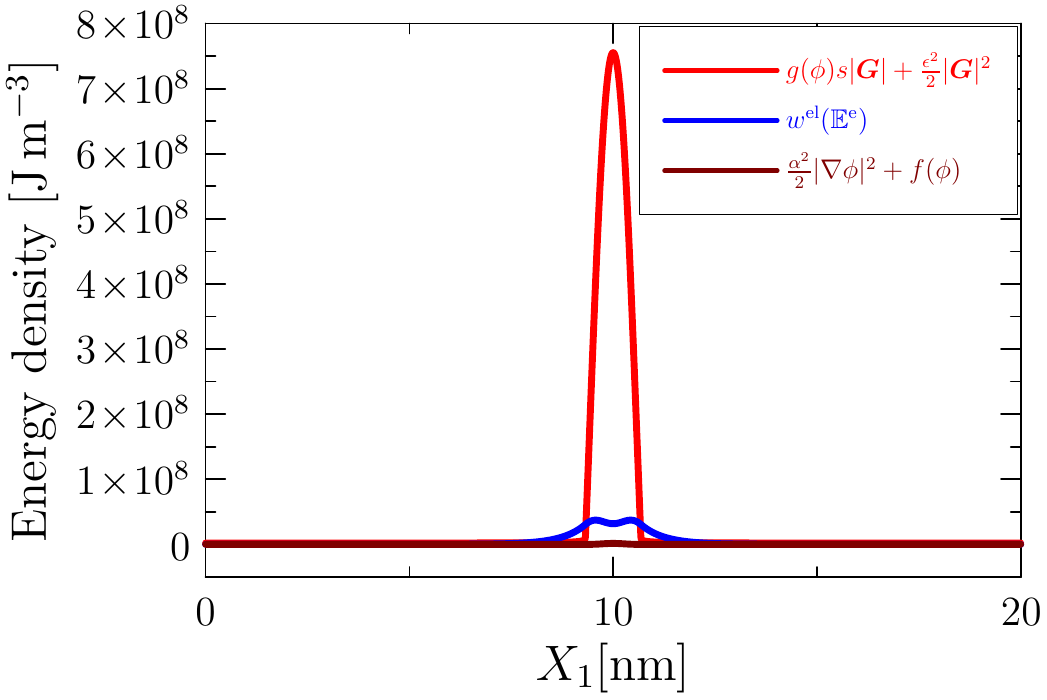}
        \label{fig:w}
    }
    \caption{Plots of the 1-d simulation modeling a flat grain boundary under
        steady state with $g(\phi)=\phi^2$, and material parameters as listed in
        Table \ref{table:kwc}.}
    \label{fig:1d}
\end{figure}
\begin{figure}[t]
    \centering
    \subfloat[$g(\phi)=\phi^2$] {\includegraphics[scale=0.8]{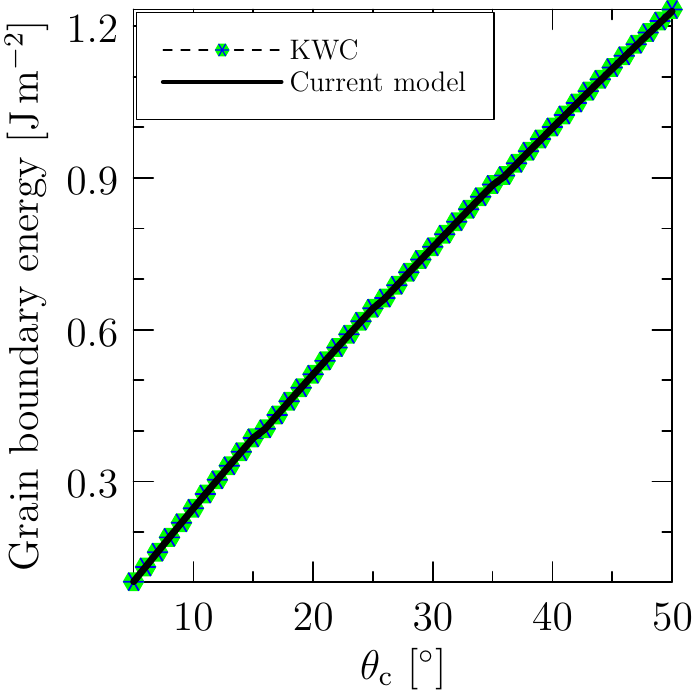}
    \label{fig:energy_parametric_squared}}
    \subfloat[$g(\phi)=-2(\log(\phi-1)-\phi)$] {\includegraphics[scale=0.8]{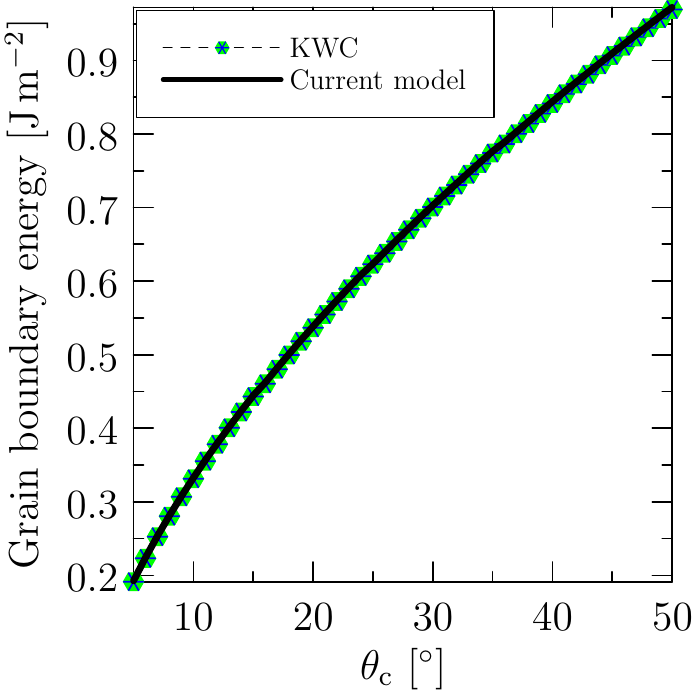}
    \label{fig:energy_parametric_log}}
    \caption{Plot of the grain boundary energy versus misorientation for
    different choices of $g$.}
    \label{fig:energy_parametric}
\end{figure}
\begin{table}[t]
    \centering
    \begin{tabular}{|c|l|}
        \hline
        $\epsilon^2$     &      \SI{3.1999e-4}{\femto\joule\per\nm} \\
        $\alpha^2$       &      \SI{7.95e-3}{\femto\joule\per\nm} \\
        $s$              &      \SI{0.0085}{\femto\joule\per\square\nm}\\
        $e$              &      \SI{0.00035}{\femto\joule\per\cubic\nm}\\
        $L$               &     \SI{20}{nm}\\
        \hline
    \end{tabular}
    \caption{Parameters used in the implementation of the current model with
    $g(\phi)=-2(\log(\phi-1)-\phi)$.}
    \label{table:log_current}
\end{table}
In this section, we present a simulation of a symmetric tilt grain boundary in a
bicrystal modeled as a one-dimensional domain $\Omega=[0,L]$. The aim of this
simulation is to study the steady state solution corresponding to a flat grain
boundary, and the corresponding energy as a function of the misorientation
angle.

The unknowns here are the two displacements $u_1$ and $u_2$, the four 
slip rates, the order parameter $\phi$, and the four components $\fp_{11}$,
$\fp_{12}$, $\fp_{21}$, and $\fp_{22}$ of $\Fp$.                        
The system is initialized as:
\begin{align}
    \bm u(X_1,0) = \bm 0, \quad
    \phi(X_1,0) = 1, \quad 
    \Fp(X_1,0) = \bm R^{0\rm T}(\widetilde \theta(X_1)),
\end{align}
where
\begin{align}
    \label{eqn:smooth_theta_1d}
    \widetilde\theta(X_1) =
    -\frac{\theta_0}{2}+\frac{\theta_0}{(1+\exp(-2.5(X_1-10)))},
\end{align}
and $\bm R^{0}(\widetilde\theta)$ is the rotation corresponding to $\widetilde \theta$.
Equation \eref{eqn:smooth_theta_1d} describes a diffuse-interface grain
boundary at $t=0$ with the lattice orientation changing smoothly from
$-\theta_0/2$ to $\theta_0/2$ across the grain boundary. 
The boundary conditions are given by
\begin{align}
    \phi(0,t) = \phi(L,t) = 1, &\quad \bm v(0,t)=\bm v(L,t) = \bm 0,
\end{align}
which indicate perfect crystalline order, and no slip at the boundaries. The
inverse mobilities $b^\alpha$ ($\alpha=1,\dots,A$) and $b^\phi$, corresponding
to the slip rates and $\phi$ respectively, are
assumed to be constant, and equal to $\SI{1}{(\femto\joule.\nano \second) \per
\nm \cubed}$. The system is evolved for $10^{-5}$ ns, long enough to reach a
steady state.

\fref{fig:1d} shows the plots obtained at the end of the simulation. We make the
following observations on the steady state solution:
\begin{enumerate}
    \item \fref{fig:Fp} shows the plots of the four components of $\Fp$, with dotted lines
    corresponding to the initial condition. From the plots we observe that 
    the gradient $\Fp$ is not identically equal to zero in the interior of the
    grains. This is a result of approximating $|\bm G|$ in the term
    $g(\phi)s |\bm G|$ of \eref{eqn:psi_gb} with $p(|\bm G|)$, an approximation
    used in the numerical implementation of the KWC model, as noted in \sref{sec:kwc}.
\item \fref{fig:G} shows the plots of the non-zero components $G_{31}$ and
    $G_{32}$ of $\bm G$. As noted in eq.~\eref{eqn:nonzeroG}, a non-zero $G_{32}$ in a
    symmetric tilt grain boundary signifies the diffuse nature of the grain
    boundary.
\item The plots of the three
    components of $\Ee$ in \fref{fig:Ee}, show the presence of a maximum
    strain of $0.4\%$ at steady state.\footnote{Recall that at $t=0$, the lattice
    Lagrangian strain is zero since $\Fe(\bm X,0) \in SO(3)$.} On the other hand, in
    real materials with no defects (other than grain boundaries) under no
    external loads, the lattice Lagrangian strain is oscillatory about zero in
    the vicinity of the grain boundary. In the current model, with grain boundaries
    represented by a continuum dislocation distribution --as opposed to a discrete array of
    dislocations--, no Lagrangian strain is expected near the grain boundaries. The
    presence of a nonzero $\Ee$ can be attributed to the construction
    $\psi_{\rm{gb}}$ (see \eref{eqn:psi_gb}), which uses $\bm G$ as opposed to
    the exact lattice gradient tensor $\bm R^{\rm{LT}} \curl \bm R^{\rm{LT}}$
    discussed at the end of \sref{sec:cn}.
\item \fref{fig:w} shows the plots of the three parts of the total energy
    density. It is clear that the major contribution is due to the GND density,
    followed by the elastic contribution and then the contribution due to the
    order parameter $\phi$.
\item In the KWC model, the approximation in
    eq.~\eref{eqn:approx} is used for numerical
    convenience, to tackle the singular diffusive term. However, the
    approximate functional --in particular the length constant $\gamma$-- has an
    interesting physical interpretation in the current model. As $\gamma$
    increases the gradient of the plastic distortion tends to zero in the bulk,
    resulting in negligible dislocation density in the bulk. Therefore, $\gamma$
    can be interpreted as the propensity of bulk dislocations to agglomerate and
    form grain boundaries. This effect will be discussed in more depth in \sref{sec:recov}
\end{enumerate}
Next, we study the variation of grain boundary energy with respect to
misorientation. \citet{kobayashi1999} have noted that $g(\phi)=\phi^2$ results in a
linear dependence of the grain boundary energy on misorientation, while $g(\phi)
=-2(\log(\phi-1)-\phi)$ results in non-convex grain boundary energy as
predicted by \citet{read1950dislocation}. Our model
confirms this observation as well, as shown in \fref{fig:energy_parametric}, from simulations performed with the material parameters given in Table
\ref{table:log_current}. In fact, the figure shows an
excellent agreement between the energy predicted by the KWC model and the
current polycrystal plasticity model for the different choices of $g$. In
addition, it also reveals the following limitation: 
although the current model is kinematically nonlinear, i.e. a
grain boundary with an arbitrary misorientation can be constructed, the energy
functional cannot detect the equivalence of zero and  $\SI{90}{\degree}$ 
misorientations. Therefore, the energy monotonically
increases as the magnitude of the misorientation increases. This limitation
appears to arise due to the diffuse-interface nature of grain boundaries. 

\subsection{Shear-induced grain boundary motion}
\label{sec:numerics_shear}
The goal of this section is to model the motion of a flat grain boundary in a
bicrystal subjected to shear stress. The absence of curvature in this
simulation results in stress being the only driving force for plastic
distortion. 

Various experimental and molecular dynamics simulations have identified two
different mechanisms by which a bicrystal plastically deforms when subjected to a shear stress.
In the first mechanism, commonly referred to as ``coupled'', the flat grain
boundary translates perpendicular to its normal, plastically distorting (i.e.
$\Fp \not\equiv \bm I$) and macroscopically deforming (i.e. $\bm F \not\equiv
\bm I$) the material in the swept volume. In addition, $\bm F$ and $\Fp$ are
uniquely determined by the misorientation of the bicrystal. On the other hand,
in the second mechanism commonly called ``sliding'', the grain boundary
remains stationary while the two grains slide with respect to each other
tangential to the grain boundary,  resulting in a plastic distortion
concentrated on the grain boundary. We now give a
precise definition of a coupled motion of a symmetric-tilt grain boundary,
followed by grain boundary sliding. We
use the notation $\mathbbm 1_{A}(x,t)$ where $x \in \mathbb R^d$ ($d=1,2$) and
$A \subset \mathbb R^{d+1}$, to denote the \emph{indicator} function defined as
\begin{align}
    \mathbbm 1_A(x,t) = 
    \begin{cases}
        1 \text{ if } (x,t) \in A, \\
        0 \text{ otherwise},
    \end{cases}
\end{align}
and $\delta_x$ to denote a Dirac delta distribution with support at $x$. We also
use the notation $(\Box)$ to denote a set whose elements satisfy the
inequality/equality $\Box$.

Consider a bicrystal defined on a one-dimensional domain $\mathcal B=[-L,L]$,
with:
$$\Fe = \mathbbm 1_{(X<0)} \bm R \left( \frac{\theta_0}{2}\right) +\mathbbm
1_{(X \ge 0)} \bm R \left(-\frac{\theta_0}{2}\right)$$ 
and 
$$\Fp=(\Fe)^{-1}.$$
In other words, $\mathcal B$ is a lattice strain-free bicrystal, with the
    symmetric-tilt grain boundary at $X=0$.
\begin{table}[t]
    \centering
    \begin{tabular}{|c|c|c|c|}
        \hline
        &
        \multicolumn{2}{|c|}{Flat}
        &
        \multicolumn{1}{|c|}{Circular} \\
        \cline{2-4}
        & Coupled & Sliding & Shrink \\
        \hline
        $G_{31}$  &  
        $-2\sin\left( \frac{\theta_0}{2} \right) \delta_{(X_1=t)}$
        &   
        $-2\sin\left( \frac{\theta_0}{2} \right) \delta_{(X_1=0)}$
        &   
            $-\frac{2 X_1}{d(\bm X)}
            \sin\left(\frac{\theta_0}{2}\right)\delta_{(d(\bm X)=r_0-t)}$
        \\
        \hline
        $G_{32}$  &  
        $0$
        & 
        $0$  
        &   
        $\frac{2 X_2}{d(\bm X)}
            \sin\left(\frac{\theta_0}{2}\right)\delta_{(d(\bm X)=r_0-t)}$
        \\
        \hline
        $\Lp$  &  
        {
            $\!\begin{aligned}
                \begin{bmatrix}
                    0   &   2\tan\left(\frac{\theta_0}{2} \right) \\
                    0   &   0
                \end{bmatrix} \delta_{(X_1=t)}.
            \end{aligned}$
        }
        &  
            $\!\begin{aligned}
                \cos^2 \left (\frac{\theta_0}{2} \right )
                \begin{bmatrix}
                    0   &   0  \\
                    \delta_{(X_1=0)}   &   0
                \end{bmatrix}
            \end{aligned}$
        &   
        {
            $\!\begin{aligned}
                \begin{bmatrix}
                    0 & -\sin(\theta_0) \\
                    \sin(\theta_0) & 0
                \end{bmatrix}
                \delta_{(d(\bm X)=r_0-t)}
            \end{aligned}$
        }
        \\
        \hline
        Slips ($\bm s^\alpha$)  &  
            $\!\begin{aligned}
                &(1,0) \\
                &(1/\sqrt{2},1/\sqrt{2}) \\
                &(1/\sqrt{2},-1/\sqrt{2})
            \end{aligned}$
        &   
            $\!\begin{aligned}
                &(0,1) \\
                &(1/\sqrt{2},1/\sqrt{2}) \\
                &(1/\sqrt{2},-1/\sqrt{2})
            \end{aligned}$
        &
            $\!\begin{aligned}
                &(1,0) \\
                &(0,1) 
            \end{aligned}$
        \\
        \hline
    \end{tabular}
    \caption{A catalog of time-dependent GND tensor and $\Lp$ fields for the 
        motion of a flat and circular sharp-interface grain boundaries. The three
        columns correspond to the evolution of coupled and sliding flat grain
        boundaries, and a shrinking circular grain boundary. The last row lists
        the slip systems used to simulate the three cases.}
    \label{table:catalog}
\end{table}
\begin{definition}[Coupled grain boundary motion]
    \label{def:coupled}
    A time-dependent coupled grain boundary motion in the bicrystal $\mathcal B$ 
    is given by the following plastic and elastic distortion fields:
    \begin{align}
        \Fe(X_1,t) &= 
         \mathbbm 1_{(X_1<t)} \bm R \left( \frac{\theta_0}{2}\right)
        +\mathbbm 1_{(X_1\ge t)} \bm R \left(-\frac{\theta_0}{2}\right),
        \label{eqn:FeCoupled}\\
        \Fp(X_1,t) &=
        \mathbbm 1_{(X_1<0)} \bm R \left(-\frac{\theta_0}{2}\right) +
        \mathbbm 1_{(X_1\ge t)} \bm R \left( \frac{\theta_0}{2}\right) +
        \mathbbm 1_{(0\le X_1<t)} \bm S \bm R \left(\frac{\theta_0}{2}\right),
        \label{eqn:FpCoupled}
    \end{align}
    where
    \begin{align}
        \bm S = 
        &\begin{bmatrix}
            1   &   2\tan\left(\frac{\theta_0}{2} \right) \\
            0   &   1
        \end{bmatrix}.
    \end{align}
    Moreover, the resulting deformation gradient $\bm F=\Fe\Fp$ is the gradient of the
    continuous deformation field given by
    \begin{align}
        \label{eqn:uCoupled}
        u_1 \equiv 0, \quad u_2(X_1,t) = 2 \tan
        \left(\frac{\theta_0}{2}\right )
        \left (\mathbbm 1_{(0 \le X_1 < t)} X_1+ \mathbbm 1_{(X_1 \ge t)}t \right ).
    \end{align}
\end{definition}
By construction, Definition~\ref{def:coupled} applies to a sharp-interface grain
boundary. Equations~\eref{eqn:FeCoupled}, \eref{eqn:FpCoupled} and
\eref{eqn:uCoupled} can be appropriately mollified to yield an analogous
definition for a diffuse-interface grain boundary. 
The time-dependent non-zero
components of the GND tensor for a coupled grain boundary motion corresponding
to the plastic distortion field in \eref{eqn:FpCoupled} is given in Table
\ref{table:catalog}.                                
The expression for $G_{31}$ clearly shows that a coupled
sharp-interface grain boundary motion involves the translation of a single
family of necessary edge dislocations. Note that there was no
mention of slip systems of the crystal until this point. The translation of
GNDs is made possible by the availability of the crystal slip systems. This is
seen by evaluating\footnote{In order to compute $\dot \Fp$, we note that
    $\dot{\mathbbm 1}_{(X_1\ge t)} = -\delta_{(X_1=t)}$. Moreover, to compute
    $\dot\Fp \invFp$, we define the
    product of a Dirac delta distribution $\delta_{(X=0)}$ and a function $f(X)$ with
    discontinuity at $X=0$ as $\frac{1}{2} \delta_{(X=0)} (f(0^+)+f(0^-))$.}
$\Lp=\dot \Fp \invFp$ (see \eref{eqn:flow}), and shown in Table
\ref{table:catalog}.
The expressions for $\bm G$ and $\Lp$ shown in Table \ref{table:catalog} suggest
that the single slip system 
\begin{align}
    \bm s=(1,0)
\end{align}
is sufficient to translate the GNDs in a
sharp-interface grain boundary. Conversely, we know from
\eref{eqn:nonzeroG} that for a diffuse-interface grain boundary
$G_{32}\not\equiv 0$. Therefore, in order for the grain boundary to translate, the GNDs corresponding
to $G_{32}$ should undertake a mechanism resulting in an ``apparent'' dislocation climb,
made possible by the availability of two additional slip systems shown in Table
\ref{table:catalog}.
The translation of GND corresponding to $G_{32}$ along with the grain boundary
can be interpreted as dissociating into dislocations in the second and third slip systems,
translating, and recombining. Such a mechanism has been proposed by \citet{cahn2006coupling}.

\begin{definition}[Grain boundary sliding]
    \label{def:sliding}
    A time-dependent grain boundary sliding in $\mathcal B$ is given by the
    following plastic and elastic distortion fields:
    \begin{align}
        \Fe(X_1,t) &= 
         \mathbbm 1_{(X_1<0)} \bm R \left( \frac{\theta_0}{2}\right)
        +\mathbbm 1_{(X_1 \ge 0)} \bm R \left(-\frac{\theta_0}{2}\right),
        \label{eqn:FeSliding}\\
        \Fp(X_1,t) &=
        \left (
            \mathbbm 1_{(X_1<0)} \bm R \left(-\frac{\theta_0}{2}\right) +
            \mathbbm 1_{(X_1\ge 0)} \bm R \left(\frac{\theta_0}{2}\right)
        \right) \bm S,
        \label{eqn:FpSliding}
    \end{align}
    where
    \begin{align}
        \bm S =
        \begin{bmatrix}
            1   &   0  \\
            t\delta_{(X_1=0)}   &   1
        \end{bmatrix},
    \end{align}
    and $\delta_{(X_1=0)}$ is the Dirac delta distribution with support at the
    origin. The resulting deformation gradient given by
    \begin{align}
        \bm F = \Fe \Fp = 
        &\begin{bmatrix}
            1   &   0  \\
            t\delta_{(X_1=0)}   &   1
        \end{bmatrix}
        \label{eqn:FSliding}
    \end{align}
    is the gradient of the following discontinuous deformation map
    \begin{align}
        u_1 \equiv 0, \quad u_2(X_1,t) = t \mathbbm 1_{[0,L]}.
        \label{eqn:uSliding}
    \end{align}
\end{definition}
Similar to Definition~\ref{def:coupled}, the above described grain boundary
sliding applies to a sharp-interface grain
boundary. Equations~\ref{eqn:FeSliding}, \ref{eqn:FpSliding} and
\ref{eqn:uSliding} can be appropriately mollified to yield an analogous
definition for a diffuse-interface grain boundary. It is interesting to note
that the non-zero component $G_{31}$ of the GND tensor for grain boundary
sliding, shown in Table \ref{table:catalog},
is independent of time
which means the grain boundary remains stationary. The expression
for $\Lp$, given in Table \ref{table:catalog}, clearly shows that the $2-1$
component is the only non-zero component.
Note that this is applicable only for a sharp-interface symmetric tilt grain boundary. 
For a diffuse grain boundary, all components of $\Lp$ are
nonzero, although the $2-1$ component is the dominant one.
Therefore, the slip system
\begin{align}
    \bm s=(0,1)
\end{align}
is responsible for sharp-interface symmetric tilt grain boundary sliding,
and two additional slip systems, shown in Table \ref{table:catalog}, are necessary in
the diffuse case.
In general, in the presence of all four slips systems (see equations in
\ref{eqn:s}), the motion of the grain boundary involves
a combination of coupling and sliding motions.

\begin{figure}[t]
    \centering
    \includegraphics[scale=1.0]{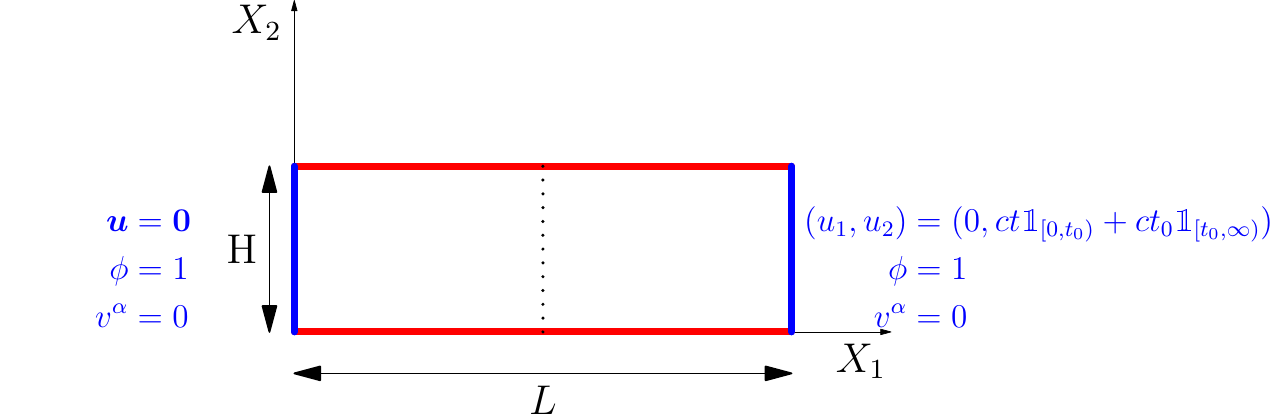}
    \caption{A schematic of a bicrystal used to study grain boundary
        sliding and coupled motion. The grain boundary is shown as a dotted line.
        Dirichlet boundary conditions are shown
    in blue, while periodic boundary conditions are shown in red. The right
surface $X_1=L$ is translated upwards with a velocity $c$ for a time $t_0$,
and subsequently held fixed in that position.}
    \label{fig:domainShear}
\end{figure}
\begin{figure}
    \centering
    \subfloat[$t=\SI{0}{\second}$]{\includegraphics[scale=0.5]{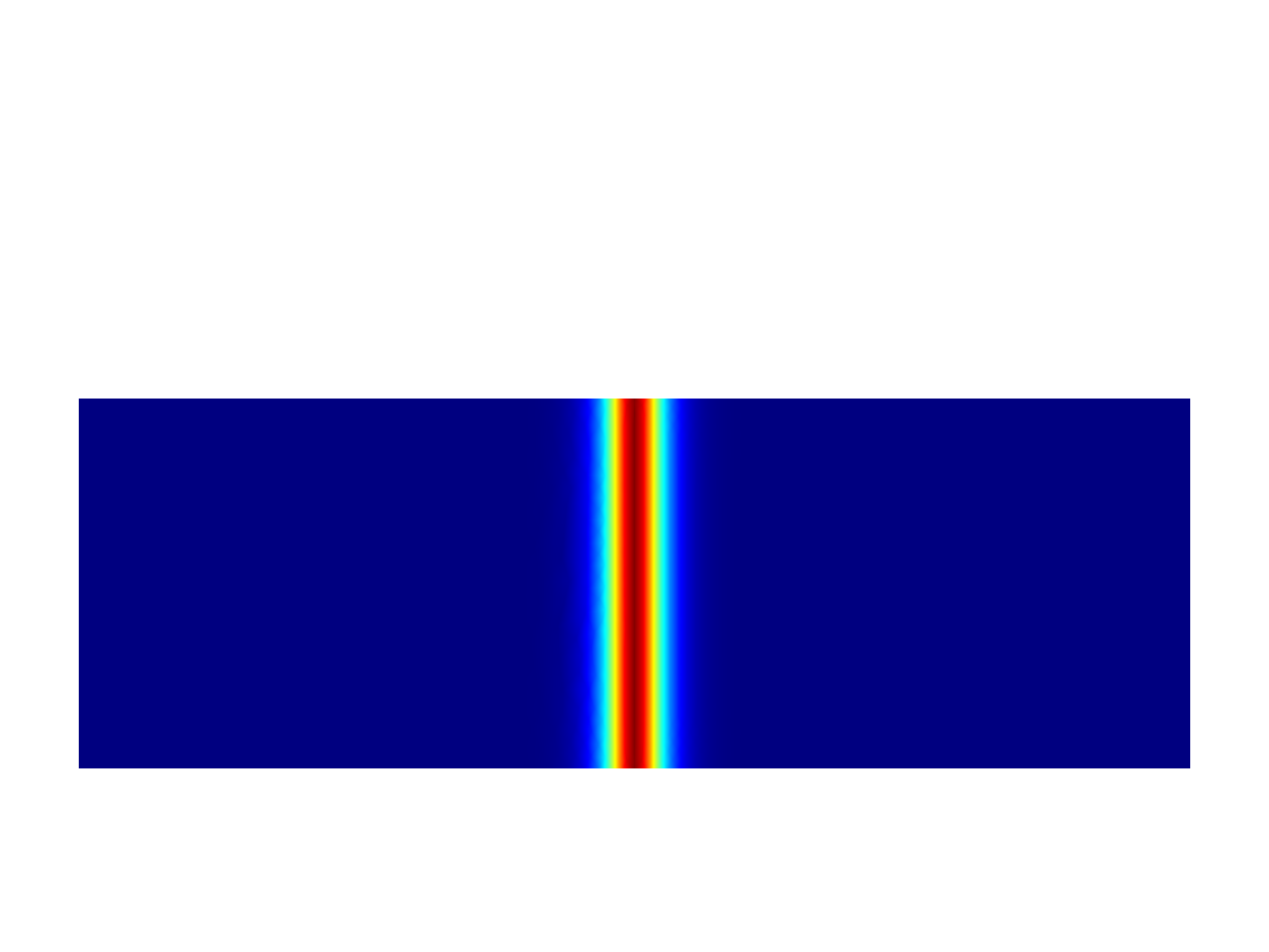}
    \label{fig:pcp_2d_GNDI}}
    \subfloat[$t=\SI{1e-4}{\second}$]{\includegraphics[scale=0.5]{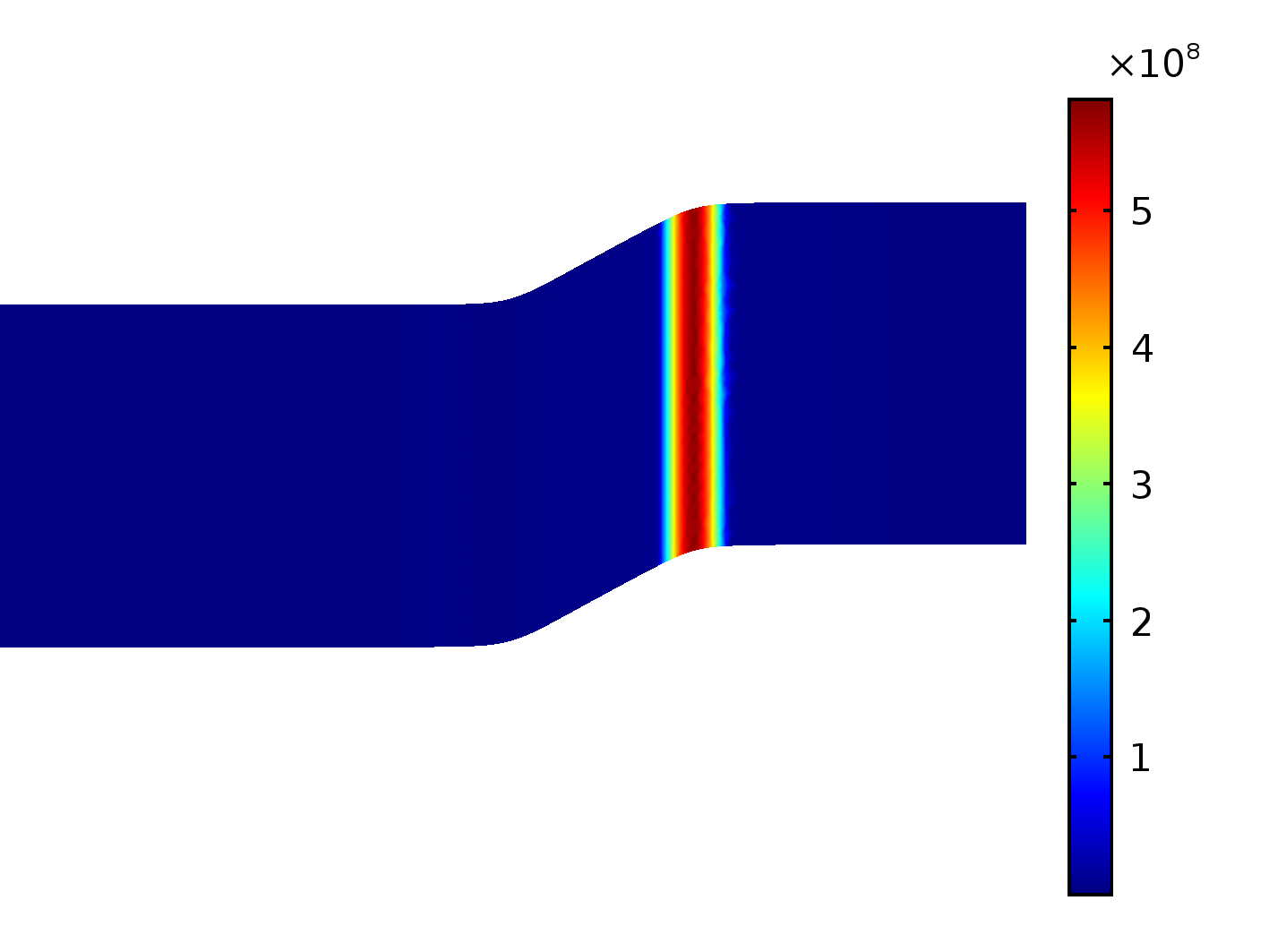}
    \label{fig:pcp_2d_shear_GNDF}}
    \caption{Coupled grain boundary motion: Color plots of the norm of the
    GND density in undeformed and the deformed configurations in units of
    \si{\meter^{-1}}.}
    \label{fig:pcp_2d_shear_GNDIF}
\end{figure}
\begin{figure}[t]
    \centering
    \subfloat{\includegraphics[scale=0.5]{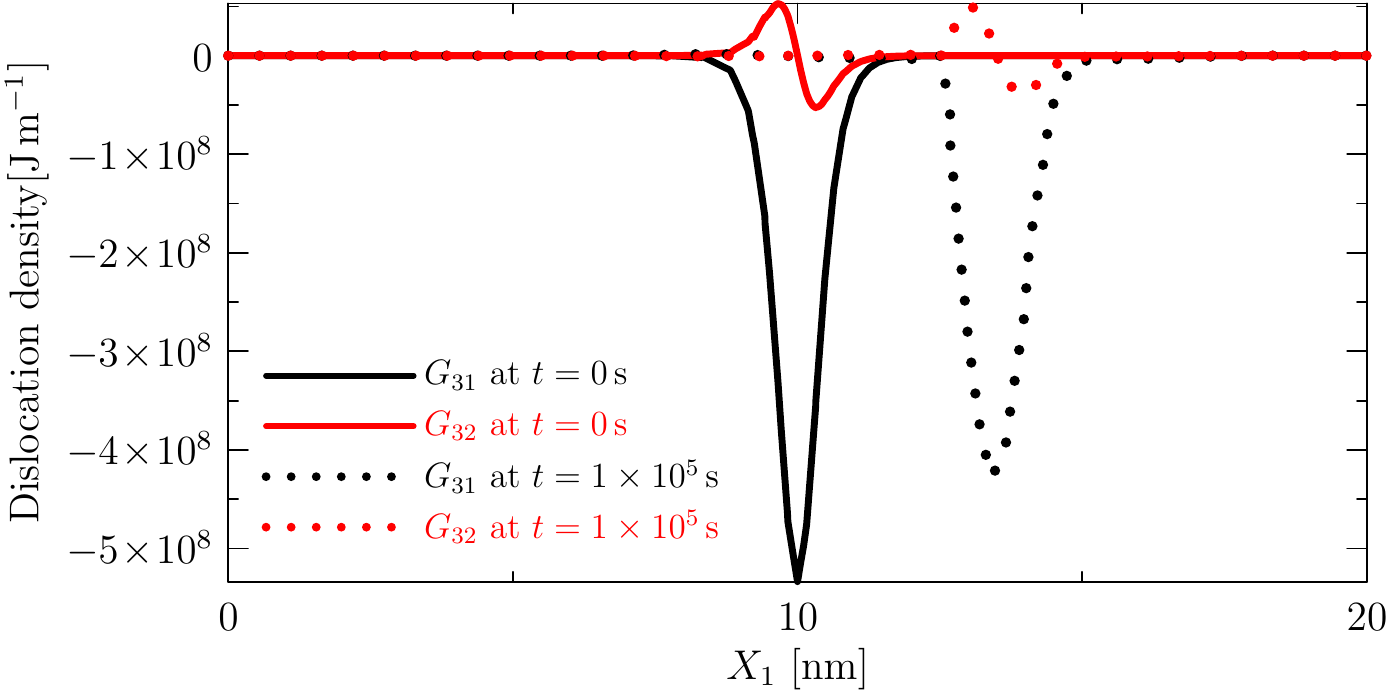}
    \label{fig:pcp_2d_shear_GND}}
    \subfloat{\includegraphics[scale=0.5]{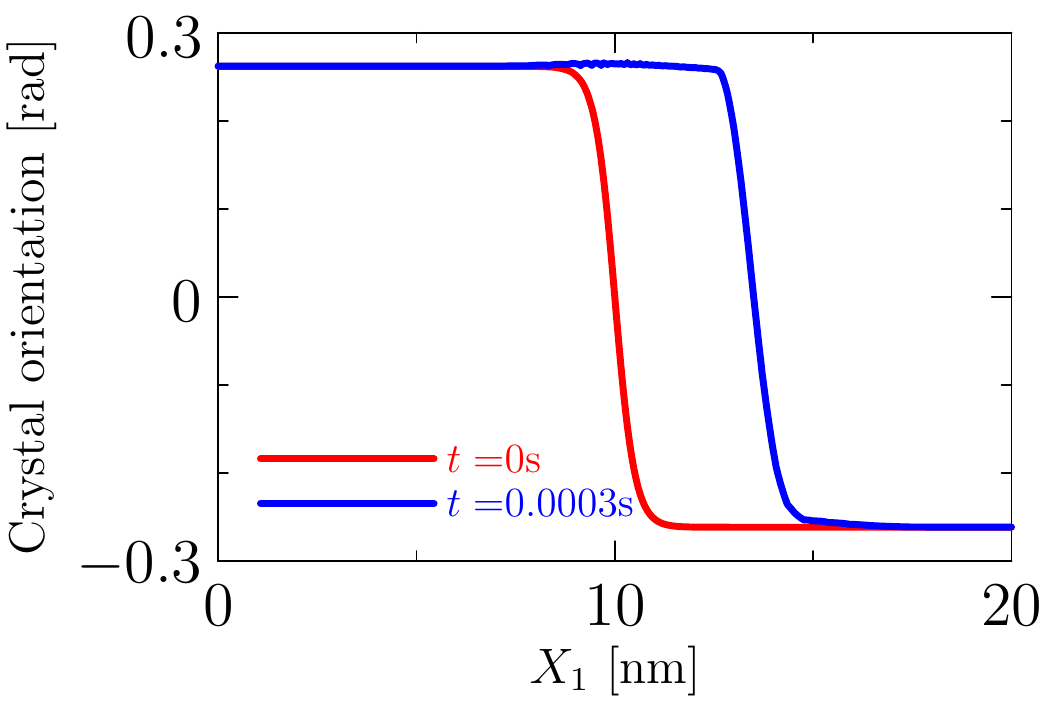}
    \label{fig:pcp_2d_shear_thetae}}
    \caption{Coupled grain boundary motion:
        \protect\subref{fig:pcp_2d_shear_GND} Plots of GND density components $G_{31}$ and
        $G_{32}$, and \protect\subref{fig:pcp_2d_shear_thetae} plots of the
        lattice orientation corresponding to the rotation $\bm R^{\rm L}$
    obtained from the polar decomposition of $\Fe$, along the horizontal central
    line $X_2 = \SI{20/6}{\nm}$.}
    \label{fig:pcp_2d_shear}
\end{figure}
\begin{figure}[t]
    \centering
    \subfloat[$t=\SI{0}{\second}$]{\includegraphics[scale=0.5]{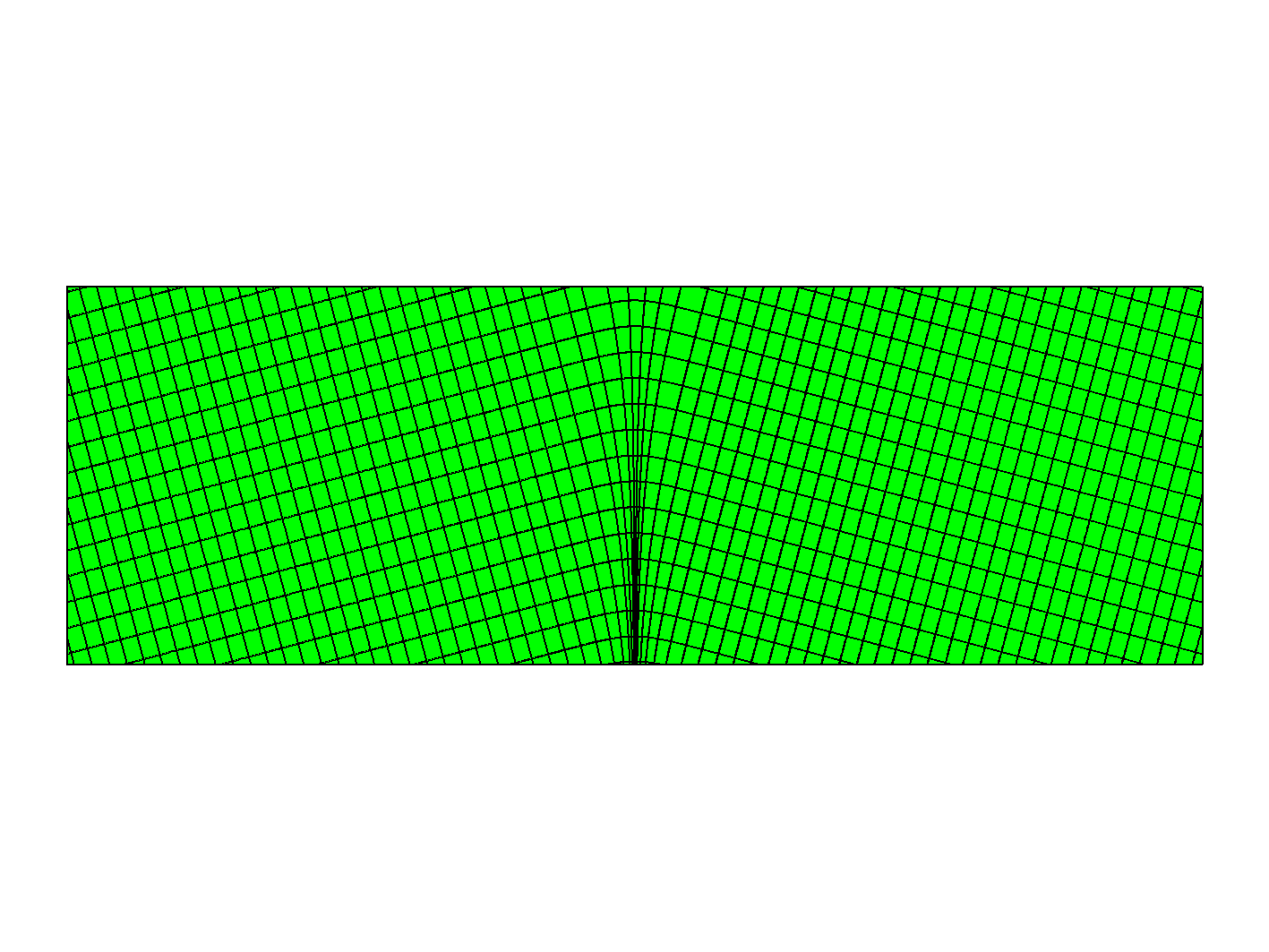}
    \label{fig:pcp_2d_shear_latticei}}
    \subfloat[$t=\SI{1e5}{\second}$]{\includegraphics[scale=0.5]{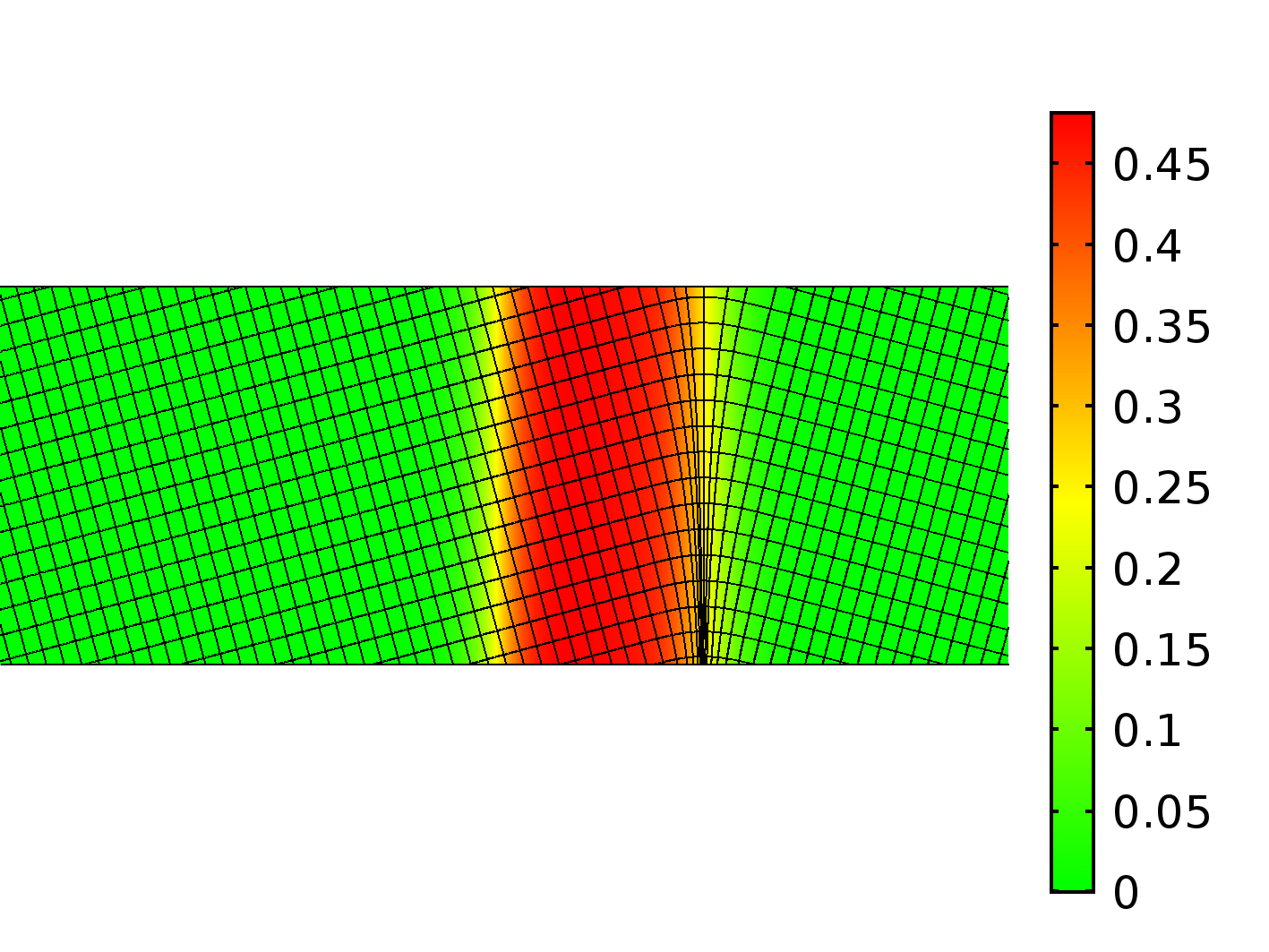}
    \label{fig:pcp_2d_shear_latticef}}
    \caption{Coupled grain boundary motion: Streamlines tangential
        to the vector fields $\Fe \bm e^1$ and $\Fe \bm e^2$ depicting the
        variations in the lattice, plotted in the reference configuration. The
        color density corresponds to the $1-2$ component of the plastic stain
    $(\FpT\Fp-\bm I)/2$.}
    \label{fig:pcp_2d_shear_lattice}
\end{figure}
\begin{figure}[t]
    \centering
    \subfloat{\includegraphics[scale=0.7]{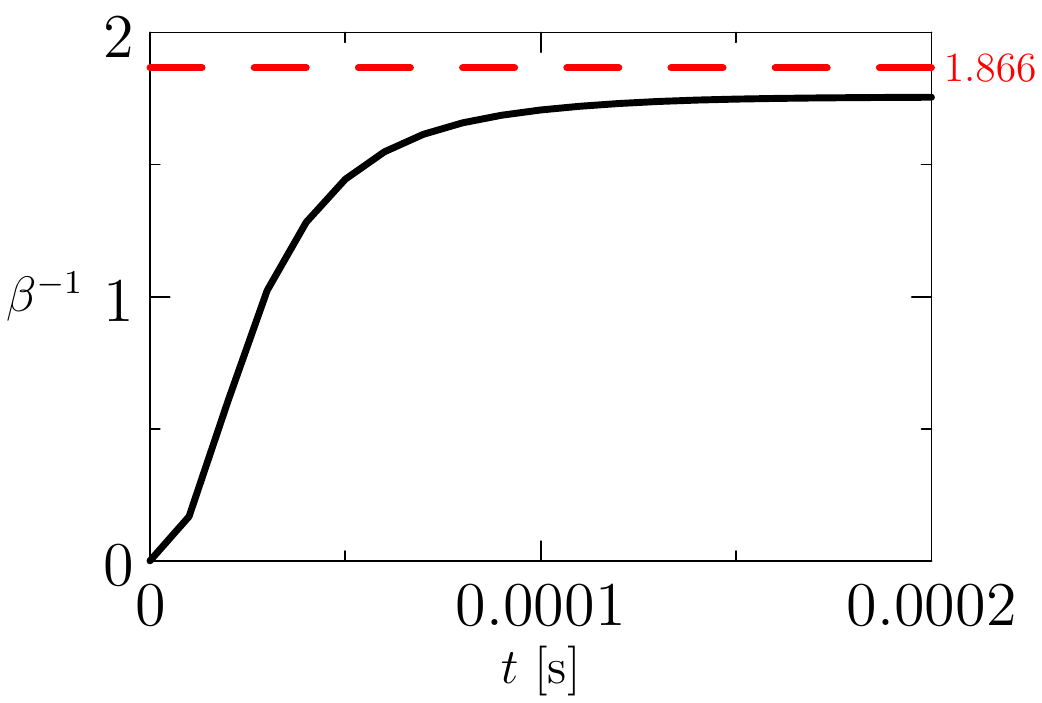}}
    \caption{Coupled grain boundary motion: Plot showing the convergence of the
        inverse coupling factor $\beta^{-1}$ with respect to time. The value
        $1.866$ is the theoretical $\beta^{-1}$ corresponding to the
        misorientation of $\SI{30}{\degree}$.}
    \label{fig:pcp_2d_shear_cf}
\end{figure}
We now simulate grain boundary coupling and sliding in a rectangular bicrystal
of size $L=\SI{20}{\nm}$ and $H=\SI{20/3}{\nm}$, with a $\SI{30}{\degree}$ symmetric tilt grain boundary. The bicrystal, and the
boundary conditions used in the simulation are depicted in
\fref{fig:domainShear}. A shear stress is imposed on the bicrystal by
translating the right surface $X_1=L$ upwards at a constant velocity
$c=\SI{1e-4}{\meter\per\second}$ for a time $t_0=\SI{2e-5}{\second}$, and then holding it in this position for the rest of the
simulation. The initial conditions are taken to be
\begin{align}
    \bm u(\bm X,0) = \bm 0, \quad
    \phi(\bm X,0) = 1, & \quad 
    \Fp(\bm X,0) = \bm R^{0\rm T}(\widetilde \theta(X_1)),
\end{align}
where
\begin{align}
    \widetilde\theta(X_1) =
    -\frac{\theta_0}{2}+\frac{\theta_0}{(1+\exp(-4(X_1-10))},
\end{align}
$\theta_0=\SI{30}{\degree}$, and $\bm R^{0}(\widetilde\theta)$ is the rotation corresponding to
$\widetilde \theta$. We begin with the simulation of grain boundary coupled motion by having
the three slip systems shown in Table \ref{table:catalog}. The mobility for all
slip systems is chosen as
\begin{align}
    (b^\alpha)^{-1} &=
    m^\alpha_{\rm{min}}+(1-\phi^3(10-15\phi+6\phi^2))(m^\alpha_{\rm{max}}-m^\alpha_{\rm{min}}),
    \label{eqn:invMobility}
\end{align}                                           
where $m^\alpha_{\rm{min}}=\SI{1e-9}{\nm\cubed\per\femto\joule\per\ns}$ and
$m^\alpha_{\rm{max}}=\SI{1}{\nm\cubed\per\femto\joule\per\ns}$ are the minimum and maximum
mobilities attained when $\phi=1$ and $\phi=0$ respectively. In other words,
\eref{eqn:invMobility} is constructed such that the material shows greater
resistance to slip in the bulk compared to the grain boundary which is in
agreement with experimental observations.\footnote{The construction of
    mobilities in \eref{eqn:invMobility} also ensures that the dislocations 
    are more likely to nucleate in the grain boundary compared to the bulk.} 
The mobility corresponding to the order parameter $\phi$ is chosen as 
$(b^\phi)^{-1} = \SI{1}{\nm \cubed \per(\femto\joule.\nano \second) }$.

The results of the simulation are shown in
\frefs{fig:pcp_2d_shear_GNDIF}{fig:pcp_2d_shear_lattice}.
The color plots of the initial and final GND density $G_{31}$ shown in \fref{fig:pcp_2d_shear_GNDIF}
clearly demonstrate coupled motion.
\fref{fig:pcp_2d_shear_GND} shows the plots of $G_{31}$ and $G_{32}$ along the
horizontal line $y=\SI{20/6}{\nm}$ passing through the center of the domain. Compared to the
plots at $t=0$, the dislocation density is more diffused at $t=\SI{1e5}{\ns}$
which can be attributed to the stressed state of the material. The translation
of the grain boundary during the
coupled motion is more explicit in \fref{fig:pcp_2d_shear_thetae} which shows the plots of
the lattice orientation for the initial and final configurations. The lattice
orientation is the angle corresponding to the unique rotation tensor $\bm R^{\rm L}$ that
is obtained through the polar decomposition of $\Fe$, i.e. $\Fe=\bm R^{\rm L}
\bm U^{\rm L}$. In order to visualize the changes in the lattice, we plot the
streamlines of the vector fields $\Fe \bm e^1$ and $\Fe \bm e^2$, where $\bm
e^1=(1,0)$ and $\bm e^2=(0,1)$ in \fref{fig:pcp_2d_shear_lattice}. In addition, the
color density of the $1-2$ component of plastic shear $(\FpT\Fp-\bm I)/2$, shown in
\fref{fig:pcp_2d_shear_lattice}, clearly demonstrates that the grain boundary
plastically distorts the material as it sweeps through the material. The extent
of coupling in a grain boundary motion is quantified using the 
coupling factor $\beta$ which is defined as the ratio of the distance covered 
by the grain boundary in the normal direction to that in the tangential direction.
From \eref{eqn:FeCoupled}, \eref{eqn:uCoupled}, \eref{eqn:FeSliding} and
\eref{eqn:uSliding}, it is clear that for a
coupled motion of a sharp-interface grain boundary, the coupling factor is equal to $2\tan\left(
\theta_0/2 \right)$, while it is zero for grain boundary sliding. In the simulation
of coupled grain boundary motion, $\beta^{-1}$ is measured as the ratio of
vertical displacement of the boundary $X_1=L$ and the distance traversed by the
grain boundary measured using \fref{fig:pcp_2d_shear_thetae}. The plot of the
inverse coupling factor versus time is shown in \fref{fig:pcp_2d_shear_cf}.
Since the measurement of $\beta^{-1}$ is not a local measurement, it takes times
to converge. From \fref{fig:pcp_2d_shear_cf}, it is clear that the
converged value of $\beta^{-1}$ is below the theoretical value of $1.866$. This
can be attributed to the diffuse nature of the grain boundary which results in some sliding
during the predominantly coupled grain boundary motion shown in 
\frefs{fig:pcp_2d_shear_GNDIF}{fig:pcp_2d_shear_cf}.

\begin{figure}[t]
    \centering
    \subfloat{\includegraphics[scale=0.5]{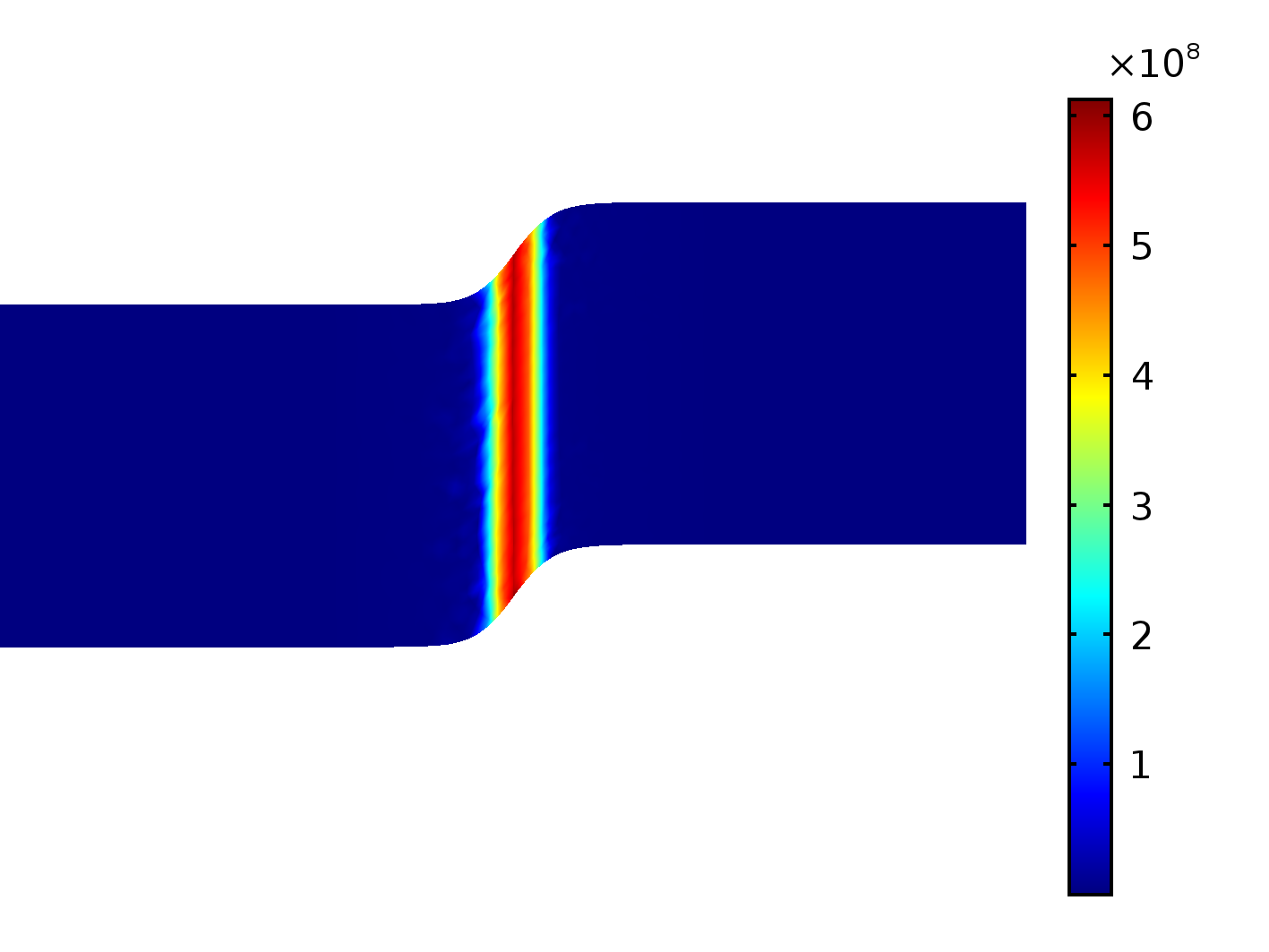}
    \label{fig:pcp_2d_shear_sliding_GNDF}}
    \subfloat{\includegraphics[scale=0.5]{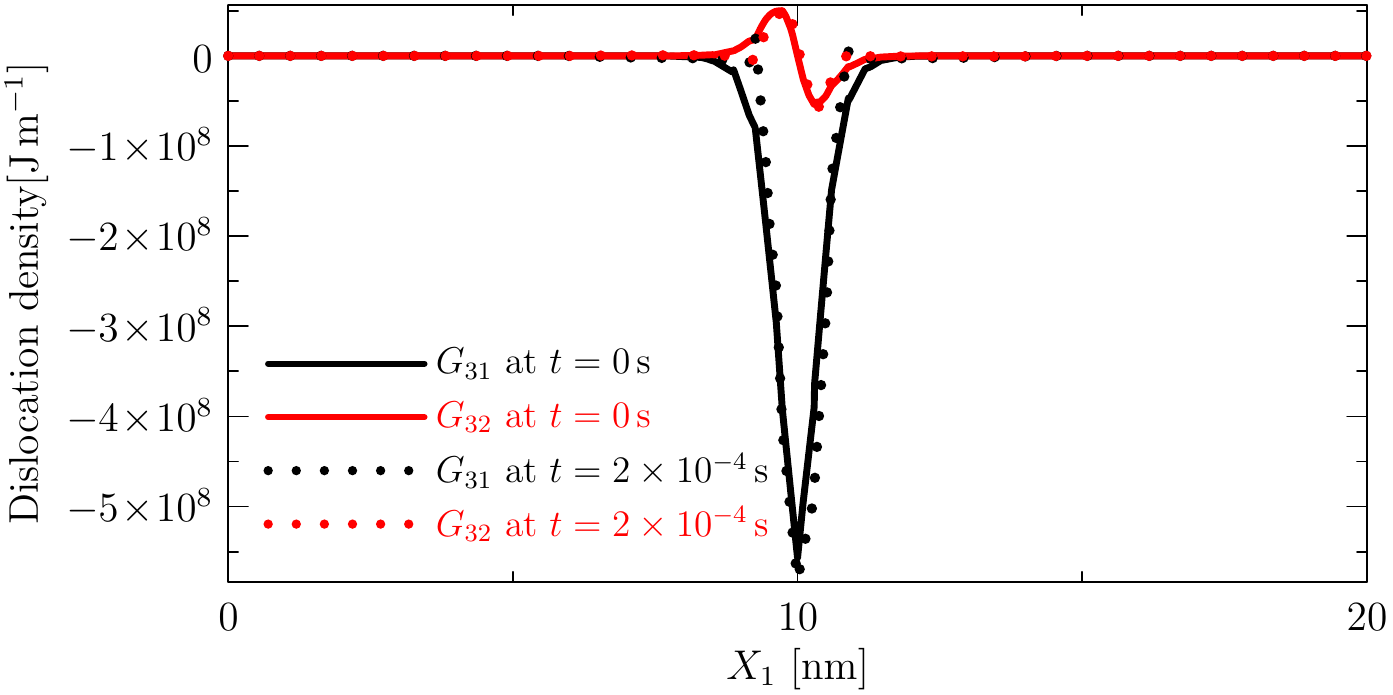}
    \label{fig:pcp_2d_shear_sliding_GND}}
    \caption{GB Sliding: a) Color plot of the norm of the GND density in the
        deformed configuration, and b) Plots of $G_{31}$ and
        $G_{32}$ as functions of the material coordinate $X_1$.}
    \label{fig:pcp_2d_shear_sliding}
\end{figure}

Next, we simulate grain boundary sliding by replacing one of the slip system
$(1,0)$ in the earlier simulation with $(0,1)$.
The color plot of GND density $G_{31}$ in 
\fref{fig:pcp_2d_shear_sliding} shows a stationary grain boundary, clearly
demonstrating grain boundary sliding.
\fref{fig:pcp_2d_shear_sliding_GND} show the plots of $G_{31}$ and $G_{32}$ along the
horizontal line $y=\SI{20/6}{\nm}$ in the center of the domain for grain
boundary sliding. From \fref{fig:pcp_2d_shear_sliding_GND}, we note
a small offset in the final position of the grain boundary relative to its
initial position. This can be attributed to a small degree of coupling due to
the diffuse nature of the grain boundary.

Summarizing the results of the analysis and simulations presented in this
section, we note that grain boundary
coupling and sliding are two independent mechanisms which can be activated
depending on the choice of corresponding slip systems. Interestingly, the above
discussion sheds light on an alternate kinematic mechanism for the translation
of a flat grain boundary wherein $\Fp$ remains a piecewise-constant rotation
field, with the discontinuity translating with time, and $\Fe = (\Fp)^{\rm T}$,
which implies $\bm F(\bm X,t) = \bm I$ for all time. In this mechanism, the grain boundary
translates along its normal with no macroscopic deformation. Clearly, there is
no driving force to activate this mechanism for a stressed/unstressed
elastically isotropic bicrystal with a flat grain boundary. On the other hand,
we postulate that for a stressed elastically \emph{anisotropic} bicrystal with
a flat grain boundary, a driving force
exists which could activate this mechanism. In the next section, we show that
the above mentioned mechanism is responsible for the shrinking of a circular
grain, where the driving force originates from the nonzero curvature.

\subsection{Curvature-induced grain boundary motion and grain rotation}
\label{sec:curvature}
In this section, with study grain boundary motion with curvature as the only
driving force. Prior studies on the evolution of a circular grain have
identified primarily three different kinds of grain evolution: 1)  Grain
rotation with no shrinkage, 2) grain shrinkage with no rotation, and 3)
simultaneous grain rotation and shrinkage.  In the latter, the interior grain can
rotate either to increase or decrease the misorientation depending on whether
the dislocations are conserved or not respectively. Recall from
\sref{sec:numerics_shear} that
grain boundary coupling and sliding are defined for a flat grain boundary.
Applying these definitions locally for a circular grain boundary, it can be
easily shown that a coupled grain boundary motion involves the conservation of
the dislocation content resulting in grain shrinkage and an increase in the
misorientation. On the other hand, grain rotation with no shrinkage and a
decreasing misorientation results in maximum rate at which dislocations are
annihilated, and this corresponds to grain boundary sliding. Therefore, the rate
of dislocation annihilation during grain shrinkage with no rotation lies in
between that observed in coupling and sliding motions.

\begin{figure}[t]
    \centering
    \subfloat[]
    {\includegraphics[scale=0.4]{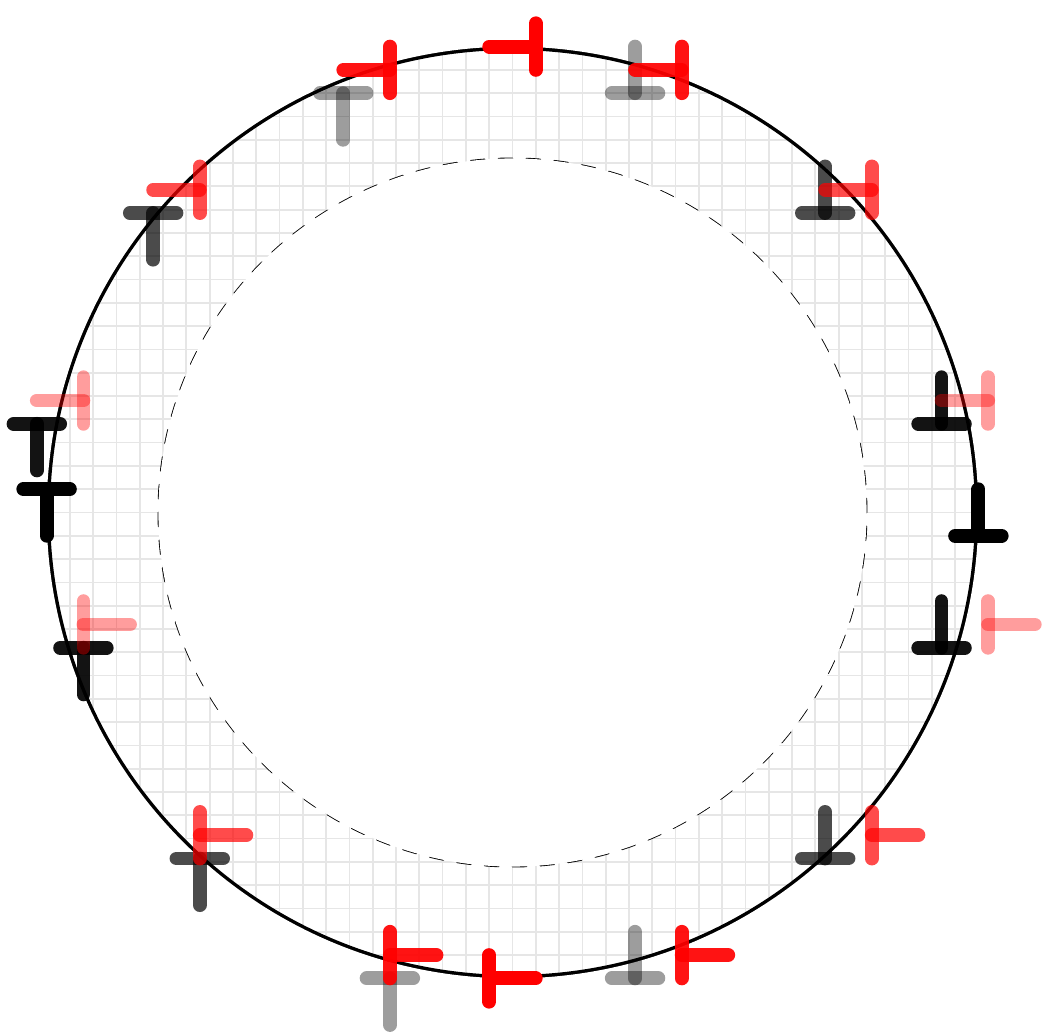}\label{fig:pcp_2d_shrink_grid1}}
    \quad
    \subfloat[]
    {\includegraphics[scale=0.4]{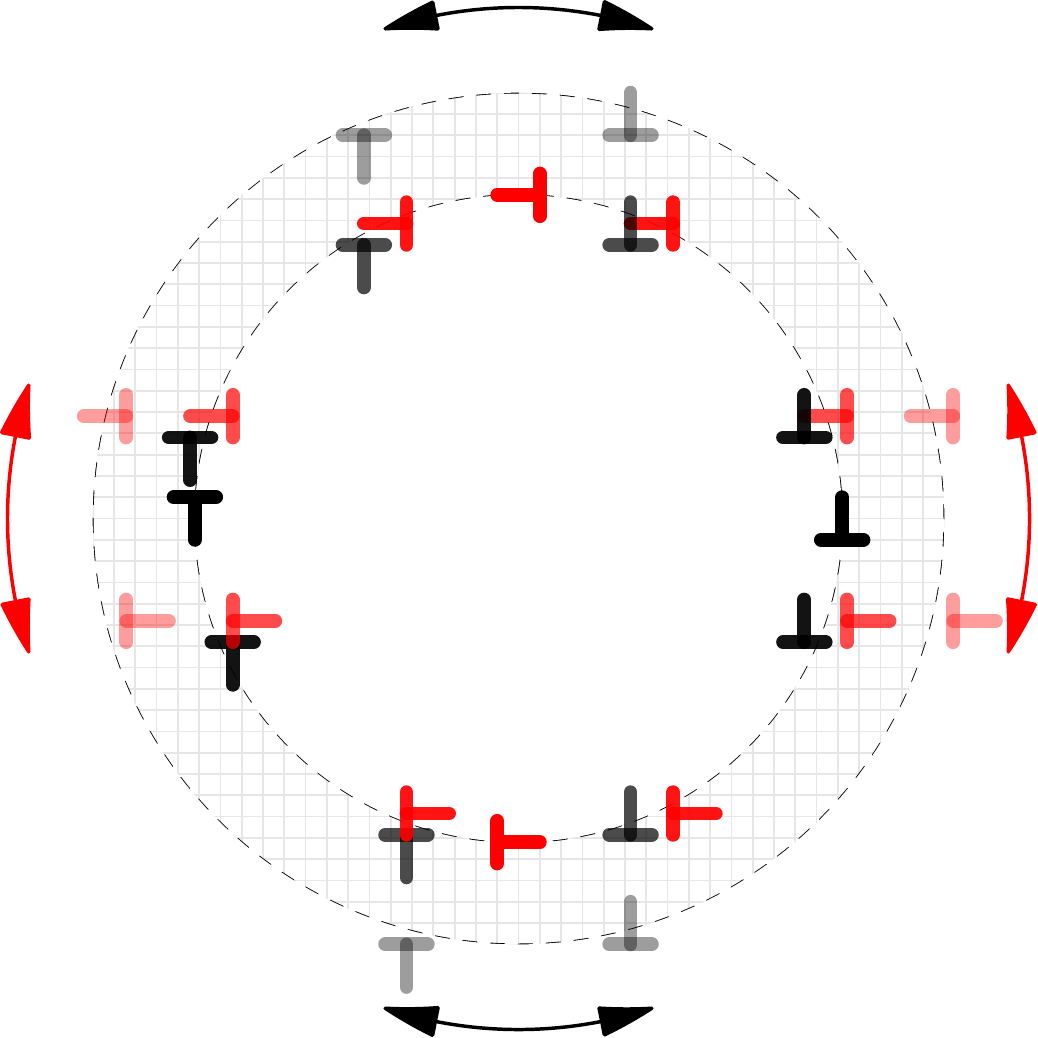}\label{fig:pcp_2d_shrink_grid2}}
    \quad
    \subfloat[]
    {\includegraphics[scale=0.4]{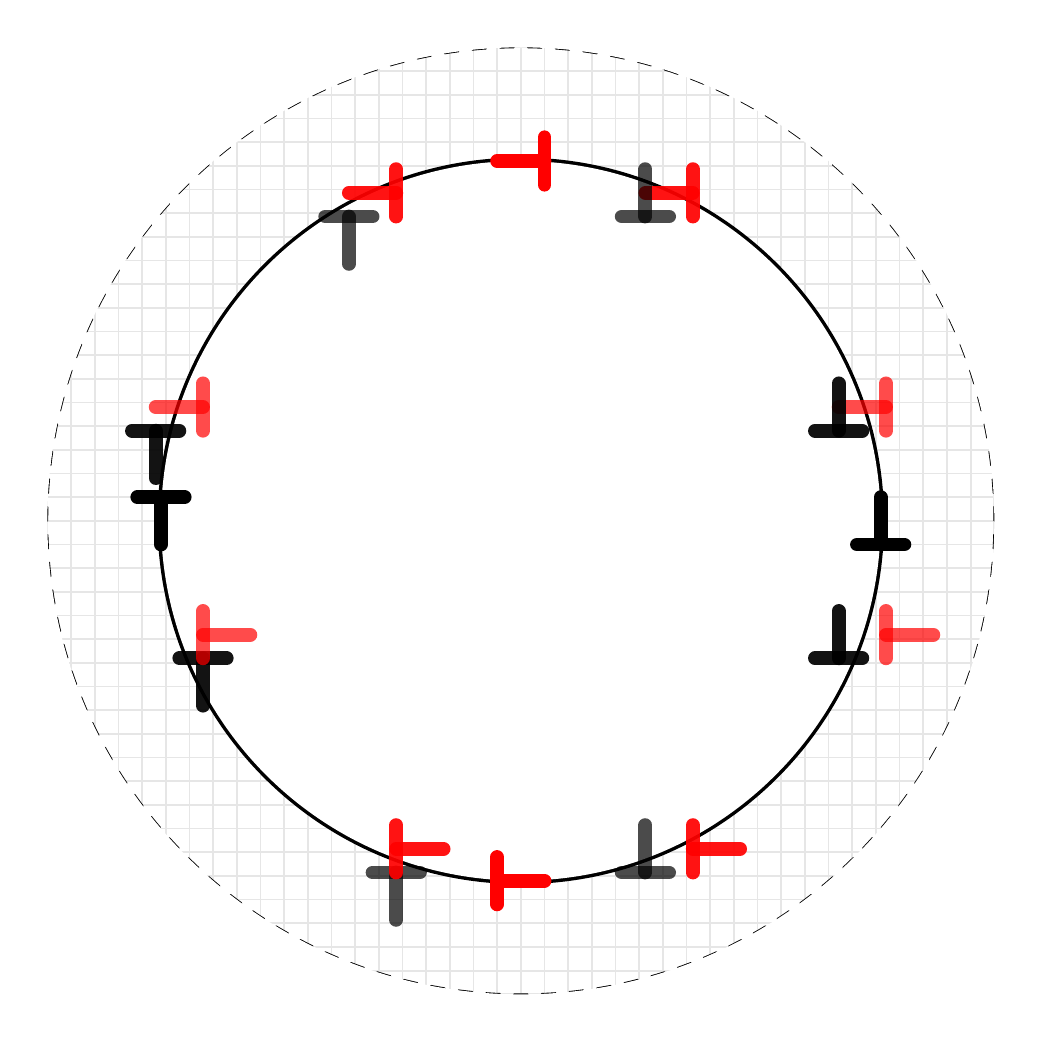}\label{fig:pcp_2d_shrink_grid3}}
    \caption{Mechanism for grain shrinkage with no rotation shown in the reference
        configuration, with the grid describing the slip planes. The circle
        drawn in solid line describes the position of the
        grain boundary. The GNDs on the grain boundary corresponding to the two
        slip systems are shown in red and black with the color intensity
        describing the magnitude of the dislocation density. Grain shrinkage occurs as the
        GNDs glide on their respect slip planes towards the interior of the
        grain. The four dislocations  on the original grain boundary, as shown
        in \protect\subref{fig:pcp_2d_shrink_grid2}, form pairs (marked by
        arrows) and move tangential to the grain boundary to
        annihilate, resulting in grain boundary shrinkage as shown in
        \protect\subref{fig:pcp_2d_shrink_grid3}.}
    \label{fig:pcp_2d_shrink_grid}
\end{figure}

In this numerical study we perform two simulations to demonstrate (i) grain
shrinkage with no rotation, and (ii) simultaneous grain shrinkage and rotation to
decrease the misorientation angle. For simplicity, we consider a circular grain
of initial radius $r_0$, with lattice orientation $\theta_0/2$ embedded inside a medium with lattice
orientation $-\theta_0/2$. We begin by postulating that the mechanism involved in
grain shrinkage is given by the following time-dependent plastic and elastic
distortion fields:
\begin{subequations}
\begin{align}
    \Fe(X_1,t) &= 
    \mathbbm 1_{(d(\bm X)<r_0-t)} \bm R \left( \frac{\theta_0}{2} \right)
    +\mathbbm 1_{(d(\bm X) \ge r_0-t)} \bm R \left(-\frac{\theta_0}{2}\right),
    \label{eqn:FeShrink}\\
    \Fp(X_1,t) &=
    \mathbbm 1_{(d(\bm X)<r_0-t)} \bm R \left(-\frac{\theta_0}{2}\right) +
    \mathbbm 1_{(d(\bm X) \ge r_0-t)} \bm R \left( \frac{\theta_0}{2}\right),
    \label{eqn:FpShrink}
\end{align}
\label{eqn:Shrink}
\end{subequations}
where $d(\bm X):=\sqrt{X_1^2+X_2^2}$. 
The equations in \eref{eqn:Shrink} result in $\bm F=\bm I$,
which implies there is no macroscopic deformation. Additionally, the
time-dependent GND tensor shown in Table \ref{table:catalog}
suggests the net dislocation content is directly proportional to the
radius of the grain. As noted in \sref{sec:numerics_shear},
the evolution of GNDs is made possible by the availability of the slip planes.
The expression for $\Lp$ for the mechanism described in eq.\ \eref{eqn:Shrink}
is shown in Table \ref{table:catalog}. Since the $1-2$ and $2-1$ components are
the only non-zero components, the two slip systems given in Table
\ref{table:catalog} are sufficient for the evolution of GNDs. Moreover,
it can also be shown that the same two slip systems also suffice for a diffuse
grain boundary. Based on the mechanism given in \eref{eqn:Shrink},
\fref{fig:pcp_2d_shrink_grid} depicts the motion of GNDs in a shrinking circular
grain boundary, and shows how dislocations are transported and
annihilated along the grain boundary.
\begin{figure}[t]
    \centering
    \subfloat[]
    {\includegraphics[scale=0.6]{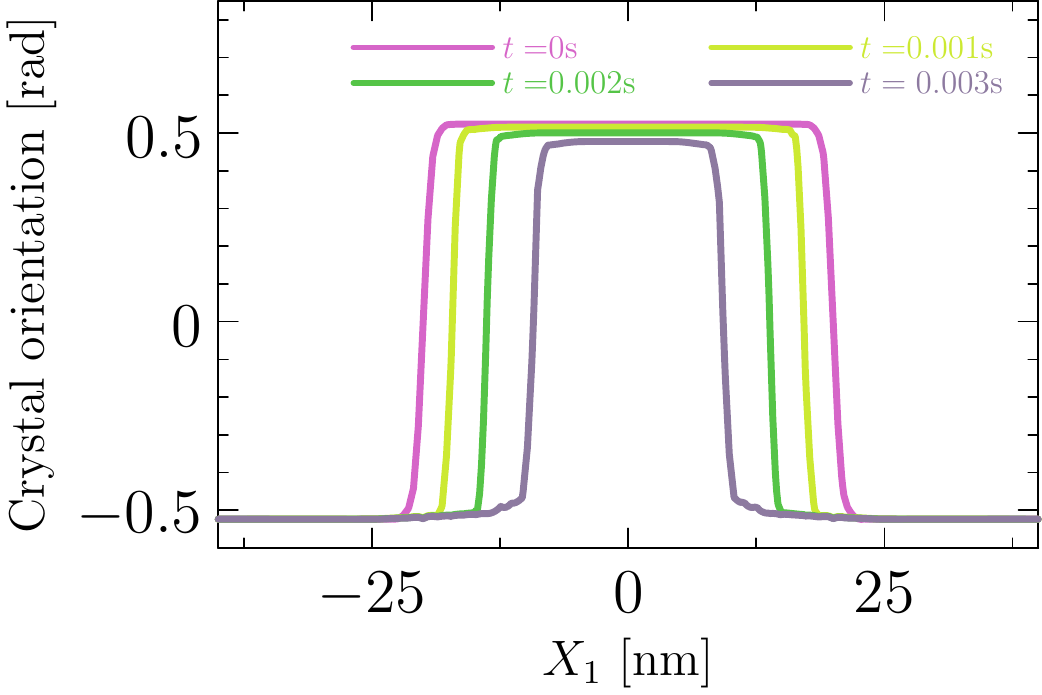}\label{fig:pcp_2d_shrink_thetae}}\\
    \subfloat[$t=\SI{0}{\second}$]{\includegraphics[scale=0.5]{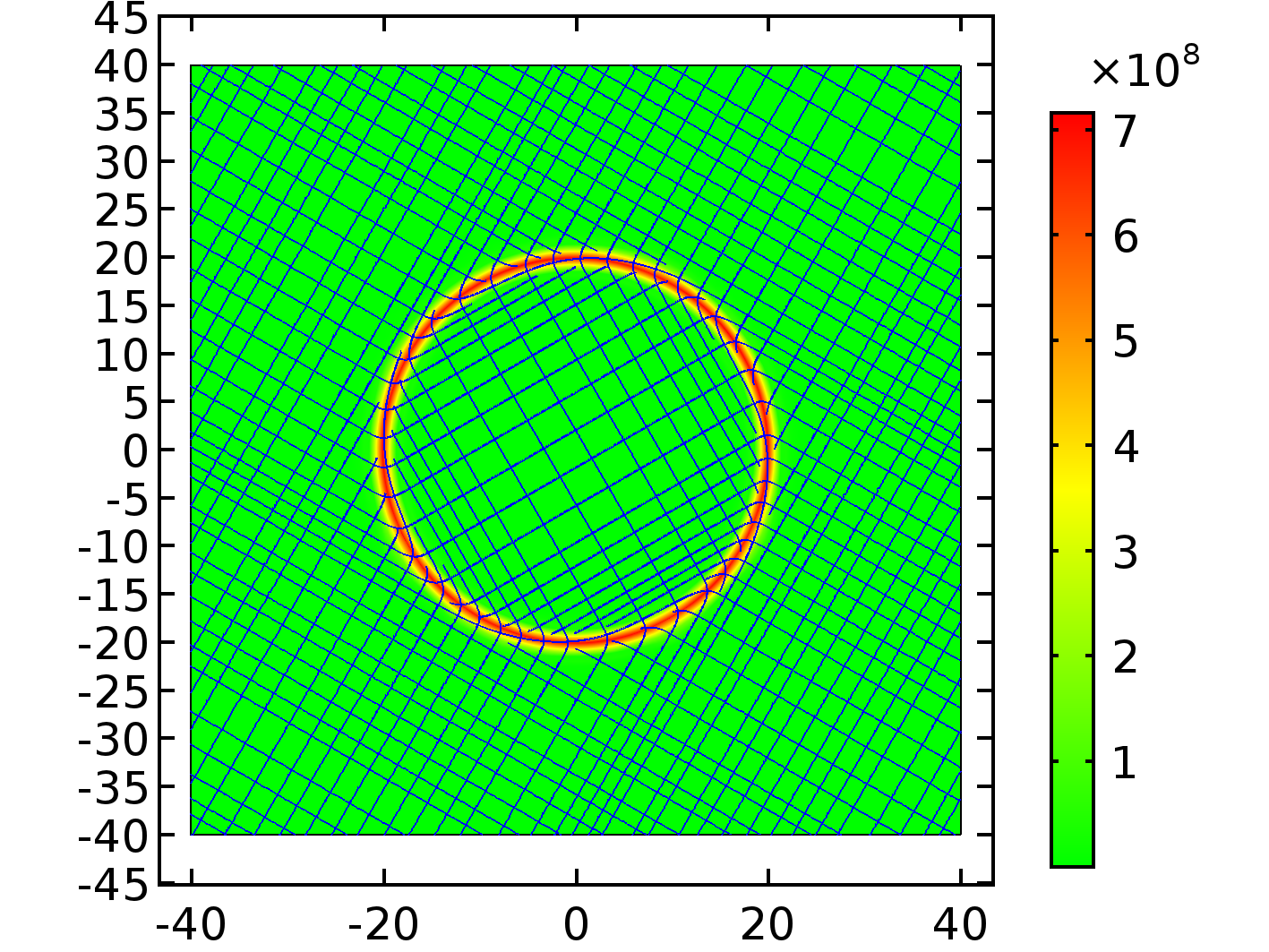}
    \label{fig:pcp_2d_shrink_lattice1}}
    \subfloat[$t=\SI{3e-3}{\second}$]{\includegraphics[scale=0.5]{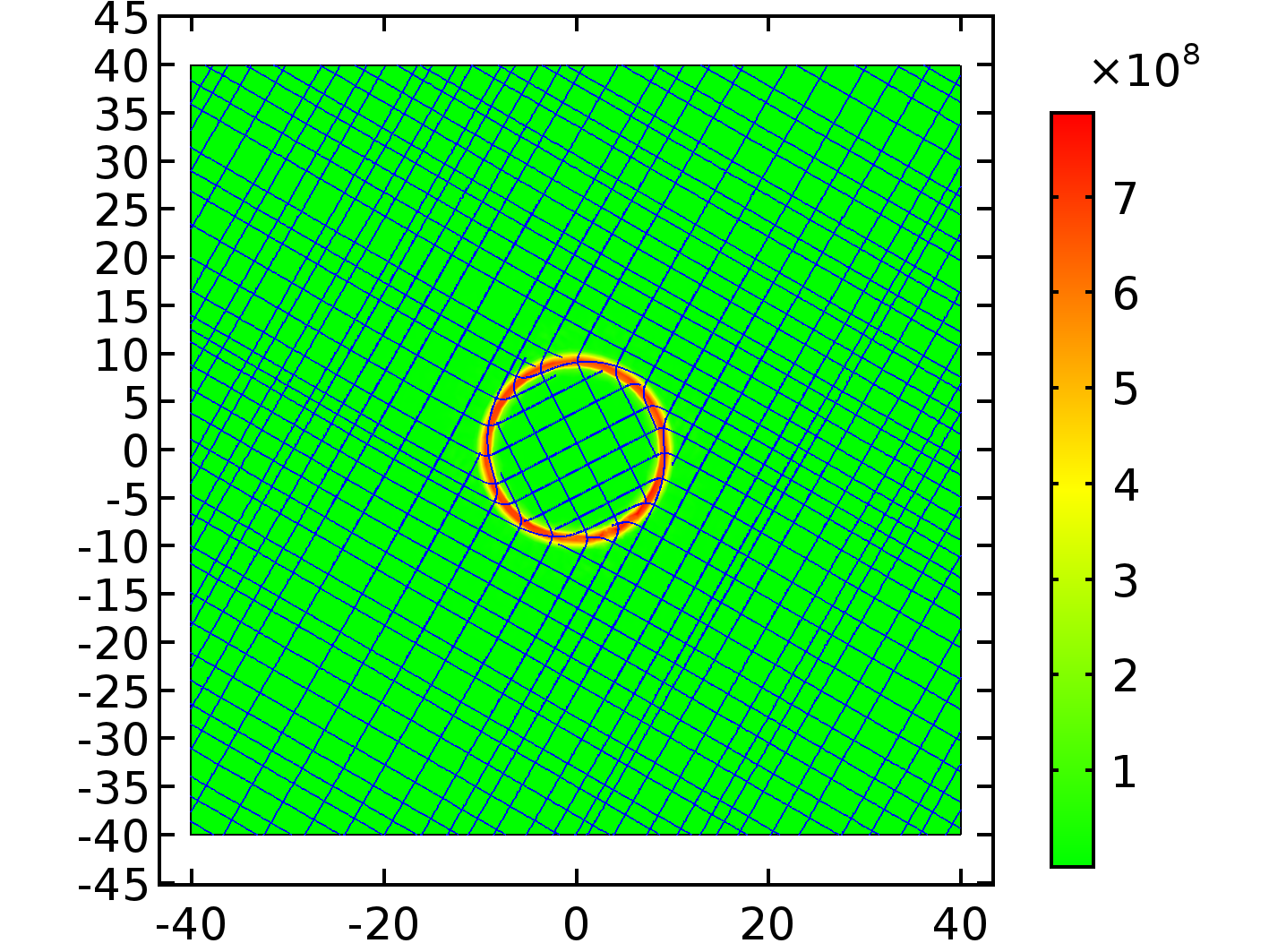}
    \label{fig:pcp_2d_shrink_lattice}}
    \caption{Grain shrinkage with no rotation: \protect\subref{fig:pcp_2d_shrink_thetae} Plots of the
        lattice orientation corresponding to the rotation $\bm R^{\rm L}$
    obtained from the polar decomposition of $\Fe$, along the horizontal central
    line $X_2 = 0$; \protect\subref{fig:pcp_2d_shrink_lattice1},
    \protect\subref{fig:pcp_2d_shrink_lattice} Streamlines tangential
        to the vector fields $\Fe \bm e^1$ and $\Fe \bm e^2$ plotted in
        in the reference configuration depicting grain shrinkage with no rotation. The
    color density corresponds to the norm of GND density in units of
    $\si{\per\meter}$. The variations
in the distances between parallel streamlines is an artefact of the algorithm
used to plot them, and should not be interpreted as lattice stretches.}
    \label{fig:pcp_2d_shrink}
\end{figure}
\begin{figure}[t]
    \centering
    \subfloat[]{\includegraphics[scale=0.6]{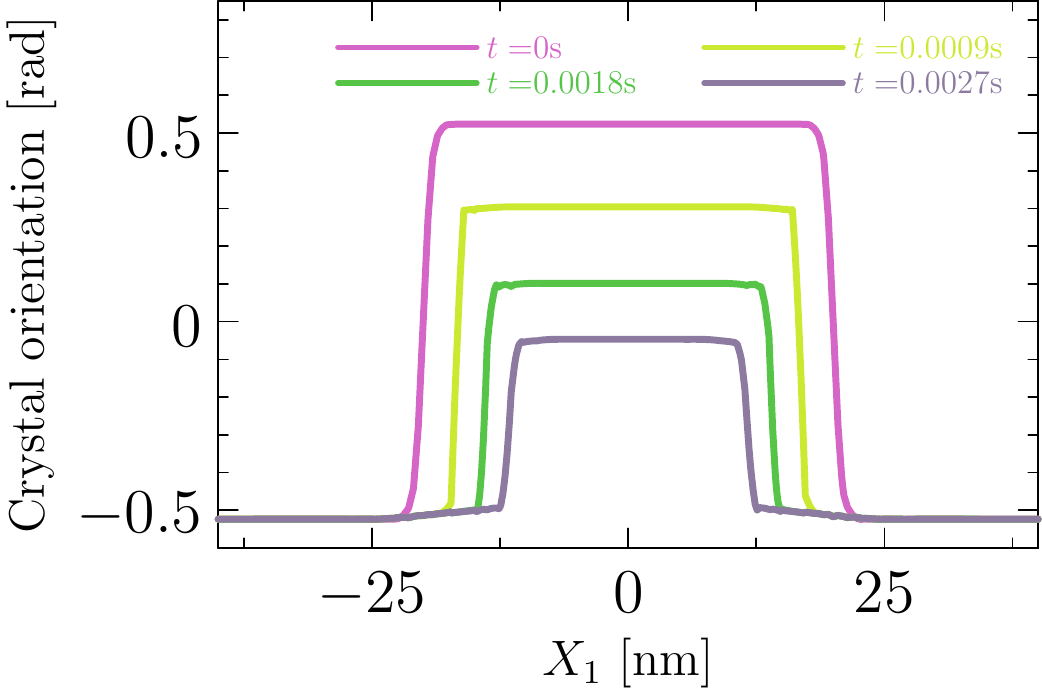}\label{fig:pcp_2d_rotate_thetae}}\\
    \subfloat[$t=\SI{0}{\second}$]{\includegraphics[scale=0.5]{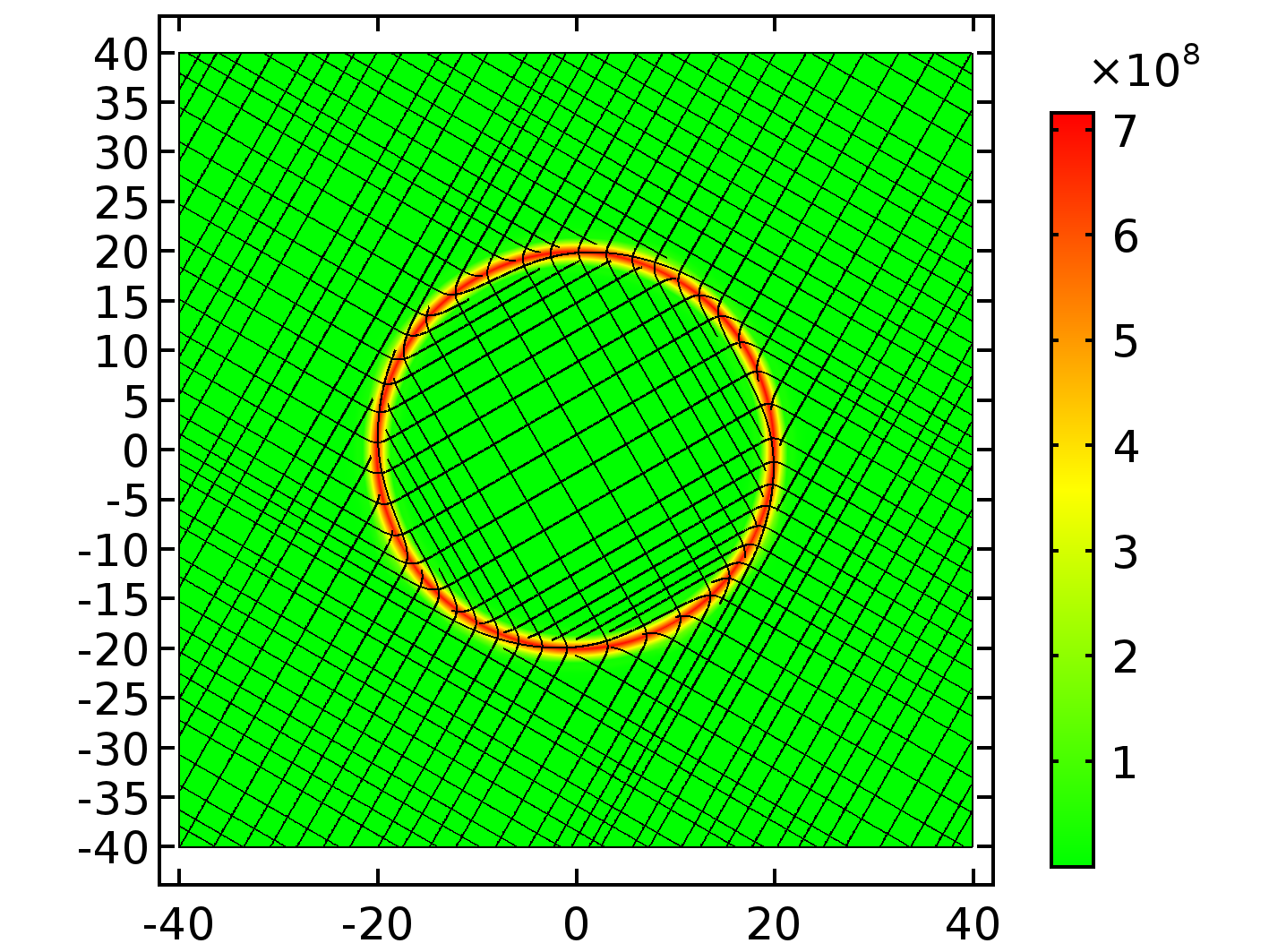}
    \label{fig:pcp_2d_rotate_lattice1}}
    \subfloat[$t=\SI{2.7e-3}{\second}$]{\includegraphics[scale=0.5]{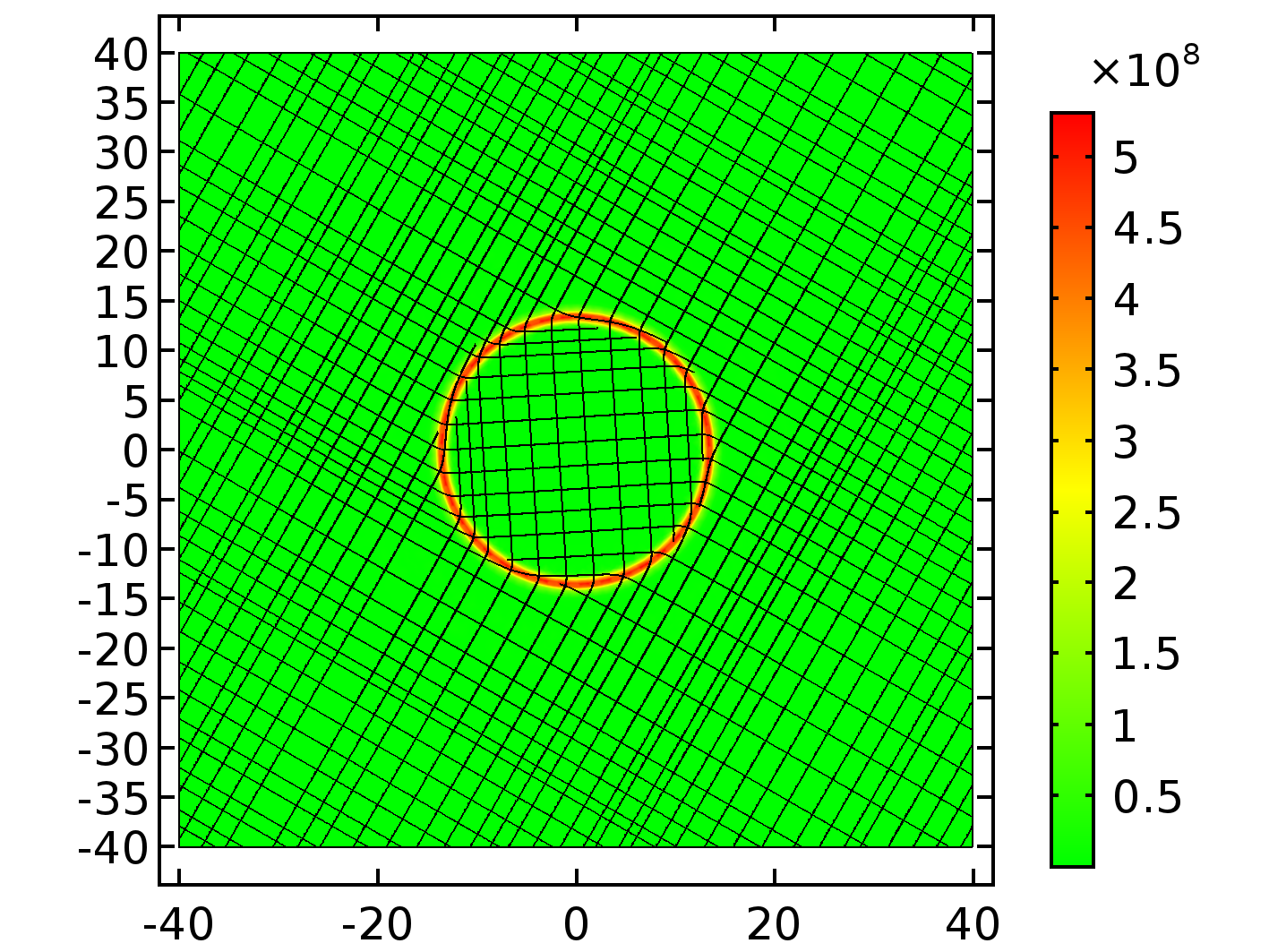}
    \label{fig:pcp_2d_rotate_lattice}}
    \caption{Grain shrinkage and rotation: \protect\subref{fig:pcp_2d_rotate_thetae} Plots of the
        lattice orientation corresponding to the rotation $\bm R^{\rm L}$
    obtained from the polar decomposition of $\Fe$, along the horizontal central
    line $X_2 = 0$; \protect\subref{fig:pcp_2d_rotate_lattice1},
    \protect\subref{fig:pcp_2d_rotate_lattice} Streamlines tangential
        to the vector fields $\Fe \bm e^1$ and $\Fe \bm e^2$, plotted
        in the reference configuration, depicting grain shrinkage and rotation. The
    color density corresponds to the norm of GND density in units of
    $\si{\per\meter}$. The spacings between parallel streamlines is an artifact of the algorithm
used to plot them, and should not be interpreted as lattice stretches.}
    \label{fig:pcp_2d_rotate}
\end{figure}
\begin{figure}[t]
    \centering
    \subfloat[]{\includegraphics[scale=0.6]{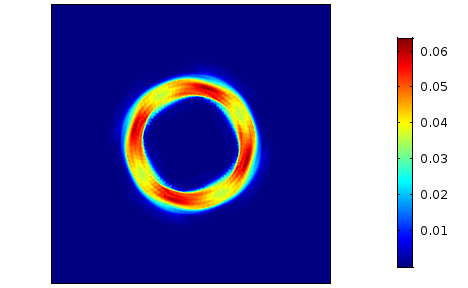}\label{fig:pcp_2d_shrink_ps}}
    \subfloat[]{\includegraphics[scale=0.6]{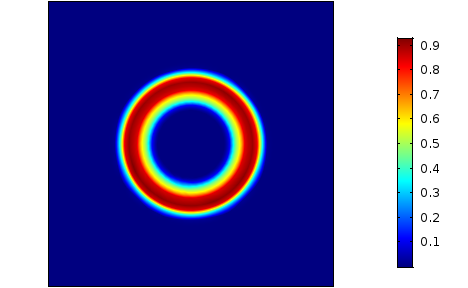}\label{fig:pcp_2d_rotate_ps}}
    \caption{Comparison of color density plots of plastic strain $|(\FpT\Fp-\bm I)/2|$
    for \protect\subref{fig:pcp_2d_shrink_ps} grain shrinkage with no rotation,
    and \protect\subref{fig:pcp_2d_rotate_ps} grain shrinkage with rotation,
showing negligible plastic strain for pure shrinkage. Both plots confirm no
dislocation activity in the grain interiors.}
\label{fig:pcp_2d_ps}
\end{figure}

We now simulate grain boundary shrinkage with no grain rotation in a square domain of size
$\SI{40}{\nm}$ with an embedded circular grain of radius $r_0=\SI{20}{\nm}$ centered
at the origin with radius $r_0=\SI{20}{\nm}$. Based on the discussion following
\eref{eqn:Shrink}, we enable two slip systems with slip directions
\begin{align*}
    (1,0), \quad (0,1).
\end{align*}
The initial conditions for the simulation are
\begin{align}
    \bm u(\bm X,0) = \bm 0, \quad
    \phi(\bm X,0) = 1, & \quad 
    \Fp(\bm X,0) = \bm R^{0\rm T}(\widetilde \theta(X_1)),
\end{align}
where
\begin{align}
    \widetilde\theta(X_1) =
    -\frac{\theta_0}{2}+\frac{\theta_0}{(1+\exp(-2.5(d(\bm X)-20))},
\end{align}
$\theta_0=\SI{60}{\degree}$, and $\bm R^{0}(\widetilde\theta)$ is
the rotation corresponding to $\widetilde \theta$. Dirichlet boundary conditions
are enforced on the boundary of the square domain:
\begin{align}
    \bm u(\partial \mathcal B,t) = \bm 0, \quad 
    \phi(\partial \mathcal B,t) = 1, \quad 
    v^\alpha(\partial \mathcal B,t) = 0.
\end{align}
The functional form of mobilities $b^\alpha$ ($\alpha=1,2$), and $b^\phi$ chosen in
\sref{sec:numerics_shear} are left unchanged. The results of the simulations are
shown in \fref{fig:pcp_2d_shrink}. \fref{fig:pcp_2d_shrink_thetae}
shows a plot of the lattice orientation $\theta^{\rm L}(\bm X,t)$ along the
$X_2=0$, clearly demonstrating grain boundary shrinkage with negligible
rotation. Figures \ref{fig:pcp_2d_shrink_lattice1} and
\ref{fig:pcp_2d_shrink_lattice} display the structure of the grain at two
different instants of time, $0$ and $3\times10^{-3}$ s. The color density in
these figures corresponds to the norm of GND density in units of
$\si{\per\meter}$.

Next, we simulate simultaneous grain
shrinkage and rotation by activating all the four slip systems given in
\eref{eqn:s}. The results of the simulation are shown in
\fref{fig:pcp_2d_rotate}, where the plots clearly demonstrate simultaneous GB
shrinkage and rotation. The two simulations described above highlight the
relationship between slip systems and the mechanisms of grain boundary
translation and rotation. Finally, \fref{fig:pcp_2d_ps} compares the norm of the
plastic strain, i.e.
$|(\FpT\Fp-\bm I)/2|$ for the two simulations described above.
\fref{fig:pcp_2d_shrink_ps} shows negligible plastic strain for the case of grain shrinkage
with no rotation, which is close to the zero plastic strain predicted by the
mechanism given in \eref{eqn:Shrink}. In addition, \fref{fig:pcp_2d_rotate_ps}
confirms grain rotates without any plastic activity in its interior.

\subsection{Dynamic recovery}\label{sec:recov}
\begin{figure}[t]
    \centering
    \subfloat{\includegraphics[scale=1.0]{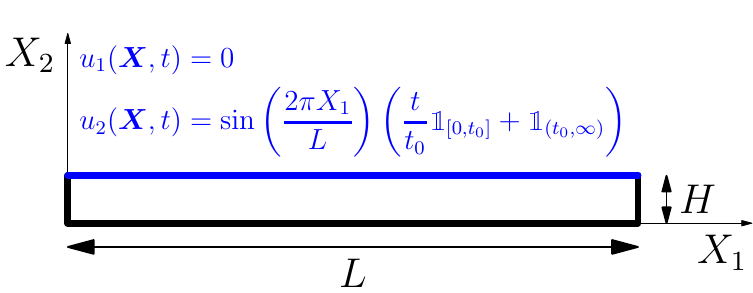}}
    \caption{A schematic of a rectangular slab with $L=\SI{40}{\nm}$ used to
    study dynamic recovery.
    The boundaries are subjected to zero-flux boundary condition in variables $\bm
    u$, $v^\alpha$ and $\phi$, except for a time-dependent Dirichlet boundary
    condition in the displacement variable on the top surface, with
    $t_0=\SI{0.2}{\second}$.}
    \label{fig:domainRecovery}
\end{figure}
\begin{figure}[t]
    \centering
    \subfloat[$t=\SI{0.16}{\second}$]{\includegraphics[scale=0.55]{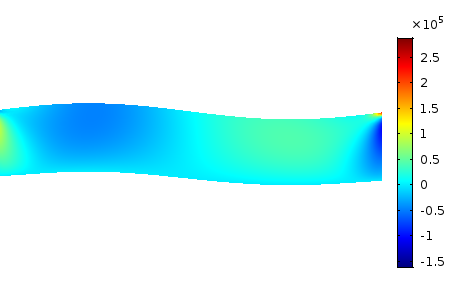}\label{fig:t1}}
    \subfloat[$t=\SI{1.22}{\second}$]{\includegraphics[scale=0.55]{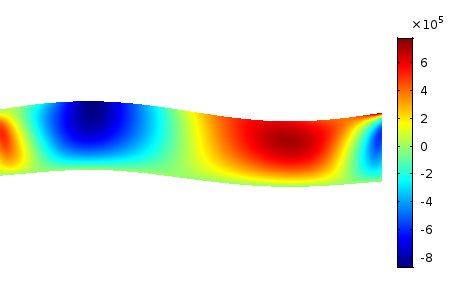}\label{fig:t2}} \\
    \subfloat[$t=\SI{1.96}{\second}$]{\includegraphics[scale=0.55]{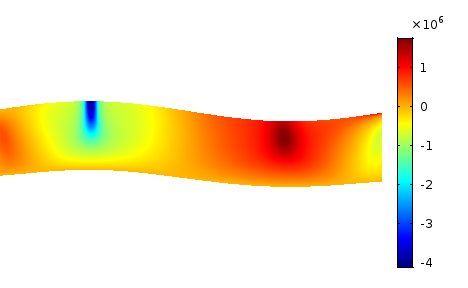}\label{fig:t3}}
    \subfloat[$t=\SI{2.37}{\second}$]{\includegraphics[scale=0.55]{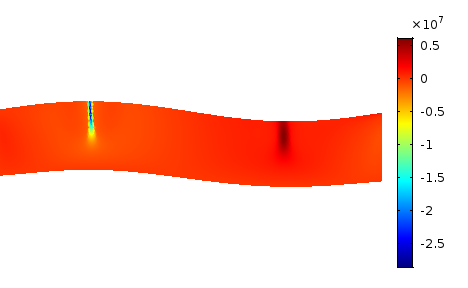}\label{fig:t4}}  \\
    \subfloat[$t=\SI{5}{\second}$]{\includegraphics[scale=0.55]{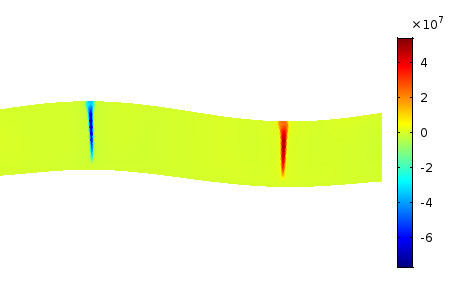}\label{fig:t5}}
    \subfloat[$t=\SI{8.11}{\second}$]{\includegraphics[scale=0.55]{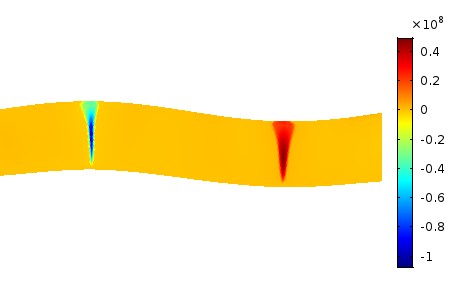}\label{fig:t6}}
    \caption{Color density plots of $|\bm G|$ at different times, plotted in the
    deformed configuration, and in units of \si{\meter^{-1}}.}
    \label{fig:polygonG}
\end{figure}
\begin{figure}[t]
    \centering
    \subfloat{\includegraphics[scale=0.5]{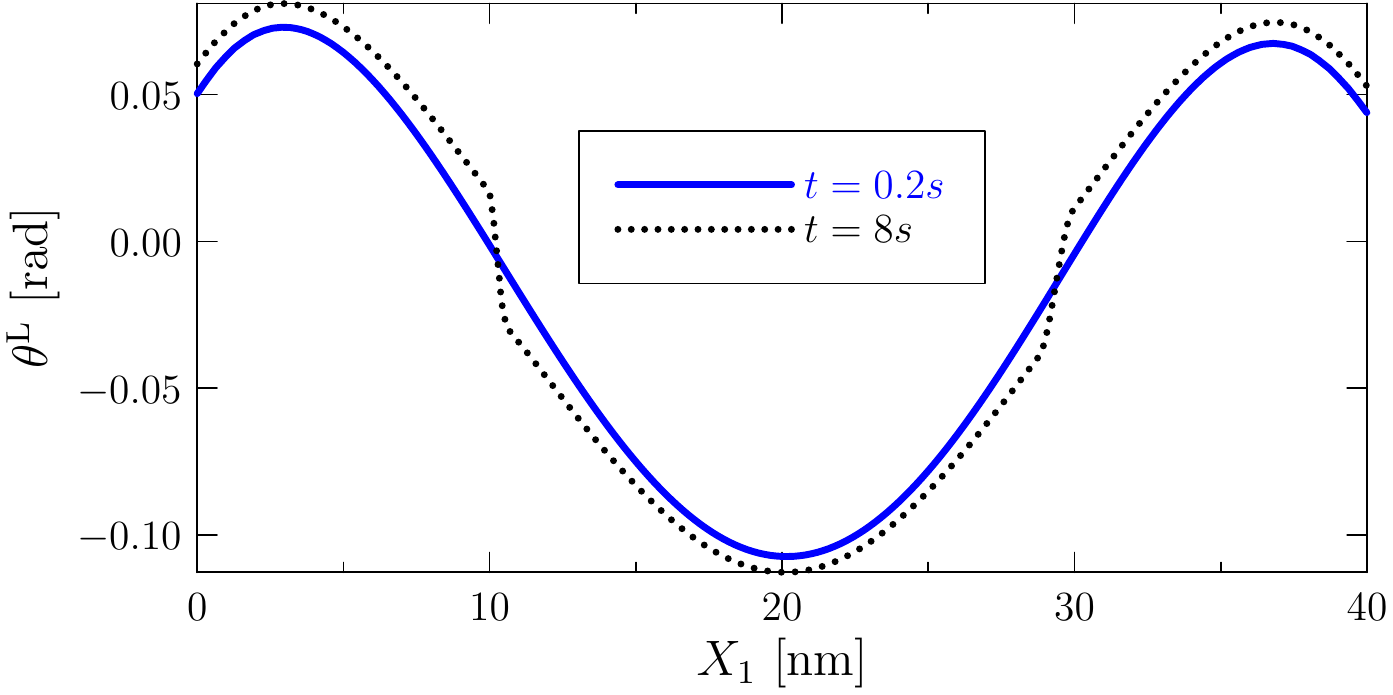}\label{fig:thetae}}
    \subfloat{\includegraphics[scale=0.5]{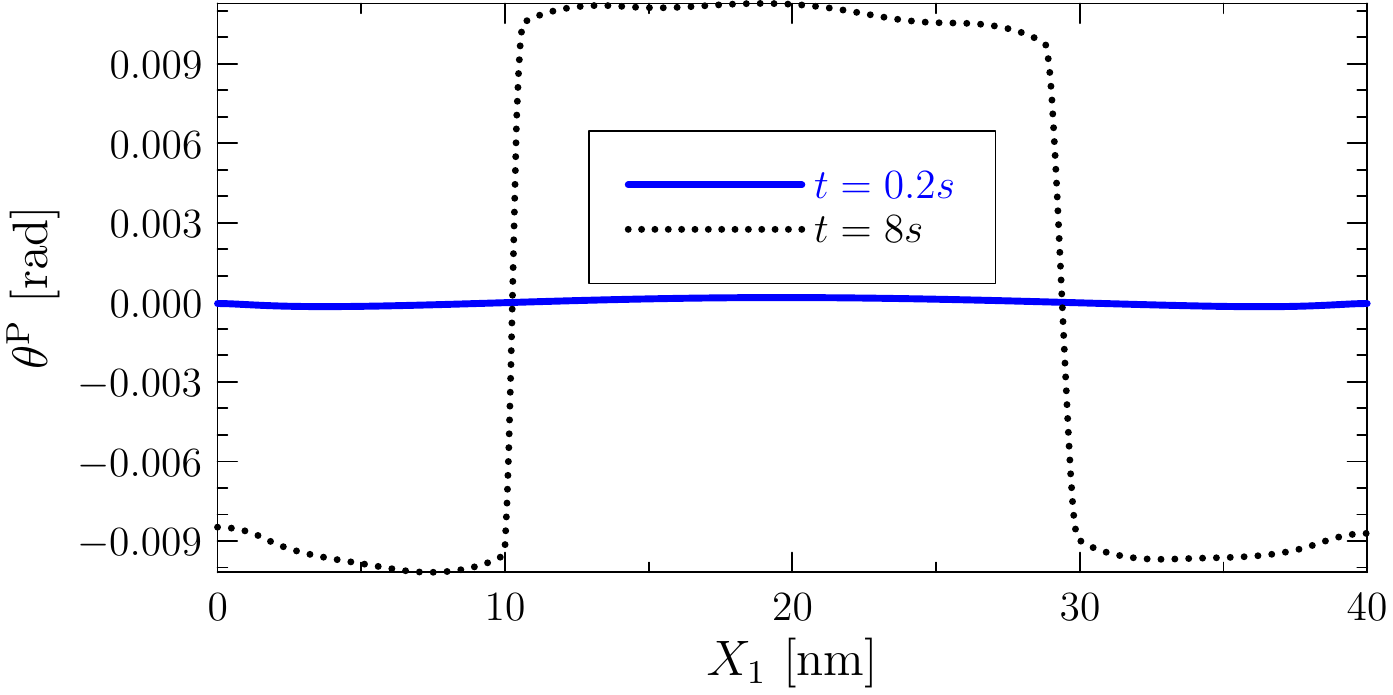}\label{fig:thetap}}
    \caption{Plots of lattice rotation, and plastic rotation plotted as 
        functions of the material coordinate
    $X_1$, along the line $X_2=\SI{20/12}{\nm}$. $\theta^{\rm L}$ and $\theta^{\rm
    P}$ are angles corresponding to $\bm R^{\rm L}$ and $\Rp$ in $SO(3)$ obtained from the
polar decomposition of $\Fe$ and $\Fp$ respectively.}
    \label{fig:polygonOrientation}
\end{figure}
In this section, we simulate the phenomenon of subgrain formation commonly
referred to as \emph{dynamic recovery}. During plastic deformation, the
dislocation density in a material increases rapidly in stage II resulting in hardening.
Following hardening, the existing dislocations consolidate forming
dislocation cell walls. The cells are interpreted as ``nucleated subgrains''
because the dislocation density within each cell is smaller than in its
boundary. Recovery precedes the stage of recrystallization which involves
spontaneous growth of a cell at the expense of others into a dislocation free
grain. Recovery during loading conditions is referred to as dynamic recovery.

Accordingly, the simulations are designed to deform a body and hold it in its deformed
configuration enabling the buildup of dislocation content due to plastic
distortion. The aim here is to study the long time evolution of the dislocation
network to eventually nucleate new subgrains. The simulated domain is a rectangular
single crystal of length $\SI{40}{\nm}$ with
an aspect ratio of $6:1$ as shown in \fref{fig:domainRecovery}. The top surface is deformed
gradually in the $X_2$ direction using a sinusoidal function with the maximum
amplitude of $\SI{1}{\nm}$ attained at $t=\SI{0.2}{\second}$. The top
surface is then held in this position for the rest of the simulation. The
remaining boundary conditions are all of the zero-flux kind. Unlike in previous
simulations, by allowing zero-flux slip rate on the boundaries we have allowed
for the buildup of a net non-zero dislocation content within the
body.\footnote{On the other hand, if $v^\alpha=0$ on the boundary, then the net
    dislocation content in the body does not change, and it remains zero since
we start with a perfect crystal.} See \fref{fig:domainRecovery} for the setup used in the simulation. 
The initial conditions are:
\begin{align}
    \bm u(\bm X,0) = \bm 0, \quad \phi(\bm X,0) = 1, \quad \Fp(\bm X,0) = \bm I,
    \quad v^\alpha(\bm X,0) = 0.
\end{align}
In this case, the system is equipped with three slip systems with slip
directions
\begin{align*}
(1,0), \quad (1,1), \quad (-1,1),
\end{align*}
and inverse mobility $b^1$ is given by \eref{eqn:invMobility}, while
$b^2=b^3=10^6 \times b^1$. In other words, dislocation activity in the first slip system is allowed
while it is hindered in the second and third slip systems, which are included
due to the diffuse nature of the model.
The remaining material parameters for this simulation remain unchanged except for $\gamma$
which is set to $\SI{2000}{\nm}$. This change is made to increase the
tendency of bulk dislocations to form grain boundaries.\footnote{See
    \sref{sec:numerics_ss} for the physical interpretation of $\gamma$ in the
current model.} The results of the simulation are shown in
\frefs{fig:polygonG}{fig:polygonOrientation}. \fref{fig:polygonG} shows the
color density plots of $|\bm G|$ for different times, plotted in the deformed
configuration. The figure clearly demonstrates the
buildup of bulk dislocations in the beginning of the simulation, which later
agglomerate to form two grain boundaries. \fref{fig:polygonOrientation} shows
plots of the lattice rotation and plastic rotation which are obtained from the
polar decomposition of $\Fe$ and $\Fp$ respectively. As seen in
\fref{fig:thetae}, the formation of grain boundaries leads to a discontinuity in
lattice rotation, resulting in a decrease in the gradient of lattice rotation
in the interior of the newly-formed grain. \fref{fig:thetap} demonstrates the
convergence of $\theta^{\rm P}$ to a step function conveying the formation of
grain boundaries.

\section{Discussion}

One of the most important issues in materials design is to understand the link
between microstructure and properties
\citep{kumar2002microstructural,mcdowell2007plasticity,mcdowell2008concurrent,MCDOWELL20101280}.
This link often connects several orders of magnitude in space and time, which
makes the formulation of physical models capable of spanning the relevant
spatio-temporal gap an extremely challenging problem. While this topic has been
attracting significant attention over the last two or so decades, our knowledge
of the mechanisms that govern materials evolution under a number of important
scenarios still presents many voids.

In terms of mechanical behavior, the study of polycrystal plasticity is one of
the essential pillars supporting the development of new structural materials.
Theory, modeling, and simulation has been consistently contributing to our
understanding of the connection between microstructure and strength, fracture,
or ductility. While tremendous progress has been made in the realm of `static'
properties, e.g.~single dislocation properties \citep{bulatov2006computer},
defect energetics and structure \citep{gehlen2012interatomic}, grain boundary
energies \citep{shenderova1998multiscale,TSCHOPP2008191}, strengthening mechanisms
\citep{osetsky2003atomic,erhart2013thermodynamic}, etc., our understanding of
the dynamic behavior of materials under stress at finite temperatures has been
relatively lagging. This includes processes such as recovery, annealing,
recrystallization, grain growth, etc., which are absolutely essential processes
in materials synthesis and fabrication \citep{tasan2015overview}. Under this
same general category can be included processes such as superplasticity or Coble
creep. This is the context within which we develop the present model: a
thermodynamically-consistent approach that can simulate the dynamic evolution of
polycrystals under the combined effects of temperature and mechanical
deformation. The key points of our development are several. First, we utilize a
special decomposition of the deformation gradient that allows us to define grain
boundaries as a geometric link between a single crystal and a polycrystal. That
is, rather than being ad-hoc structures introduced externally, grain boundaries
appear naturally within our formulation, as a necessity to preserve the
compatibility of the lattice in a multiple-grain configuration. 
Second, deformation and temperature-driven processes can be considered in unison
in our model. This means that we can study mechanically-driven processes, as
well as thermal processes, in conjunction. This is of course one of the
essential premises to model dynamic recovery and recrystallization during
high-temperature deformation. Third, this dislocation-based definition of GBs
allows for a seamless consideration of (dislocation-induced) bulk slip and grain
boundary-mediated plasticity. This is guaranteed by the very definition of GBs
within our formulation, which is done precisely in terms of a special class of
dislocations. Lastly, the free energy expression employed here follows --at its
core-- a standard form commonly used in crystal plasticity, which is why our
model is so easily integrable into existing crystal plasticity formulations.

These elements provide our model with a generality that we believe is quite
unique. As discussed in Section \ref{sec:intro}, the state of the art at present
in the modeling of coupled bulk and GB-mediated plasticity involves evolving the
GB microstructure and the dislocation network separately, and linking them
externally via some penalty function. Since our model originates from a single
free energy density, it truly permits simulating the
\emph{co-evolution} within the same framework of both sub-structures, enabling
the consideration of dynamic processes under deformation. Specifically, we have
looked at stress-driven processes and curvature-driven processes. From among the
stress-driven transformations, we have studied the two mechanisms: shear-induced
grain boundary motion and GB sliding. Their range of operation is thought to be
clearly differentiated by temperature with coupled motion occurring primarily at
low temperatures, and sliding at higher temperatures. 
Our model can yield the so-called coupling factor $\beta$, relating GB motion and grain translation, which makes it amenable to comparison to molecular dynamics simulations
\citep{cahn2006coupling,cahn2006duality,suzuki2005atomic,ivanov2008dynamics,trautt2012grain}.
In fact, for a coupled GB motion, $\beta$ is know to be a geometric
factor that depends only on the misorientation, and not on the 
inclination of the grain boundary. While various atomistic simulations have
demonstrated coupled boundary motion in symmetric tilt GBs,
the same is hard to reproduce for asymmetric tilt GBs
\citep{trautt2012coupled}. 
However, it can be shown in a similar fashion as in \sref{sec:numerics_shear} that normal stresses are needed to induce the necessary plastic distortion in asymmetric tilt boundaries. We conjecture that imposing the necessary stress boundary conditions
predicted by the current framework in an MD simulation of an asymmetric tilt GB, would result in a coupled grain boundary motion.
Sliding, for its part, occurs primarily at higher temperatures and low strain
rates, and contributes for instance to creep, superplasticity, failure of
ceramic materials at high temperature, etc
\citep{raj1971grain,gifkins1976grain,van2001grain}.

Regarding curvature-driven processes, they appear to decrease the excess energy
of a polycrystalline body\footnote{W.r.t.~a single crystal, i.e., the product of
the grain boundary area and the grain boundary energy}. From a fundamental point
of view, grains can shrink to reduce the GB area while keeping the
misorientation constant (grain shrinkage), or maintain their size while changing
their misorientation to lower GB energy values (grain rotation)
\cite{margulies2001situ}. Evidently, both phenomena are most commonly found
simultaneously during materials deformation, and both involve interactions
between grain boundary dislocations, as we have shown here. All the processes mentioned above
have been profusely investigated computationally in recent years
\citep{gottstein2009grain,upmanyu2006simultaneous,kobayashi1998,kobayashi2000},
and our approach draws in fact on knowledge acquired from these works. 

A different issue is the phenomenon of dislocation subgrain formation, commonly found at the end of dynamic recovery at elevated stress and/or temperature. Such process is defined by the self-assembly of stored dislocations above a critical density into lower-energy dislocation wall configurations. These \emph{walls} demarcate so-called subgrains, which have been theorized to be the precursors to kinetic hardening and recrystallization \citep{mughrabi1983dislocation,bay1992overview,sedlavcek2002subgrain}.
The driving force behind this fragmentation of the polycrystal into a collection
of subgrains has long been unclear, although it is thought to be related to slip
and hardening inhomogeneities across different crystal orientations in the
grains. These inhomogeneities must preserve the compatibility of the deformation
nonetheless, resulting in different grains suffering different lattice
rotations, thus constituting subgrains. It is then that the dislocations stored
at the boundaries of the regions change their character to GNDs, forming
subgrain boundaries without long-range stresses. A recent study by
\citet{xia2015computational} has demonstrated the formation of
subgrains using a mean field continuum dislocation dynamics model, with
cross slip playing a pivotal role.
However, it is important to note that the model presented by
\citet{xia2015computational} is a geometrically linear model, and the 
only driving force on dislocations is due to the resolved shear stress. On the
other hand, our current geometrically non-linear kinematic framework clearly
shows that one can construct a smooth dislocation density field (using arbitrary
rotation fields) that results in zero stress. In such a scenario, in the
presence of no resolved shear stress, there are no driving forces on the
dislocations to form subgrains. 
This clearly highlights the importance of additional driving forces arising from higher-order stresses ($\bm \xi^\alpha$) that are responsible for subgrain formation.
As shown in Section \ref{sec:recov}, we have induced
the formation of dislocation walls by bending stresses, which has been suggested
as the intrinsic cause behind the formation of dislocation cells and subgrains
\citep{sedlavcek2002subgrain}. Our simulations show a very clean assemblage of
dislocations into walls and the formation of misoriented grain boundaries. While
we do not allege to solve but a small aspect of a complex and rich process, we
believe that our model is capable of simulating the relevant mechanisms of
polygonization, and we continue to further study this process.

The elementary GB phenomena discussed here have been simulated for model demonstration purposes. Ultimately, the goal is to combine all these processes under a single simulation scenario, to study phenomena such as dynamic recovery, dynamic recrystallization, polygonization, etc., and provide an avenue for validation and benchmarking. In addition, we intend to parameterize the approach using atomistic calculations, following recent trends in the community worldwide. This is the subject of ongoing studies by the authors for which significant efforts in numerical efficiency and optimization of the implementation will be required. However, we are confident that our model will open up new opportunities to investigate the complex phenomena associated with polycrystal plasticity.


\section{Final summary}
As a final summary, the main features of the present work are listed below.
\begin{enumerate}
\item The framework used to construct the initial lattice strain-free
polycrystal begins with a single crystal as the reference
configuration. Using the decomposition $\Fe = \bm R^0(\bm
X)$ and $\Fp = \FeT$ of $\bm F(\bm X,0)\equiv \bm I$, where $\bm R^0$ is
a smoothened piecewise constant rotation field, we obtain a
lattice strain-free diffuse-interface polycrystal. The misorientation
between adjacent grains is a consequence of the
presence of GNDs, described by the tensor $\bm G=\Fp \Curl \Fp$, concentrated at
the grain boundaries. The 
framework described above is kinematically nonlinear and holds for arbitrary
misorientations. Geometric nonlinearity plays a key role in
obtaining lattice strain-free polycrystal because, in a linear theory,
a non-zero dislocation density always results in a non-zero lattice strain.

\item The model includes grain boundary and bulk elastic
energies. The bulk energy is the usual classical elastic energy which is a
function of the lattice Lagrangian strain. The construction of grain
boundaries using GNDs enables us to
formulate the grain boundary energy as a function of $\bm G$. The exact
form of this function is the most non-trivial part of this paper.
Inspired by the non-standard energy functional of the KWC model, which results in
a singular diffusive equation for its order parameter $\theta$ that represents
grain orientations, we construct
the grain boundary energy by replacing $\nabla \theta$ appearing in the KWC
energy density, with the GND density tensor $\bm G$. 

\item The model has been applied to the following fundamental grain boundary
    processes (discussed in Section \ref{sec:numerics}): planar grain boundary
    sliding and coupled motion, curvature-driven grain shrinkage, and
    curvature-driven grain rotation. As well, we have shown that our approach is
    capable of simulating the essential elements of the well-known process of
    polygonization, i.e. the self-assembly of dislocations into cell walls that
    gives rise to the formation of subgrains. A length scale parameter $\gamma$
    in the free energy functional describes the propensity of the dislocations to
    agglomerate to form grain boundaries. From a numerical viewpoint, increasing
    $\gamma$ increases the stiffness of the governing equations. Although the
    unified framework presented in this paper is dimension-independent, all our
    simulations are implemented in 2-d. A three-dimensional implementation would
    differ only in the representation of the plastic rotation, used to solve the
    flow rule. In our recent work \citep{admal:po:marian:2017}, where we
    introduced the abstract kinematic
    framework discussed in section 3, we have implemented a 3-d simulation of a
    polycrystal using an angle-axis representation of the plastic rotation. A
    similar strategy can be adopted for a 3-d simulation of the current model.
    
\item The approximation in the equivalence of $\bm G$ with $\nabla \theta$
results in residual lattice strains in the vicinity of the
grain boundary of an unstressed bicrystal in steady state. While this artifact does not 
limit the model's capacity to
simulate the various grain boundary-mediated plastic phenomena, we prove
an interesting identity which states that the exact gradient of the lattice
rotation can be additively decomposed into $\bm G$ and a term that depends only
on lattice stretch and its gradients. This lays the groundwork for a generalization
of the current model, where $\nabla \theta$ of the KWC energy density is
replaced by the exact lattice rotation gradient, thus resulting in a more
accurate steady state solution with zero lattice strain in an unstressed bicrystal.
Evidently, due to the presence of gradients in the lattice stretch, such a
generalization results in a lattice strain gradient model which we will pursue
in the near future.
\end{enumerate}

\section{Acknowledgements}
Useful discussions with Timofey Frolov and Eliot Fried are acknowledged. NA and
JM's work has been supported by the US Department of Energy's Office of Fusion Energy Sciences, grant
DE-SC0012774:0001. Computer time allocations at UCLA's IDRE Hoffman2
supercomputer are acknowledged.
\appendix
\section{Results on the 1-d KWC problem on $\Omega_0=[-L,L]$}
\label{sec:kwc_1d}

In this section, for the sake of completeness, we present a collection of
results \citep{lobkovsky2001} for the solution to the 1-D
KWC boundary value problem with boundary conditions
\begin{align}
    \phi(\pm L) = 1, \quad \theta(\pm L) = \pm \theta_0.
\end{align}
As an ansatz, it is assumed that there exists a region in $\Omega_0$ given by
$\Omegagb=(-l,l)$ where $\nabla \theta \ne \bm 0$, while $\nabla \theta
\equiv \bm 0$ in the region $(\Omegagbc)$, and $\phi$ and
$\theta$ are symmetric and anti-symmetric respectively about the origin.
Therefore, in the region $\Omegagbc$, $\phi$
satisfies the equation
\begin{align}
    \alpha^2 \phi_{,XX} - f_{,\phi} = 0.
    \label{eqn:phi_gbc}
\end{align}
Multiplying \eref{eqn:phi_gbc} by $\phi_{,X}$ and integrating, we obtain
\begin{align}
    \frac{\alpha^2}{2} \phi_{,X}^2 - f(\phi) = c, \quad
    (X \in \Omegagbc)
    \label{eqn:phix_gbc}
\end{align}
where $c$ is an integration constant. Integrating \eref{eqn:phix_gbc}, we obtain
the solution for $\phi$ outside $\Omegagb$ in its inversed form as
\begin{align}
    L-X=\int_\phi^1 \frac{\alpha}{\sqrt{2(f(\tilde\phi)+c)}} \, d\tilde\phi.
    \quad
    (X \in \Omegagbc)
    \label{eqn:phigbc}
\end{align}
In the region $\Omegagb$, since $\theta_{,X} \ne 0$, we have
$\theta,{_X}/|\theta,{_X}|\equiv 1$. Therefore, $\phi$ and $\theta$ satisfy the
equations
\begin{align}
    \begin{rcases}
        \alpha^2 \phi_{,XX} - f_{,\phi} - sg_{,\phi} |\theta_{,X}| = 0, \\
        \left ( 
        \epsilon^2 \theta_{,X} + sg(\phi)
        \right )_{,X} = 0.
    \end{rcases}
    (X \in \Omegagb)
    \label{eqn:gb}
\end{align}
Therefore, the term $(\epsilon^2 \theta_{,X} + sg(\phi))$ is constant in
$\Omegagb$. Since
$\theta_X(X)=0$ at $X=\pm l$, it follows that
\begin{align}
    \theta_{,X} = \frac{s(g(\phi_2)-g(\phi))}{\epsilon^2},\quad
    (X \in \Omegagb)
    \label{eqn:thetax_gb}
\end{align}
where $\phi_2 : = \phi(\pm l)$. Substituting \eref{eqn:thetax_gb} into
\eref{eqn:gb}, and multiplying \eref{eqn:gb} by $\phi,_X$ results in 
\begin{align}
    \left (
        \frac{\alpha^2}{2} \phi,_X - f(\phi) + \frac{s^2}{2
        \epsilon^2}(g(\phi_2)-g(\phi))^2
    \right ),_X = 0. \quad
    (X \in \Omegagb)
    \label{eqn:phixx}
\end{align}
Integrating \eref{eqn:phixx}, and noting from \eref{eqn:phix_gbc} that 
\begin{align}
    \frac{\alpha^2}{2} \phi_{,X}^2(\pm l) - f(\phi_2) = c,
\end{align}
it follows that
\begin{align}
    \frac{\alpha^2}{2} \phi,_X - f(\phi) + \frac{s^2}{2
    \epsilon^2}(g(\phi_2)-g(\phi))^2 = c.\quad
    (X \in \Omegagb)
    \label{eqn:phix}
\end{align}
Further integrating \eref{eqn:phix} results in 
\begin{align}
    X(\phi) = \int_{\phi_1}^{\phi} \frac{\alpha}{\sqrt{2(f(\tilde
    \phi)+c-\left (\frac{s}{\epsilon}\right )^2(g(\phi_2)-g(\tilde \phi))^2}}\,
    d\tilde\phi,
    \quad (\phi<\phi_2, X\in \Omegagb)
    \label{eqn:phigb}
\end{align}
where $\phi_1:=\phi(0)$. The solution for $\theta$ in $\Omegagb$ is obtained by
integrating \eref{eqn:thetax_gb} resulting in 
\begin{align}
    \theta(\phi) &= \frac{1}{\epsilon^2}\int_{\phi_1}^{\phi}
    \frac{s(g(\phi_2)-g(\tilde \phi))}{\phi,_X} \, d\tilde \phi, \quad 
    (\phi<\phi_2) \notag \\
    &=\frac{\alpha s}{\epsilon^2} \int_{\phi_1}^{\phi}
    \frac{g(\phi_2)-g(\tilde \phi)}{\sqrt{2(f(\tilde
    \phi)+c)-\left (\frac{s}{\epsilon}\right )^2(g(\phi_2)-g(\tilde\phi))^2 }} \,
    d\tilde\phi \quad 
    (\phi<\phi_2)
    \label{eqn:thetagb}
\end{align}
where in the last equality we have used the expression for $\phi,_X$ given in
\eref{eqn:phix}. Summarising the solution, \eref{eqn:phigbc} and
\eref{eqn:phigb} describe $\phi$ in the regions $\Omegagbc$ and $\Omegagb$
respectively, while \eref{eqn:thetagb} describes $\theta$ in the region
$\Omegagb$. The solution is expressed in terms of constants $c$, $\phi_1$ and
$\phi_2$ which are obtained implicitly from the relations
\begin{align}
    \theta_0 &= \frac{\alpha s}{\epsilon^2} \int_{\phi_1}^{\phi_2}
    \frac{g(\phi_2)-g(\tilde \phi)}{\sqrt{2(f(\tilde
    \phi)+c)-\left (\frac{s}{\epsilon}\right )^2(g(\phi_2)-g(\tilde\phi))^2 }} \,
    d\tilde\phi,  \notag \\
    c&=-f(\phi_1) + \frac{s^2}{2\epsilon^2}(g(\phi_2)-g(\phi_1))^2,\notag \\
    L&=\int_{\phi_1}^{\phi_2} \frac{\alpha}{\sqrt{2(f(\tilde
    \phi+c-\left (\frac{s}{\epsilon}\right )^2(g(\phi_2)-g(\tilde \phi))^2}}\,
    d\tilde\phi + \int_{\phi_2}^1 \frac{\alpha}{\sqrt{2(f(\tilde \phi)+c)}}\,
    d\tilde \phi.
\end{align}
Finally, the grain boundary energy is defined as the minimum of the KWC free
energy functional. It is instructive to observe the role of $\epsilon$ in the
KWC energy functional. In the limit $\epsilon \to 0$, it is easy to see that
$\phi_2\to \phi_1$, and $\theta$ converges to a step function with discontinuity
at $X=0$. Therefore, the quadratic term in $\nabla \theta$ in the KWC energy
functional serves as a regularization parameter.

\subsection*{Motion by mean curvature and mobilities}
\citet{lobkovsky2001} have shown that the sharp interface limit of equations
\eref{eqn:phi_dot} and \eref{eqn:theta_dot} gives rise the grain rotation and
grain boundary motion by curvature. In particular, the sharp interface limit
was obtained by studying the scaled KWC
Euler--Lagrange equations 
\begin{align}
    \varepsilon b^\phi \dot \phi &= 
    \varepsilon \alpha^2 \triangle \phi-\frac{1}{\varepsilon}f'(\phi)-
    sg,_\phi |\nabla \theta|, \notag\\
    \varepsilon b^\theta \dot \theta &=  \Div \left [
        \varepsilon \epsilon^2 \nabla \theta +
        sg \frac{\nabla \theta}{|\nabla \theta|}
    \right ],
\end{align}
in the limit $\varepsilon \to 0$ using the method of matched asymptotics. The
resulting sharp interface model is given by
\begin{align}
    \mathfrak v = -\gamma \kappa M,
\end{align}
where $\mathfrak v$, $\kappa$ and $M$ are the normal velocity, mean
curvature, and mobility of the interface describing the grain boundary, and
$\gamma$ is the grain boundary energy. The constants $M$ and $\gamma$ are given
in terms of the solution to \eref{eqn:y_dot} described above for $L=\infty$ as
\begin{align}
    M^{-1} = \int_0^\infty (b^\phi \phi,_X^2+b^\theta \theta,_X^2) \, dX,
\end{align}
and $\gamma$ is the corresponding evaluation of the KWC energy functional. Note
that in the limit $\epsilon\to 0$, the mobility tends to zero. This implies, in
addition to being a regularization parameter, $\epsilon$ plays an important role
in rendering positive mobility to the phase field model.

\subsection*{Numerical implementation of the KWC model in 1-d}
\begin{table}[t]
    \centering
    \begin{tabular}{|c|l|}
        \hline
        $\epsilon^2$     &      \SI{2.1333e-4}{\femto\joule\per\nm} \\
        $\alpha^2$       &      \SI{5.3e-3}{\femto\joule\per\nm} \\
        $s$              &      \SI{0.0017}{\femto\joule\per\square\nm}\\
        $e$              &      \SI{0.0021}{\femto\joule\per\cubic\nm}\\
        $\gamma$         &      \SI{500}{\nm}\\
        $\theta_0$       &      \ang{30}\\
        $L$               &      \SI{10}{nm}\\
        \hline
    \end{tabular}
    \caption{Parameters used in the implementation of the KWC model}
    \label{table:kwc}
\end{table}
We now present details of the numerical implementation of the KWC model in 1-d.
The simulation plots are presented in \sref{sec:kwc}. The governing equations
given in \eref{eqn:y_dot_approx} are
numerically solved on the domain $[-L,L]$, with $L=\SI{20}{\nm}$, using the finite element
method. The initial and boundary conditions used for the simulation are
\begin{align}
    \phi(X,0) = 1, &\quad \theta(X,0) =
    -\frac{\theta_0}{2}+\frac{\theta_0}{(1+\exp(-2.5X))}, \notag \\
    \phi(L,t) = \phi(-L,t) = 1, &\quad \theta(L,t) = -\theta(-L,t) = \theta_0/2,
\end{align}
where $\theta_0$ denotes the jump in the lattice orientation across the grain
boundary. The KWC parameters used for the simulation are listed in Table
\ref{table:kwc}. In addition, the inverse mobilities $b^\phi$ and $b^\theta$ are assumed to be constant,
and equal to $\SI{1}{(\femto\joule.\nano \second) \per \nm
\cubed }$. The variables $\theta$ and $\phi$ are
interpolated using the Lagrange quadratic
finite elements. Equations in \eref{eqn:y_dot_approx} are solved using the MUMPS direct solver, and
BDF (Backward Differential Formula) time stepping algorithm implemented in
\texttt{COMSOL 5.2}. The system is evolved until a steady state is reached.
\fref{fig:kwc_phi_theta} shows a comparison of the steady state solution obtained analytically
and numerically. It is clear that $\nabla \theta$ in the numerical solution is
not identically equal to zero. This is a result of simulating the approximate
model governed by \eref{eqn:y_dot_approx} as opposed to \eref{eqn:y_dot}. It
can the easily observed, although not shown here, that as $\gamma$ increases
the two solutions converge.\footnote{Of course, we cannot increase $\gamma$
    indefinitely as this increases the stiffness of the resulting
equations, thus resulting in higher computational cost.}

Next, we demonstrate grain shrinking and rotation of a circular grain of radius
$\SI{20}{\nm}$ with a misorientation of $\ang{60}$ embedded inside a square
domain of size $\SI{80}{\nm}$.
The initial and boundary conditions are given by 
\begin{align}
    \theta(\Omega,0) = -\frac{\theta_0}{2} +
    \frac{\theta_0}{1+\exp(-2.5(\sqrt{X_1^2+X_2^2}-20))},& \quad \phi(\Omega,0) = 1, \notag \\
    \theta(\partial \Omega,t) = \theta(\partial \Omega,0),& \quad \phi(\partial \Omega,t) = 1
\end{align}
respectively. The system lower its free energy by a combination of grain
shrinking and rotation. The two modes can be explored independently by appropriately
choosing the mobilities. Grain rotation is a result of choosing constant
mobilities, equal to $\SI{1}{(\femto\joule.\nano \second) \per \nm
\cubed }$, while 
\begin{align*}
    (b^\phi)^{-1} &= \SI{1}{\nm \cubed \per(\femto\joule.\nano \second) },\\
    (b^\theta)^{-1} &=
    m^\theta_{\rm{min}}+(1-\phi^3(10-15\phi+6\phi^2))(m^\theta_{\rm{max}}-m^\theta_{\rm{min}}),
\end{align*}                                           
where $m^\theta_{\rm{max}}= \SI{1}{\nm \cubed \per(\femto\joule.\nano \second) }$ and 
$m^\theta_{\rm{min}}= \SI{1e-9}{\nm \cubed \per(\femto\joule.\nano \second) }$ 
results in pure shrink. The two modes are shown in \fref{fig:kwc_rotshr}.

\section{The Coleman--Noll procedure}
\label{sec:cn_app}
In this section, we use the Coleman--Noll procedure to arrive at a
thermodynamically-consistent constitutive law for our model. As mentioned in
\sref{sec:cn}, the free energy is assumed to be a function of  $s=(T,\nabla
T,\Ee,\bm G,\phi,\dot \phi,\nabla \phi)$, $\bm v = (v^1,\dots,v^\alpha)$ and
$\nabla \bm v = (\nabla v^1,\dots,\nabla v^\alpha)$, and the fields $\eta$, $\bm
q$, $\bm \xi^\alpha$, $\pi$, $\Pi^\alpha$ and $\bm S$ are assumed to be
functions of $s$, $\bm v$, $\nabla \bm v$ and $\Fp$.
Substituting these functional forms into the inequality in
\eref{eqn:dissipation}, we obtain
\begin{align}
    (\psi,_T &+ \eta ) \dot T + 
    \psi,_{\nabla T} \cdot \nabla \dot T + 
    (\psi,_{\Ee}-\bm S) \cdot\dot{\mathbb E}^{\rm L} +
    (\psi,_\phi - \pi ) \dot \phi +
    \notag \\
    & (\psi,_{\nabla \phi} - \Ph )\cdot \nabla \dot \phi -
    \sum_{\alpha=1}^A 
    \left (
        \Pi^\alpha v^\alpha + \tau^\alpha v^\alpha +\bm \xi^\alpha \cdot \nabla v^\alpha 
    \right ) + 
    \psi,_{\bm G} \dot{\bm G} +  \notag \\
    & \psi,_{\bm v} \cdot \dot{\bm v} +
    + \psi,_{\nabla \bm v} \cdot \nabla \dot{\bm v} +
    + \psi,_{\dot \phi} \ddot \phi +
    \frac{\bm q \cdot \nabla T}{T} \le 0.
    \label{eqn:dissipation_full_app}
\end{align}
The above inequality must hold for all material points $\bm X \in \BO$.
Since $\nabla \dot T$ occurs in exactly one term in
\eref{eqn:dissipation_full_app}, and its coefficient is independent of $\nabla \dot
T$, the inequality in \eref{eqn:dissipation_full_app} can be violated unless
$\psi,_{\nabla \dot T}\equiv 0$. By a similar argument, we have $\psi,_{\dot
\phi}\equiv0$, $\psi,_{\bm v}\equiv \bm 0$, and $\psi,_{\nabla \bm v}\equiv \bm
0$. Therefore, $\psi$ does not depend
on $\nabla T$, $\dot \phi$, $\bm v$ and $\nabla \bm v$, resulting in a 
simplification of the inequality in \eref{eqn:dissipation_full_app} to
\begin{align}
    (\psi,_T + \eta ) \dot T &+ 
    (\psi,_{\Ee} - \bm S) \cdot \dot{\mathbb E}^{\rm L} +
    (\psi,_\phi - \pi ) \dot \phi +
    (\psi,_{\nabla \phi} - \Ph )\cdot \nabla \dot \phi - \notag \\
    &\sum_{\alpha=1}^A 
    \left (
        \Pi^\alpha v^\alpha +\bm \xi^\alpha \cdot \nabla v^\alpha 
        +\tau^\alpha v^\alpha
    \right ) + 
    \psi,_{\bm G} \dot{\bm G} +
    \frac{\bm q \cdot \nabla T}{T} \le 0.
    \label{eqn:dissipation_full_new_app}
\end{align}
Using an argument similar to the one following \eref{eqn:dissipation_full_app}, we
conclude that $\eta$, $\bm S$ and $\Ph$ are
independent of $\nabla T$, $\Fp$, $\bm v$, $\nabla \bm v$ and $\dot \phi$, and 
\begin{subequations}
    \label{eqn:relations_app}
    \begin{align}
        \eta(T,\Ee,\bm G,\phi,\nabla \phi) &= -\psi,_T,\label{eqn:relations1_app} \\
        \bm S(T,\Ee,\bm G,\phi,\nabla \phi)&= \psi,_{\Ee},
        \label{eqn:relations2_app}\\
        \Ph(T,\Ee,\bm G,\phi,\nabla \phi)&= \psi,_{\nabla \phi}.
        \label{eqn:relations3_app}
    \end{align}
\end{subequations}
The term $\dot{\bm G}$ appearing in \eref{eqn:dissipation_full_new_app} can be
expressed in terms of the slip rates using the evolution equation (see Section
11.3 in \citet{cermelli2002geometrically})
\begin{align}
    \dot{\bm G} &=
        \Lp \bm G + \bm G (\Lp)^{\rm T} + 
        \Jp \sum_{\alpha=1}^A ((\Fp)^{-T} \nabla v^\alpha \wedge \bm
        m^\alpha) \otimes \bm s^\alpha, \notag
\end{align}
which implies
\begin{align}
    \psi,_{\bm G} \cdot \dot{\bm G} &= \sum_{\alpha=1}^A \left [
        \psi_{\bm G} \cdot (\mathbb S^\alpha \bm G + \bm G (\mathbb S^\alpha)^{\rm T}) v^\alpha +
        \Jp (\Fp)^{-1} (\bm m^\alpha \wedge \psi,_{\bm G} \bm s^\alpha) \cdot
        \nabla v^\alpha
    \right ].
    \label{eqn:gnd_evolve_app}
\end{align}
Substituting \eref{eqn:relations_app} and \eref{eqn:gnd_evolve_app} into the dissipation
inequality \eref{eqn:dissipation_full_app}, we obtain
\begin{align}
    \label{eqn:dissipation_cancelled_app}
    (\psi,_\phi - \pi ) \dot \phi +
    \sum_{\alpha=1}^A 
    &\left [
        \left (
            \psi,_{\bm G} : (\mathbb S^\alpha \bm G + \bm G (\mathbb S^\alpha)^{\rm T})
            -\Pi^\alpha -\tau^\alpha\right) v^\alpha + \right. \notag \\ &\left.
        \left (
        \Jp (\Fp)^{-1} (\bm m^\alpha \wedge \psi,_{\bm G} \bm s^\alpha) -
        \bm \xi^\alpha \right ) \cdot \nabla v^\alpha
    \right ] + 
    \frac{\bm q \cdot \nabla T}{T} \le 0.
\end{align}
We now assume that the microscopic stress $\bm \xi^\alpha$ is additively decomposed
into an energetic part, which is independent of $\nabla \bm v$, and a
dissipative part that depends on $\nabla \bm v$:
\begin{align}
    \bm \xi^\alpha = \bm \xi^\alpha_{\rm{en}} + \bm \xi^\alpha_{\rm{d}}.
    \label{eqn:xi_split_app}
\end{align}
Substituting \eref{eqn:xi_split_app} into \eref{eqn:dissipation_cancelled_app}, and
using the Coleman--Noll procedure, results in 
\begin{align}
    \bm \xi ^\alpha_{\rm{en}} = \Jp (\Fp)^{-1} (\bm m^\alpha \wedge
    \psi,_{\bm G} \bm s^\alpha),
    \label{eqn:xi_relation_app}
\end{align}
which implies $\bm \xi^\alpha_{\rm{en}}$ is the distributed Peach--Koehler force
due to the pile up of dislocations, and $\bm \xi^\alpha_{\rm d}$ is the
dissipative microstress conjugate to the gradient in slip rate.
Substituting \eref{eqn:xi_split_app} and \eref{eqn:xi_relation_app} into
\eref{eqn:dissipation_cancelled_app}, the dissipation inequality reduces to
\begin{align}
    (\psi,_\phi - \pi ) \dot \phi +
    \sum_{\alpha=1}^A 
    \left [
        \left (
            \psi,_{\bm G} : (\mathbb S^\alpha \bm G + \bm G (\mathbb S^\alpha)^{\rm T})
            -\Pi^\alpha -\tau^\alpha
            \right) v^\alpha - \bm \xi^\alpha_{\rm d} \cdot \nabla v^\alpha
    \right ] +
    \frac{\bm q \cdot \nabla T}{T} \le 0.
    \label{eqn:dissipation_nopeach_app}
\end{align}
A solution to \eref{eqn:dissipation_nopeach_app} is given by
\begin{subequations}
    \begin{align}
        \Pi^\alpha &= \psi,_{\bm G} : (\mathbb S^\alpha \bm G + \bm G
        \mathbb S^{\alpha\rm T})-\tau^\alpha+b^\alpha(s,\bm v,\nabla \bm v)
        v^\alpha,\label{eqn:dissipative_relations1_app}\\
        \bm q &= - \bm K(s,\bm v,\nabla \bm v) \nabla T,
        \label{eqn:dissipative_relations2_app}\\
        \pi &= \psi,_\phi + b^\phi(s,\bm v,\nabla \bm v) \dot
        \phi,\label{eqn:dissipative_relations3_app} \\
        \bm \xi^\alpha_{\rm d} &= B^\alpha(s,\bm v,\nabla \bm v) \nabla v^\alpha, \label{eqn:xi_d_app}
    \end{align}
\end{subequations}
where $\bm K$ is the positive-definite thermal conductivity tensor, and the functions $b^\alpha$,
$B^\alpha$ and $b^\phi$ are positive-valued inverse mobility functions associated with
$v^\alpha$, $\nabla \bm v$ and $\dot \phi$ respectively.

\section{Gradient of lattice rotation}
\label{sec:theorem}
In this section, we derive the following relation (see \eref{eqn:decompose}) which expresses the gradient
of lattice rotation in terms of the GND tensor, the stretch tensor $\Ue$ obtained
from the polar decomposition $\Fe=\bm R^{\rm L} \Ue$, and the gradient of $\Ue$:
\begin{align}
    \ReT (\curl \ReT) = (J^{\rm L})^{-1}(\Ue \bm G \UeT + \Ue \ocurl \Ue),
    \label{eqn:decompose_app}
\end{align}
where $\curl$ and $\ocurl$ denote the curl operators with respect to the
deformed and lattice configurations respectively. 

We begin with an alternate representation of $\bm G$ in terms of $\Fe$,
given by
\begin{align}
    \bm G = J^{\rm L} \invFe \curl \invFe.
    \label{eqn:G_Fe}
\end{align}
Substituting the polar decomposition of $\Fe$ into \eref{eqn:G_Fe}, and
simplifying the resulting expression using indicial notation, we obtain
\begin{align}
    G_{\bar i \bar j} &= J^{\rm L} \invfe_{\bar i k} \left [
        \curl (\invFe)
    \right ]_{k\bar j} \notag \\
    &= J^{\rm L} \invue_{\bar i \bar k} \invre_{\bar k k} \epsilon_{krs}
    \invfe_{\bar j s,r} \notag \\
    &= J^{\rm L} \invue_{\bar i \bar k} \invre_{\bar k k} \epsilon_{krs}
    \left [
        \invue_{\bar j \bar l,r} \invre_{\bar l s} +
        \invue_{\bar j \bar l} \invre_{\bar l s,r}
    \right ], \label{eqn:g_rot}
\end{align}
where letters in subscript appearing with a bar indicate
components corresponding to the lattice configuration, while those in upper and
lower case indicate components corresponding to the reference and deformed
configurations respectively.
Since $\invue_{\bar j \bar l} \ue_{\bar l \bar t} = \delta_{\bar j \bar l}$,
its gradient with respect to the spatial coordinate is identically equal to
zero. In other words,
\begin{align}
    \invue_{\bar j\bar l,r} \ue_{\bar l \bar t} + \invue_{\bar j \bar l}
    \ue_{\bar l \bar t,r} = 0.
\end{align}
This implies
\begin{align}
    \invue_{\bar j \bar l,r} &= -\invue_{\bar j \bar p} \invue_{\bar t \bar l}
    \ue_{\bar p \bar t, r} \notag \\
    &=-\invue_{\bar j \bar p} \invue_{\bar t \bar l}
    \ue_{\bar p \bar t, \bar r} \invfe_{\bar r,r}  \notag \\
    &=-\invue_{\bar j \bar p} \invue_{\bar t \bar l}
    \ue_{\bar p \bar t, \bar r} \invue_{\bar r,\bar s} \invre_{\bar s r},
    \label{eqn:invUeGrad}
\end{align}
where in the second equality we have expressed the gradient of the stretch
tensor with respect to the coordinate in the lattice configuration.
Substituting \eref{eqn:invUeGrad} into \eref{eqn:g_rot}, we obtain
\begin{align}
    [\bm G]_{\bar i \bar j} &=J^{\rm L} \left [
        -\underbrace{\epsilon_{krs} \re_{k \bar k} \re_{r \bar s} \re_{s
        \bar l}}_{\epsilon_{\bar k \bar s \bar l} (\det \bm R^{\rm L})}
        \invue_{\bar i \bar k} \invue_{\bar j \bar p} \invue_{\bar t \bar l} \invue_{\bar r \bar s} 
        \ue_{\bar p \bar t, \bar r} +
        \invue_{\bar i \bar k} \re_{k \bar k} \epsilon_{krs}  \invre_{\bar l s,r}
        \invue_{\bar j \bar l} 
    \right ] \notag \\
    &=-J^{\rm L} \underbrace{\epsilon_{\bar k \bar s \bar l}
    \invue_{\bar i \bar k} \invue_{\bar r \bar s} \invue_{\bar t \bar l}}_{\epsilon_{\bar i \bar r \bar t} (\det \invUeT)} 
    \invue_{\bar j \bar p} \ue_{\bar p \bar t, \bar r} +
    J^{\rm L} [\invUe \ReT (\curl \ReT) \invUeT]_{\bar i \bar j} \notag \\ 
    &=-\epsilon_{\bar i \bar r \bar t} \ue_{\bar p \bar t, \bar r} 
    \invue_{\bar j \bar p}+
    J^{\rm L} [\invUe \ReT (\curl \ReT) \invUeT]_{\bar i \bar j}
    \notag \\ 
    &=[-(\ol{\curl}\, \Ue) \invUeT + J^{\rm L} \invFe (\curl \ReT)
    \invUeT]_{\bar i \bar j}.
    \label{eqn:decompose_ind}
\end{align}
Expressing \eref{eqn:decompose_ind} in direct notation, we have
\begin{align}
    \bm G = -(\ol \curl \, \Ue) \invUeT + J^{\rm L} \invFe (\curl \ReT) \invUeT,
\end{align}
which implies \eref{eqn:decompose_app}.

\section*{References}

\bibliography{references,refs_JM}

\begin{thebibliography}{72}
\expandafter\ifx\csname natexlab\endcsname\relax\def\natexlab#1{#1}\fi
\providecommand{\url}[1]{\texttt{#1}}
\providecommand{\href}[2]{#2}
\providecommand{\path}[1]{#1}
\providecommand{\DOIprefix}{doi:}
\providecommand{\ArXivprefix}{arXiv:}
\providecommand{\URLprefix}{URL: }
\providecommand{\Pubmedprefix}{pmid:}
\providecommand{\doi}[1]{\href{http://dx.doi.org/#1}{\path{#1}}}
\providecommand{\Pubmed}[1]{\href{pmid:#1}{\path{#1}}}
\providecommand{\bibinfo}[2]{#2}
\ifx\xfnm\relax \def\xfnm[#1]{\unskip,\space#1}\fi
\bibitem[{Cotterill and Mould(1976)}]{cotterill1976recrystallization}
\bibinfo{author}{P.~Cotterill}, \bibinfo{author}{P.~R. Mould},
  \bibinfo{title}{Recrystallization and grain growth in metals},
  \bibinfo{publisher}{Krieger Pub Co}, \bibinfo{year}{1976}.
\bibitem[{Humphreys and Hatherly(2012)}]{humphreys2012recrystallization}
\bibinfo{author}{F.~J. Humphreys}, \bibinfo{author}{M.~Hatherly},
  \bibinfo{title}{Recrystallization and related annealing phenomena},
  \bibinfo{publisher}{Elsevier}, \bibinfo{year}{2012}.
\bibitem[{Van~Swygenhoven(2002)}]{van2002grain}
\bibinfo{author}{H.~Van~Swygenhoven},
\newblock \bibinfo{title}{Grain boundaries and dislocations},
\newblock \bibinfo{journal}{Science} \bibinfo{volume}{296}
  (\bibinfo{year}{2002}) \bibinfo{pages}{66--67}.
\bibitem[{Evers et~al.(2004)Evers, Brekelmans, and Geers}]{evers2004scale}
\bibinfo{author}{L.~Evers}, \bibinfo{author}{W.~Brekelmans},
  \bibinfo{author}{M.~Geers},
\newblock \bibinfo{title}{Scale dependent crystal plasticity framework with
  dislocation density and grain boundary effects},
\newblock \bibinfo{journal}{International Journal of solids and structures}
  \bibinfo{volume}{41} (\bibinfo{year}{2004}) \bibinfo{pages}{5209--5230}.
\bibitem[{Cahn et~al.(2006)Cahn, Mishin, and Suzuki}]{cahn2006}
\bibinfo{author}{J.~W. Cahn}, \bibinfo{author}{Y.~Mishin},
  \bibinfo{author}{A.~Suzuki},
\newblock \bibinfo{title}{Coupling grain boundary motion to shear deformation},
\newblock \bibinfo{journal}{Acta Materialia} \bibinfo{volume}{54}
  (\bibinfo{year}{2006}) \bibinfo{pages}{4953 -- 4975}.
\bibitem[{Upmanyu et~al.(2006)Upmanyu, Srolovitz, Lobkovsky, Warren, and
  Carter}]{upmanyu2006simultaneous}
\bibinfo{author}{M.~Upmanyu}, \bibinfo{author}{D.~J. Srolovitz},
  \bibinfo{author}{A.~Lobkovsky}, \bibinfo{author}{J.~A. Warren},
  \bibinfo{author}{W.~Carter},
\newblock \bibinfo{title}{Simultaneous grain boundary migration and grain
  rotation},
\newblock \bibinfo{journal}{Acta Materialia} \bibinfo{volume}{54}
  (\bibinfo{year}{2006}) \bibinfo{pages}{1707--1719}.
\bibitem[{Mishin et~al.(2010)Mishin, Asta, and Li}]{mishin2010}
\bibinfo{author}{Y.~Mishin}, \bibinfo{author}{M.~Asta},
  \bibinfo{author}{J.~Li},
\newblock \bibinfo{title}{Atomistic modeling of interfaces and their impact on
  microstructure and properties},
\newblock \bibinfo{journal}{Acta Materialia} \bibinfo{volume}{58}
  (\bibinfo{year}{2010}) \bibinfo{pages}{1117--1151}.
\bibitem[{Trautt et~al.(2012)Trautt, Adland, Karma, and Mishin}]{trautt2012}
\bibinfo{author}{Z.~Trautt}, \bibinfo{author}{A.~Adland},
  \bibinfo{author}{A.~Karma}, \bibinfo{author}{Y.~Mishin},
\newblock \bibinfo{title}{Coupled motion of asymmetrical tilt grain boundaries:
  Molecular dynamics and phase field crystal simulations},
\newblock \bibinfo{journal}{Acta Materialia} \bibinfo{volume}{60}
  (\bibinfo{year}{2012}) \bibinfo{pages}{6528--6546}.
\bibitem[{Molodov et~al.(2015)Molodov, Barrales-Mora, and
  Brandenburg}]{molodov2015}
\bibinfo{author}{D.~Molodov}, \bibinfo{author}{L.~Barrales-Mora},
  \bibinfo{author}{J.~Brandenburg},
\newblock \bibinfo{title}{Grain boundary motion and grain rotation in aluminum
  bicrystals: recent experiments and simulations},
\newblock in: \bibinfo{booktitle}{IOP Conference Series: Materials Science and
  Engineering}, volume~\bibinfo{volume}{89}, \bibinfo{organization}{IOP
  Publishing}, \bibinfo{year}{2015}, p. \bibinfo{pages}{012008}.
\bibitem[{Elder et~al.(2002)Elder, Katakowski, Haataja, and
  Grant}]{elder2002modeling}
\bibinfo{author}{K.~Elder}, \bibinfo{author}{M.~Katakowski},
  \bibinfo{author}{M.~Haataja}, \bibinfo{author}{M.~Grant},
\newblock \bibinfo{title}{Modeling elasticity in crystal growth},
\newblock \bibinfo{journal}{Physical review letters} \bibinfo{volume}{88}
  (\bibinfo{year}{2002}) \bibinfo{pages}{245701}.
\bibitem[{Elder and Grant(2004)}]{elder2004modeling}
\bibinfo{author}{K.~Elder}, \bibinfo{author}{M.~Grant},
\newblock \bibinfo{title}{Modeling elastic and plastic deformations in
  nonequilibrium processing using phase field crystals},
\newblock \bibinfo{journal}{Physical Review E} \bibinfo{volume}{70}
  (\bibinfo{year}{2004}) \bibinfo{pages}{051605}.
\bibitem[{Yamanaka et~al.(2017)Yamanaka, McReynolds, and
  Voorhees}]{yamanaka2017phase}
\bibinfo{author}{A.~Yamanaka}, \bibinfo{author}{K.~McReynolds},
  \bibinfo{author}{P.~W. Voorhees},
\newblock \bibinfo{title}{Phase field crystal simulation of grain boundary
  motion, grain rotation and dislocation reactions in a bcc bicrystal},
\newblock \bibinfo{journal}{Acta Materialia} \bibinfo{volume}{133}
  (\bibinfo{year}{2017}) \bibinfo{pages}{160--171}.
\bibitem[{Taupin et~al.(2015)Taupin, Capolungo, Fressengeas, Upadhyay, and
  Beausir}]{taupin2015}
\bibinfo{author}{V.~Taupin}, \bibinfo{author}{L.~Capolungo},
  \bibinfo{author}{C.~Fressengeas}, \bibinfo{author}{M.~Upadhyay},
  \bibinfo{author}{B.~Beausir},
\newblock \bibinfo{title}{A mesoscopic theory of dislocation and disclination
  fields for grain boundary-mediated crystal plasticity},
\newblock \bibinfo{journal}{International Journal of Solids and Structures}
  \bibinfo{volume}{71} (\bibinfo{year}{2015}) \bibinfo{pages}{277--290}.
\bibitem[{Kobayashi et~al.(2000)Kobayashi, Warren, and {Craig
  Carter}}]{kobayashi2000}
\bibinfo{author}{R.~Kobayashi}, \bibinfo{author}{J.~A. Warren},
  \bibinfo{author}{W.~{Craig Carter}},
\newblock \bibinfo{title}{{A continuum model of grain boundaries}},
\newblock \bibinfo{journal}{Physica D: Nonlinear Phenomena}
  \bibinfo{volume}{140} (\bibinfo{year}{2000}) \bibinfo{pages}{141--150}.
\bibitem[{Kobayashi et~al.(1998)Kobayashi, Warren, and Carter}]{kobayashi1998}
\bibinfo{author}{Kobayashi}, \bibinfo{author}{Warren},
  \bibinfo{author}{Carter},
\newblock \bibinfo{title}{{Vector-valued phase field model for crystallization
  and grain boundary formation}} \bibinfo{volume}{119} (\bibinfo{year}{1998})
  \bibinfo{pages}{415--423}.
\bibitem[{Steinbach and Pezzolla(1999)}]{steinbach1999}
\bibinfo{author}{I.~Steinbach}, \bibinfo{author}{F.~Pezzolla},
\newblock \bibinfo{title}{A generalized field method for multiphase
  transformations using interface fields},
\newblock \bibinfo{journal}{Physica D: Nonlinear Phenomena}
  \bibinfo{volume}{134} (\bibinfo{year}{1999}) \bibinfo{pages}{385--393}.
\bibitem[{III and Chen(2002)}]{krill2002}
\bibinfo{author}{C.~K. III}, \bibinfo{author}{L.-Q. Chen},
\newblock \bibinfo{title}{Computer simulation of 3-d grain growth using a
  phase-field model},
\newblock \bibinfo{journal}{Acta Materialia} \bibinfo{volume}{50}
  (\bibinfo{year}{2002}) \bibinfo{pages}{3059 -- 3075}.
\bibitem[{Raabe(2002)}]{raabe2002}
\bibinfo{author}{D.~Raabe},
\newblock \bibinfo{title}{Cellular automata in materials science with
  particular reference to recrystallization simulation},
\newblock \bibinfo{journal}{Annual review of materials research}
  \bibinfo{volume}{32} (\bibinfo{year}{2002}) \bibinfo{pages}{53--76}.
\bibitem[{Gurtin(2000)}]{gurtin1}
\bibinfo{author}{M.~E. Gurtin},
\newblock \bibinfo{title}{On the plasticity of single crystals: free energy,
  microforces, plastic-strain gradients},
\newblock \bibinfo{journal}{Journal of the Mechanics and Physics of Solids}
  \bibinfo{volume}{48} (\bibinfo{year}{2000}) \bibinfo{pages}{989--1036}.
\bibitem[{Gurtin(2008)}]{gurtin2}
\bibinfo{author}{M.~E. Gurtin},
\newblock \bibinfo{title}{A finite-deformation, gradient theory of
  single-crystal plasticity with free energy dependent on densities of
  geometrically necessary dislocations},
\newblock \bibinfo{journal}{International Journal of Plasticity}
  \bibinfo{volume}{24} (\bibinfo{year}{2008}) \bibinfo{pages}{702--725}.
\bibitem[{Ma et~al.(2006)Ma, Roters, and Raabe}]{ma2006consideration}
\bibinfo{author}{A.~Ma}, \bibinfo{author}{F.~Roters},
  \bibinfo{author}{D.~Raabe},
\newblock \bibinfo{title}{On the consideration of interactions between
  dislocations and grain boundaries in crystal plasticity finite element
  modeling--theory, experiments, and simulations},
\newblock \bibinfo{journal}{Acta Materialia} \bibinfo{volume}{54}
  (\bibinfo{year}{2006}) \bibinfo{pages}{2181--2194}.
\bibitem[{Van~Beers et~al.(2013)Van~Beers, McShane, Kouznetsova, and
  Geers}]{van2013}
\bibinfo{author}{P.~Van~Beers}, \bibinfo{author}{G.~McShane},
  \bibinfo{author}{V.~Kouznetsova}, \bibinfo{author}{M.~Geers},
\newblock \bibinfo{title}{Grain boundary interface mechanics in strain gradient
  crystal plasticity},
\newblock \bibinfo{journal}{Journal of the Mechanics and Physics of Solids}
  \bibinfo{volume}{61} (\bibinfo{year}{2013}) \bibinfo{pages}{2659--2679}.
\bibitem[{Wulfinghoff et~al.(2013)Wulfinghoff, Bayerschen, and
  Böhlke}]{WULFINGHOFF201333}
\bibinfo{author}{S.~Wulfinghoff}, \bibinfo{author}{E.~Bayerschen},
  \bibinfo{author}{T.~Böhlke},
\newblock \bibinfo{title}{A gradient plasticity grain boundary yield theory},
\newblock \bibinfo{journal}{International Journal of Plasticity}
  \bibinfo{volume}{51} (\bibinfo{year}{2013}) \bibinfo{pages}{33 -- 46}.
\bibitem[{Wei and Anand(2004)}]{wei2004grain}
\bibinfo{author}{Y.~Wei}, \bibinfo{author}{L.~Anand},
\newblock \bibinfo{title}{Grain-boundary sliding and separation in
  polycrystalline metals: application to nanocrystalline fcc metals},
\newblock \bibinfo{journal}{Journal of the Mechanics and Physics of Solids}
  \bibinfo{volume}{52} (\bibinfo{year}{2004}) \bibinfo{pages}{2587--2616}.
\bibitem[{Gurtin and Anand(2008)}]{gurtin2008nanocrystalline}
\bibinfo{author}{M.~E. Gurtin}, \bibinfo{author}{L.~Anand},
\newblock \bibinfo{title}{Nanocrystalline grain boundaries that slip and
  separate: A gradient theory that accounts for grain-boundary stress and
  conditions at a triple-junction},
\newblock \bibinfo{journal}{Journal of the Mechanics and Physics of Solids}
  \bibinfo{volume}{56} (\bibinfo{year}{2008}) \bibinfo{pages}{184--199}.
\bibitem[{Basak and Gupta(2015)}]{basak2015}
\bibinfo{author}{A.~Basak}, \bibinfo{author}{A.~Gupta},
\newblock \bibinfo{title}{Simultaneous grain boundary motion, grain rotation,
  and sliding in a tricrystal},
\newblock \bibinfo{journal}{Mechanics of Materials} \bibinfo{volume}{90}
  (\bibinfo{year}{2015}) \bibinfo{pages}{229--242}.
\bibitem[{Frolov and Mishin(2012)}]{frolov2012}
\bibinfo{author}{T.~Frolov}, \bibinfo{author}{Y.~Mishin},
\newblock \bibinfo{title}{Thermodynamics of coherent interfaces under
  mechanical stresses. i. theory},
\newblock \bibinfo{journal}{Physical Review B} \bibinfo{volume}{85}
  (\bibinfo{year}{2012}) \bibinfo{pages}{224106}.
\bibitem[{Raabe and Becker(2000)}]{raabe2000coupling}
\bibinfo{author}{D.~Raabe}, \bibinfo{author}{R.~C. Becker},
\newblock \bibinfo{title}{Coupling of a crystal plasticity finite-element model
  with a probabilistic cellular automaton for simulating primary static
  recrystallization in aluminium},
\newblock \bibinfo{journal}{Modelling and Simulation in Materials Science and
  Engineering} \bibinfo{volume}{8} (\bibinfo{year}{2000}) \bibinfo{pages}{445}.
\bibitem[{Abrivard et~al.(2012{\natexlab{a}})Abrivard, Busso, Forest, and
  Appolaire}]{abrivard2012a}
\bibinfo{author}{G.~Abrivard}, \bibinfo{author}{E.~P. Busso},
  \bibinfo{author}{S.~Forest}, \bibinfo{author}{B.~Appolaire},
\newblock \bibinfo{title}{Phase field modelling of grain boundary motion driven
  by curvature and stored energy gradients. part ii: application to
  recrystallisation},
\newblock \bibinfo{journal}{Philosophical magazine} \bibinfo{volume}{92}
  (\bibinfo{year}{2012}{\natexlab{a}}) \bibinfo{pages}{3643--3664}.
\bibitem[{Abrivard et~al.(2012{\natexlab{b}})Abrivard, Busso, Forest, and
  Appolaire}]{abrivard2012b}
\bibinfo{author}{G.~Abrivard}, \bibinfo{author}{E.~P. Busso},
  \bibinfo{author}{S.~Forest}, \bibinfo{author}{B.~Appolaire},
\newblock \bibinfo{title}{Phase field modelling of grain boundary motion driven
  by curvature and stored energy gradients. part i: theory and numerical
  implementation},
\newblock \bibinfo{journal}{Philosophical magazine} \bibinfo{volume}{92}
  (\bibinfo{year}{2012}{\natexlab{b}}) \bibinfo{pages}{3618--3642}.
\bibitem[{Bernacki et~al.(2011)Bernacki, Log{\'e}, and Coupez}]{bernacki2011}
\bibinfo{author}{M.~Bernacki}, \bibinfo{author}{R.~E. Log{\'e}},
  \bibinfo{author}{T.~Coupez},
\newblock \bibinfo{title}{Level set framework for the finite-element modelling
  of recrystallization and grain growth in polycrystalline materials},
\newblock \bibinfo{journal}{Scripta Materialia} \bibinfo{volume}{64}
  (\bibinfo{year}{2011}) \bibinfo{pages}{525--528}.
\bibitem[{Takaki et~al.(2008)Takaki, Hirouchi, Hisakuni, Yamanaka, and
  Tomita}]{takaki2008}
\bibinfo{author}{T.~Takaki}, \bibinfo{author}{T.~Hirouchi},
  \bibinfo{author}{Y.~Hisakuni}, \bibinfo{author}{A.~Yamanaka},
  \bibinfo{author}{Y.~Tomita},
\newblock \bibinfo{title}{Multi-phase-field model to simulate microstructure
  evolutions during dynamic recrystallization},
\newblock \bibinfo{journal}{Materials transactions} \bibinfo{volume}{49}
  (\bibinfo{year}{2008}) \bibinfo{pages}{2559--2565}.
\bibitem[{Popova et~al.(2015)Popova, Staraselski, Brahme, Mishra, and
  Inal}]{POPOVA201585}
\bibinfo{author}{E.~Popova}, \bibinfo{author}{Y.~Staraselski},
  \bibinfo{author}{A.~Brahme}, \bibinfo{author}{R.~Mishra},
  \bibinfo{author}{K.~Inal},
\newblock \bibinfo{title}{Coupled crystal plasticity – probabilistic cellular
  automata approach to model dynamic recrystallization in magnesium alloys},
\newblock \bibinfo{journal}{International Journal of Plasticity}
  \bibinfo{volume}{66} (\bibinfo{year}{2015}) \bibinfo{pages}{85 -- 102}.
\bibitem[{Admal et~al.(2017)Admal, Po, and Marian}]{admal:po:marian:2017}
\bibinfo{author}{N.~C. Admal}, \bibinfo{author}{G.~Po},
  \bibinfo{author}{J.~Marian},
\newblock \bibinfo{title}{Diffuse-interface polycrystal plasticity: expressing
  grain boundaries as geometrically necessary dislocations},
\newblock \bibinfo{journal}{Materials Theory} \bibinfo{volume}{1}
  (\bibinfo{year}{2017}) \bibinfo{pages}{6}.
\bibitem[{Lobkovsky and Warren(2001)}]{lobkovsky2001}
\bibinfo{author}{A.~Lobkovsky}, \bibinfo{author}{J.~Warren},
\newblock \bibinfo{title}{{Sharp interface limit of a phase-field model of
  crystal grains}},
\newblock \bibinfo{journal}{Physical Review E} \bibinfo{volume}{63}
  (\bibinfo{year}{2001}) \bibinfo{pages}{051605}.
\bibitem[{Kobayashi and Giga(1999)}]{kobayashi1999}
\bibinfo{author}{R.~Kobayashi}, \bibinfo{author}{Y.~Giga},
\newblock \bibinfo{title}{Equations with singular diffusivity},
\newblock \bibinfo{journal}{Journal of Statistical Physics}
  \bibinfo{volume}{95} (\bibinfo{year}{1999}) \bibinfo{pages}{1187--1220}.
\bibitem[{Lee(1969)}]{lee1969elastic}
\bibinfo{author}{E.~H. Lee},
\newblock \bibinfo{title}{Elastic-plastic deformation at finite strains},
\newblock \bibinfo{organization}{ASME}, \bibinfo{year}{1969}.
\bibitem[{Kr{\"o}ner(1959)}]{kroner1959allgemeine}
\bibinfo{author}{E.~Kr{\"o}ner},
\newblock \bibinfo{title}{Allgemeine kontinuumstheorie der versetzungen und
  eigenspannungen},
\newblock \bibinfo{journal}{Archive for Rational Mechanics and Analysis}
  \bibinfo{volume}{4} (\bibinfo{year}{1959}) \bibinfo{pages}{273--334}.
\bibitem[{Reina and Conti(2014)}]{reina2014kinematic}
\bibinfo{author}{C.~Reina}, \bibinfo{author}{S.~Conti},
\newblock \bibinfo{title}{Kinematic description of crystal plasticity in the
  finite kinematic framework: a micromechanical understanding of f= f e f p},
\newblock \bibinfo{journal}{Journal of the Mechanics and Physics of Solids}
  \bibinfo{volume}{67} (\bibinfo{year}{2014}) \bibinfo{pages}{40--61}.
\bibitem[{Acharya(2008)}]{acharya2008counterpoint}
\bibinfo{author}{A.~Acharya},
\newblock \bibinfo{title}{A counterpoint to cermelli and gurtin’s criteria
  for choosing the ‘correct’geometric dislocation tensor in finite
  plasticity},
\newblock in: \bibinfo{booktitle}{IUTAM Symposium on theoretical, computational
  and modelling aspects of inelastic media}, \bibinfo{organization}{Springer},
  \bibinfo{year}{2008}, pp. \bibinfo{pages}{99--105}.
\bibitem[{Berdichevsky(2006)}]{berdichevsky2006continuum}
\bibinfo{author}{V.~L. Berdichevsky},
\newblock \bibinfo{title}{Continuum theory of dislocations revisited},
\newblock \bibinfo{journal}{Continuum Mechanics and Thermodynamics}
  \bibinfo{volume}{18} (\bibinfo{year}{2006}) \bibinfo{pages}{195--222}.
\bibitem[{Ciarlet and Geymonat(1982)}]{ciarlet1982}
\bibinfo{author}{P.~G. Ciarlet}, \bibinfo{author}{G.~Geymonat},
\newblock \bibinfo{title}{Sur les lois de comportement en {\'e}lasticit{\'e}
  non lin{\'e}aire compressible},
\newblock \bibinfo{journal}{CR Acad. Sci. Paris S{\'e}r. II}
  \bibinfo{volume}{295} (\bibinfo{year}{1982}) \bibinfo{pages}{423--426}.
\bibitem[{Nye(1953)}]{nye1953some}
\bibinfo{author}{J.~Nye},
\newblock \bibinfo{title}{Some geometrical relations in dislocated crystals},
\newblock \bibinfo{journal}{Acta metallurgica} \bibinfo{volume}{1}
  (\bibinfo{year}{1953}) \bibinfo{pages}{153--162}.
\bibitem[{Kr{\"o}ner et~al.(1981)}]{kroner1981continuum}
\bibinfo{author}{E.~Kr{\"o}ner}, et~al.,
\newblock \bibinfo{title}{Continuum theory of defects},
\newblock \bibinfo{journal}{Physics of defects} \bibinfo{volume}{35}
  (\bibinfo{year}{1981}) \bibinfo{pages}{217--315}.
\bibitem[{Read and Shockley(1950)}]{read1950dislocation}
\bibinfo{author}{W.~T. Read}, \bibinfo{author}{W.~Shockley},
\newblock \bibinfo{title}{Dislocation models of crystal grain boundaries},
\newblock \bibinfo{journal}{Physical review} \bibinfo{volume}{78}
  (\bibinfo{year}{1950}) \bibinfo{pages}{275}.
\bibitem[{Cahn et~al.(2006)Cahn, Mishin, and Suzuki}]{cahn2006coupling}
\bibinfo{author}{J.~W. Cahn}, \bibinfo{author}{Y.~Mishin},
  \bibinfo{author}{A.~Suzuki},
\newblock \bibinfo{title}{Coupling grain boundary motion to shear deformation},
\newblock \bibinfo{journal}{Acta materialia} \bibinfo{volume}{54}
  (\bibinfo{year}{2006}) \bibinfo{pages}{4953--4975}.
\bibitem[{Kumar et~al.(2002)Kumar, Schwartz, and
  King}]{kumar2002microstructural}
\bibinfo{author}{M.~Kumar}, \bibinfo{author}{A.~J. Schwartz},
  \bibinfo{author}{W.~E. King},
\newblock \bibinfo{title}{Microstructural evolution during grain boundary
  engineering of low to medium stacking fault energy fcc materials},
\newblock \bibinfo{journal}{Acta materialia} \bibinfo{volume}{50}
  (\bibinfo{year}{2002}) \bibinfo{pages}{2599--2612}.
\bibitem[{McDowell et~al.(2007)McDowell, Choi, Panchal, Austin, Allen, and
  Mistree}]{mcdowell2007plasticity}
\bibinfo{author}{D.~L. McDowell}, \bibinfo{author}{H.~J. Choi},
  \bibinfo{author}{J.~Panchal}, \bibinfo{author}{R.~Austin},
  \bibinfo{author}{J.~Allen}, \bibinfo{author}{F.~Mistree},
\newblock \bibinfo{title}{Plasticity-related microstructure-property relations
  for materials design},
\newblock in: \bibinfo{booktitle}{Key Engineering Materials}, volume
  \bibinfo{volume}{340}, \bibinfo{organization}{Trans Tech Publ},
  \bibinfo{year}{2007}, pp. \bibinfo{pages}{21--30}.
\bibitem[{McDowell and Olson(2008)}]{mcdowell2008concurrent}
\bibinfo{author}{D.~L. McDowell}, \bibinfo{author}{G.~Olson},
\newblock \bibinfo{title}{Concurrent design of hierarchical materials and
  structures},
\newblock in: \bibinfo{booktitle}{Scientific Modeling and Simulations},
  \bibinfo{publisher}{Springer}, \bibinfo{year}{2008}, pp.
  \bibinfo{pages}{207--240}.
\bibitem[{McDowell(2010)}]{MCDOWELL20101280}
\bibinfo{author}{D.~L. McDowell},
\newblock \bibinfo{title}{A perspective on trends in multiscale plasticity},
\newblock \bibinfo{journal}{International Journal of Plasticity}
  \bibinfo{volume}{26} (\bibinfo{year}{2010}) \bibinfo{pages}{1280 -- 1309}.
  \bibinfo{note}{Special Issue In Honor of David L. McDowell}.
\bibitem[{Bulatov and Cai(2006)}]{bulatov2006computer}
\bibinfo{author}{V.~Bulatov}, \bibinfo{author}{W.~Cai},
  \bibinfo{title}{Computer simulations of dislocations},
  volume~\bibinfo{volume}{3}, \bibinfo{publisher}{Oxford University Press on
  Demand}, \bibinfo{year}{2006}.
\bibitem[{Gehlen(2012)}]{gehlen2012interatomic}
\bibinfo{author}{P.~Gehlen}, \bibinfo{title}{Interatomic potentials and
  simulation of lattice defects}, \bibinfo{publisher}{Springer Science \&
  Business Media}, \bibinfo{year}{2012}.
\bibitem[{Shenderova et~al.(1998)Shenderova, Brenner, Nazarov, Romanov, and
  Yang}]{shenderova1998multiscale}
\bibinfo{author}{O.~Shenderova}, \bibinfo{author}{D.~Brenner},
  \bibinfo{author}{A.~Nazarov}, \bibinfo{author}{A.~Romanov},
  \bibinfo{author}{L.~Yang},
\newblock \bibinfo{title}{Multiscale modeling approach for calculating
  grain-boundary energies from first principles},
\newblock \bibinfo{journal}{Physical Review B} \bibinfo{volume}{57}
  (\bibinfo{year}{1998}) \bibinfo{pages}{R3181}.
\bibitem[{Tschopp and McDowell(2008)}]{TSCHOPP2008191}
\bibinfo{author}{M.~Tschopp}, \bibinfo{author}{D.~McDowell},
\newblock \bibinfo{title}{Dislocation nucleation in Σ3 asymmetric tilt grain
  boundaries},
\newblock \bibinfo{journal}{International Journal of Plasticity}
  \bibinfo{volume}{24} (\bibinfo{year}{2008}) \bibinfo{pages}{191 -- 217}.
\bibitem[{Osetsky and Bacon(2003)}]{osetsky2003atomic}
\bibinfo{author}{Y.~N. Osetsky}, \bibinfo{author}{D.~J. Bacon},
\newblock \bibinfo{title}{An atomic-level model for studying the dynamics of
  edge dislocations in metals},
\newblock \bibinfo{journal}{Modelling and simulation in materials science and
  engineering} \bibinfo{volume}{11} (\bibinfo{year}{2003})
  \bibinfo{pages}{427}.
\bibitem[{Erhart et~al.(2013)Erhart, Marian, and
  Sadigh}]{erhart2013thermodynamic}
\bibinfo{author}{P.~Erhart}, \bibinfo{author}{J.~Marian},
  \bibinfo{author}{B.~Sadigh},
\newblock \bibinfo{title}{Thermodynamic and mechanical properties of copper
  precipitates in $\alpha$-iron from atomistic simulations},
\newblock \bibinfo{journal}{Physical Review B} \bibinfo{volume}{88}
  (\bibinfo{year}{2013}) \bibinfo{pages}{024116}.
\bibitem[{Tasan et~al.(2015)Tasan, Diehl, Yan, Bechtold, Roters, Schemmann,
  Zheng, Peranio, Ponge, Koyama et~al.}]{tasan2015overview}
\bibinfo{author}{C.~C. Tasan}, \bibinfo{author}{M.~Diehl},
  \bibinfo{author}{D.~Yan}, \bibinfo{author}{M.~Bechtold},
  \bibinfo{author}{F.~Roters}, \bibinfo{author}{L.~Schemmann},
  \bibinfo{author}{C.~Zheng}, \bibinfo{author}{N.~Peranio},
  \bibinfo{author}{D.~Ponge}, \bibinfo{author}{M.~Koyama}, et~al.,
\newblock \bibinfo{title}{An overview of dual-phase steels: advances in
  microstructure-oriented processing and micromechanically guided design},
\newblock \bibinfo{journal}{Annual Review of Materials Research}
  \bibinfo{volume}{45} (\bibinfo{year}{2015}) \bibinfo{pages}{391--431}.
\bibitem[{Cahn et~al.(2006)Cahn, Mishin, and Suzuki}]{cahn2006duality}
\bibinfo{author}{J.~W. Cahn}, \bibinfo{author}{Y.~Mishin},
  \bibinfo{author}{A.~Suzuki},
\newblock \bibinfo{title}{Duality of dislocation content of grain boundaries},
\newblock \bibinfo{journal}{Philosophical Magazine} \bibinfo{volume}{86}
  (\bibinfo{year}{2006}) \bibinfo{pages}{3965--3980}.
\bibitem[{Suzuki and Mishin(2005)}]{suzuki2005atomic}
\bibinfo{author}{A.~Suzuki}, \bibinfo{author}{Y.~M. Mishin},
\newblock \bibinfo{title}{Atomic mechanisms of grain boundary motion},
\newblock in: \bibinfo{booktitle}{Materials Science Forum}, volume
  \bibinfo{volume}{502}, \bibinfo{organization}{Trans Tech Publ},
  \bibinfo{year}{2005}, pp. \bibinfo{pages}{157--162}.
\bibitem[{Ivanov and Mishin(2008)}]{ivanov2008dynamics}
\bibinfo{author}{V.~Ivanov}, \bibinfo{author}{Y.~Mishin},
\newblock \bibinfo{title}{Dynamics of grain boundary motion coupled to shear
  deformation: An analytical model and its verification by molecular dynamics},
\newblock \bibinfo{journal}{Physical Review B} \bibinfo{volume}{78}
  (\bibinfo{year}{2008}) \bibinfo{pages}{064106}.
\bibitem[{Trautt and Mishin(2012)}]{trautt2012grain}
\bibinfo{author}{Z.~Trautt}, \bibinfo{author}{Y.~Mishin},
\newblock \bibinfo{title}{Grain boundary migration and grain rotation studied
  by molecular dynamics},
\newblock \bibinfo{journal}{Acta Materialia} \bibinfo{volume}{60}
  (\bibinfo{year}{2012}) \bibinfo{pages}{2407--2424}.
\bibitem[{Trautt et~al.(2012)Trautt, Adland, Karma, and
  Mishin}]{trautt2012coupled}
\bibinfo{author}{Z.~Trautt}, \bibinfo{author}{A.~Adland},
  \bibinfo{author}{A.~Karma}, \bibinfo{author}{Y.~Mishin},
\newblock \bibinfo{title}{Coupled motion of asymmetrical tilt grain boundaries:
  Molecular dynamics and phase field crystal simulations},
\newblock \bibinfo{journal}{Acta Materialia} \bibinfo{volume}{60}
  (\bibinfo{year}{2012}) \bibinfo{pages}{6528--6546}.
\bibitem[{Raj and Ashby(1971)}]{raj1971grain}
\bibinfo{author}{R.~Raj}, \bibinfo{author}{M.~Ashby},
\newblock \bibinfo{title}{On grain boundary sliding and diffusional creep},
\newblock \bibinfo{journal}{Metallurgical and Materials Transactions B}
  \bibinfo{volume}{2} (\bibinfo{year}{1971}) \bibinfo{pages}{1113--1127}.
\bibitem[{Gifkins(1976)}]{gifkins1976grain}
\bibinfo{author}{R.~Gifkins},
\newblock \bibinfo{title}{Grain-boundary sliding and its accommodation during
  creep and superplasticity},
\newblock \bibinfo{journal}{Metallurgical and Materials Transactions A}
  \bibinfo{volume}{7} (\bibinfo{year}{1976}) \bibinfo{pages}{1225--1232}.
\bibitem[{Van~Swygenhoven and Derlet(2001)}]{van2001grain}
\bibinfo{author}{H.~Van~Swygenhoven}, \bibinfo{author}{P.~Derlet},
\newblock \bibinfo{title}{Grain-boundary sliding in nanocrystalline fcc
  metals},
\newblock \bibinfo{journal}{Physical Review B} \bibinfo{volume}{64}
  (\bibinfo{year}{2001}) \bibinfo{pages}{224105}.
\bibitem[{Margulies et~al.(2001)Margulies, Winther, and
  Poulsen}]{margulies2001situ}
\bibinfo{author}{L.~Margulies}, \bibinfo{author}{G.~Winther},
  \bibinfo{author}{H.~Poulsen},
\newblock \bibinfo{title}{In situ measurement of grain rotation during
  deformation of polycrystals},
\newblock \bibinfo{journal}{Science} \bibinfo{volume}{291}
  (\bibinfo{year}{2001}) \bibinfo{pages}{2392--2394}.
\bibitem[{Gottstein and Shvindlerman(2009)}]{gottstein2009grain}
\bibinfo{author}{G.~Gottstein}, \bibinfo{author}{L.~S. Shvindlerman},
  \bibinfo{title}{Grain boundary migration in metals: thermodynamics, kinetics,
  applications}, \bibinfo{publisher}{CRC press}, \bibinfo{year}{2009}.
\bibitem[{Mughrabi(1983)}]{mughrabi1983dislocation}
\bibinfo{author}{H.~Mughrabi},
\newblock \bibinfo{title}{Dislocation wall and cell structures and long-range
  internal stresses in deformed metal crystals},
\newblock \bibinfo{journal}{Acta metallurgica} \bibinfo{volume}{31}
  (\bibinfo{year}{1983}) \bibinfo{pages}{1367--1379}.
\bibitem[{Bay et~al.(1992)Bay, Hansen, Hughes, and
  Kuhlmann-Wilsdorf}]{bay1992overview}
\bibinfo{author}{B.~Bay}, \bibinfo{author}{N.~Hansen},
  \bibinfo{author}{D.~Hughes}, \bibinfo{author}{D.~Kuhlmann-Wilsdorf},
\newblock \bibinfo{title}{Overview no. 96 evolution of fcc deformation
  structures in polyslip},
\newblock \bibinfo{journal}{Acta metallurgica et materialia}
  \bibinfo{volume}{40} (\bibinfo{year}{1992}) \bibinfo{pages}{205--219}.
\bibitem[{Sedl{\'a}{\v{c}}ek et~al.(2002)Sedl{\'a}{\v{c}}ek, Blum, Kratochvil,
  and Forest}]{sedlavcek2002subgrain}
\bibinfo{author}{R.~Sedl{\'a}{\v{c}}ek}, \bibinfo{author}{W.~Blum},
  \bibinfo{author}{J.~Kratochvil}, \bibinfo{author}{S.~Forest},
\newblock \bibinfo{title}{Subgrain formation during deformation: physical
  origin and consequences},
\newblock \bibinfo{journal}{Metallurgical and Materials Transactions A}
  \bibinfo{volume}{33} (\bibinfo{year}{2002}) \bibinfo{pages}{319--327}.
\bibitem[{Xia and El-Azab(2015)}]{xia2015computational}
\bibinfo{author}{S.~Xia}, \bibinfo{author}{A.~El-Azab},
\newblock \bibinfo{title}{Computational modelling of mesoscale dislocation
  patterning and plastic deformation of single crystals},
\newblock \bibinfo{journal}{Modelling and Simulation in Materials Science and
  Engineering} \bibinfo{volume}{23} (\bibinfo{year}{2015})
  \bibinfo{pages}{055009}.
\bibitem[{Cermelli and Gurtin(2002)}]{cermelli2002geometrically}
\bibinfo{author}{P.~Cermelli}, \bibinfo{author}{M.~E. Gurtin},
\newblock \bibinfo{title}{Geometrically necessary dislocations in viscoplastic
  single crystals and bicrystals undergoing small deformations},
\newblock \bibinfo{journal}{International Journal of Solids and Structures}
  \bibinfo{volume}{39} (\bibinfo{year}{2002}) \bibinfo{pages}{6281--6309}.

\end{thebibliography}

\end{document}